\documentclass[12pt]{article}
\usepackage{scicite}

\newenvironment{sciabstract}{%
\begin{quote} \bf}
{\end{quote}}

\usepackage{pdfpages}
\usepackage{amsmath} 
\usepackage{amsthm} 
\usepackage{amssymb}	
\usepackage{graphicx}
\usepackage{enumerate}
\usepackage{wrapfig}
\usepackage[dvips,letterpaper,margin=1in,bottom=1in]{geometry}
\usepackage{color}
\usepackage{todonotes}
\usepackage{subcaption}
\usepackage{setspace}
\usepackage{comment}
\usepackage{cancel}
\usepackage[]{units}
\usepackage{hyperref}
\usepackage[all]{hypcap}
\usepackage{fancyhdr}
\makeatletter 
\renewcommand{\v}[1]{\ensuremath{\boldsymbol{#1}}} 

\newcommand{\abs}[1]{\left| #1 \right|} 
\newcommand{\pd}[2]{\frac{\partial #1}{\partial #2}}

\newcommand{\fd}[2]{\frac{\delta #1}{\delta #2}} 
\newcommand{\grad}{\nabla} 
\newcommand{\tr}{\mbox{tr}\,}
\let\baraccent=\= 
\renewcommand{\=}[1]{\stackrel{#1}{=}} 
\renewcommand{\(}{\left(}
\renewcommand{\)}{\right)}
\renewcommand{\[}{\left[}
\renewcommand{\]}{\right]}
\newcommand{\<}{\left<}
\renewcommand{\>}{\right>}

\theoremstyle{definition}

\theoremstyle{remark}

\newtheorem{?}{\textbf{Question}}
\newcommand{\e}[1]{\mbox{e}^{#1}} 
\renewcommand{\exp}[1]{\mbox{exp}\[#1\]} 

\newcommand{\bigO}[1]{O\(#1\)}  

\begin{document}
\title{Low rattling: a predictive principle for self-organization \\in active collectives}
\author{Pavel Chvykov,$^{1}$ Thomas A. Berrueta,$^{2}$ Akash Vardhan,$^{3}$\\ William Savoie,$^{3}$ Alexander Samland,$^{2}$ Todd D. Murphey,$^{2}$\\ Kurt Wiesenfeld,$^{3}$ Daniel I. Goldman,$^{3}$ Jeremy L. England$^{3,4\ast}$\\
\normalsize{$^{1}$Physics of Living Systems, Massachusetts Institute of Technology,}\\ \normalsize{Cambridge, Massachusetts 02139, USA}\\
\normalsize{$^{2}$Department of Mechanical Engineering, Northwestern University,}\\ \normalsize{Evanston, Illinois 60208, USA}\\
\normalsize{$^{3}$School of Physics, Georgia Institute of Technology, Atlanta, Georgia 30332, USA}\\
\normalsize{$^{4}$GlaxoSmithKline AI/ML, 200 Cambridgepark Drive, Cambridge, Massachusetts 02140, USA}\\
\normalsize{$^\ast$To whom correspondence should be addressed; E-mail: j@englandlab.com}
}
\date{}
\maketitle
\baselineskip24pt
\begin{sciabstract}
Self-organization is frequently observed in active collectives, from ant rafts to molecular motor assemblies. General principles describing self-organization away from equilibrium have been challenging to identify.  We offer a unifying framework that models the behavior of complex systems as largely random, while capturing their configuration-dependent response to external forcing. This allows derivation of a Boltzmann-like principle for understanding and manipulating driven self-organization. We validate our predictions experimentally in shape-changing robotic active matter, and outline a methodology for controlling collective behavior. Our findings highlight how emergent order depends sensitively on the matching between external patterns of forcing and internal dynamical response properties, pointing towards future approaches for design and control of active particle mixtures and metamaterials.
\end{sciabstract}
\pagebreak

Self-organization in nature is surprising because getting a large group of separate particles to act in an organized way is often difficult. By definition, arrangements of matter we call ``orderly'' are special, making up a tiny minority of all allowed configurations. For example, we find each unique, symmetrical shape of a snowflake visually striking, in contrast with any randomly-rearranged clump of the same water molecules. Thus, any theory of emergent order in many-particle collectives must explain how a small subset of configurations are spontaneously selected among the vast set of disorganized arrangements.

Spontaneous many-body order is well-understood in thermal equilibrium cases such as crystalline solids or DNA origami~\cite{Rothemund2006}, where the assembling matter is allowed to sit unperturbed for a long time at constant temperature $T$. The statistical mechanical approach proceeds by approximating the complex deterministic dynamics of the particles with a probabilistic ``molecular chaos,'' positing that the law of conservation of energy governs otherwise random behavior~\cite{kardar2007particles}. What follows is the Boltzmann distribution for the steady-state probabilities, $p_{ss}(\v{q})\propto \exp{-E(\v{q})/T}$, which shows that the degree to which special configurations $\v{q}$ of low energy $E(\v{q})$ have a high probability $p_{ss}(\v{q})$ in the long-term depends on the amplitude of the thermal noise. Orderly configurations can assemble and remain stable, so long as inter-particle attractions are strong enough to overcome the randomizing effects of thermal fluctuations.

However, there are also many examples of emergent order outside of thermal equilibrium. From ``random organization" in sheared colloids~\cite{corte2008chaikin_spun_colloids}, to phase separation in multi-temperature particle mixtures~\cite{grosberg2015phase-sep_2T}, and dynamic vortices in protein filaments~\cite{Sumino2012}, a variety of ordered behaviors arise far from equilibrium that cannot be explained in terms of simple inter-particle attraction or energy gradients~\cite{ramaswamy2010actMatt_review, bertini2015macroFluctTheory, paoluzzi16critPhenActMatt, speck16neqThermo_activeMatter}.

In all of these examples, the energy flux from external sources allows different system configurations to experience fluctuations of different magnitude~\cite{Chv18least_ratt, cates15Teff_motilityPhaseSepar}. We suggest that the emergence of such configuration-dependent fluctuations, which cannot happen in equilibrium, may be key to understanding many nonequilibrium self-organization phenomena. In particular, we introduce a measure of driving-induced random fluctuations, which we term rattling $\mathcal{R}(\v{q})$, and argue that it could play a similar role in many far-from-equilibrium systems as energy does in equilibrium. We test this in a number of systems, including a flexible active matter system of simple robots we call ``smarticles'' (smart active particles)~\cite{savoie_thesis} as a convenient test-platform~(see movie~S1) inspired by similar robo-physical emulators of collective behavior~\cite{Aguilar2018,rubenstein2014swarm1000,Werfel2014}. Despite their purely repulsive inter-robot interactions, we find that smarticles spontaneously self-organize into collective ``dances,'' whose shape and motions are matched to the temporal pattern of external driving forces~(see movies~S2~and~S3). This platform and others~\cite{Li2019,Vicsek2018,Mayya2019nonuniform}, including the nonequilibrium ordering examples mentioned above, all exhibit low-rattling ordered behaviors that echo low-energy structures emergent at equilibrium. We thus motivate and test a predictive theory based on rattling that may explain a broad class of nonequilibrium ordering phenomena.

In devising our approach, we take inspiration from the phenomenon of thermophoresis, which is the simplest example of purely nonequilibrium self-organization, and is characterized by the diffusion of colloidal particles from hot regions to cold regions~\cite{Duhr2006}. If non-interacting particles in a viscous fluid are subject to a temperature $T(\v{q})$ that varies over position $\v{q}$, their resulting density in the steady-state $p_{ss}(\v{q})$ will concentrate in the regions of low temperature. Particles diffuse to regions where thermal noise is weaker and become trapped there.  With the diffusivity landscape set by thermal noise locally according to the fluctuation dissipation relation $D(\v{q}) \propto T(\v{q})$~\cite{van1992stochastic}, the steady-state diffusion equation $\grad^2\left(D(\v{q})p_{ss}(\v{q})\right) = 0$ is satisfied by the probability density $p_{ss}(\v{q})\propto 1/D(\v{q})$. Hence, a low-entropy, ``ordered'' arrangement of particles can be stable when the diffusivity landscape has a few locations $\v{q}$ that are strongly selected by their extremely low $D(\v{q})$ values.

\begin{figure*}[ht]
	\includegraphics[width=1.\textwidth]{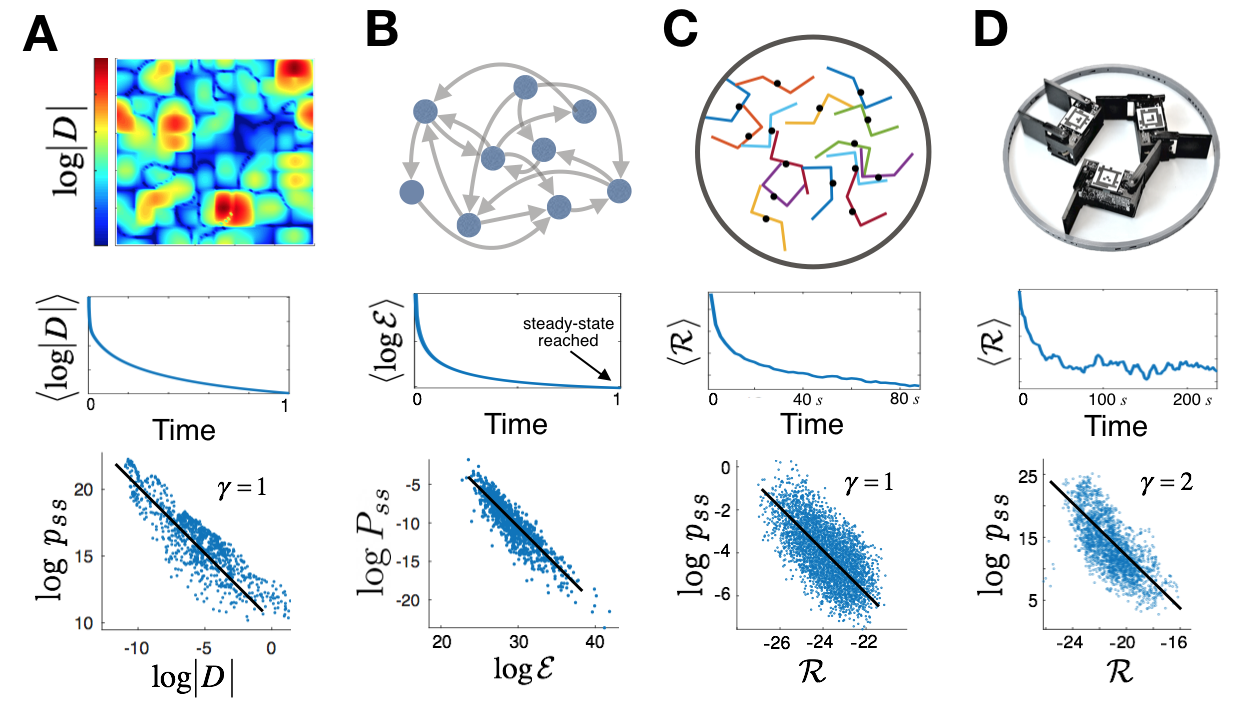}
	\caption{\textbf{Rattling $ \mathcal{R} $ is predictive of steady-state likelihood across far-from-equilibrium systems.}  (\textbf{A})~shows inhomogeneous anisotropic diffusion in 2D, where the steady-state density $p_{ss}(\v{q})$ is seen to be approximately given by the magnitude of local fluctuations $\log \abs{D(\v{q})} \propto \mathcal{R}(\v{q})$  ($ \abs{D} $---determinant of the diffusion tensor). (\textbf{B})~shows a random walk on a large random graph (1000 states), where $P_{ss}$---the probability at a state---is approximately given by $\mathcal{E}$---that state's exit rate. (\textbf{C})~shows an active matter system of shape-changing agents: an enclosed ensemble of 15 ``smarticles'' in simulation. (\textbf{D})~realizes similar agents experimentally with an enclosed three-robot smarticle ensemble. The middle row shows that relaxation to the steady-state of a uniform initial distribution is accompanied by monotonic decay in the average rattling value in all cases---analogous to free energy in equilibrium systems. The bottom row shows the validity of the nonequilibrium Boltzmann-like principle in Eq.~\ref{eq:p_ss}, where the black lines in (A, B, and C) illustrate the theoretical correlation slope for a sufficiently large and complex system (see supplementary materials). The mesoscopic regime in~(D) provides the most stringent test of rattling theory (where we observe deviations in $\gamma$ from 1), while also exhibiting global self-organization. In (A and B, middle) time units are arbitrary, and for (C and D, middle) time is in seconds, where the drive period is $2$~s.
	}
	\label{fig:systems}
\end{figure*}

We seek to extend this intuition to explain nonequilibrium self-organization more broadly. However, a straightforward mathematical extension of the idea encounters challenges in only slightly more complicated scenarios. For an arbitrary diffusion tensor landscape $\v{D}(\v{q})$, in which diffusivity can depend on the direction of motion, one can no longer find general solutions for the steady-state. Moreover, the steady-state density $p_{ss}(\v{q})$ at configuration $\v{q}$ may depend on the diffusivity $\v{D}(\tilde{\v{q}})$ at arbitrarily distant configurations $\tilde{\v{q}}$. Nonetheless, we suggest that for most typical diffusion landscapes, the local magnitude of fluctuations $|\v{D}(\v{q})|$ should statistically bias $p_{ss}(\v{q})$, and hence be approximately predictive of it. This insight, which is central to our theory, is illustrated to hold numerically in Fig.~\ref{fig:systems}A for a randomly constructed two-dimensional anisotropic landscape, and in fig.~S3 for higher dimensions. While contrived counterexamples which break the relationship may be constructed, they require specific fine-tuning (see fig.~S4).

The key assumption underlying our approach is that the complex system dynamics are so messy that only the amplitude of local drive-induced fluctuations governs the otherwise random behavior---an assumption inspired by molecular chaos at equilibrium.  We expect this to apply when the system dynamics are so complex, nonlinear, and high-dimensional that no global symmetry or constraint can be found for its simplification. Although one cannot predict a configuration's nonequilibrium steady-state probability from its local properties in the general case~\cite{Landauer1993, landauer75blowtorch}, the feat becomes achievable in practice for ``messy" systems. To illustrate this point explicitly, we consider a discrete dynamical system with random transition rates between a large number of states. Here, we can show analytically that the net rate at which we exit any given state predicts its long-term probability approximately, even though the exact result requires global system knowledge (see Fig.~\ref{fig:systems}B and supplementary materials for derivation). This result may be related to the above discussion of thermophoresis by noting that the discrete state exit rates are determined by the continuum diffusivity if our dynamics are built by discretizing the domain of a diffusion process.

To formulate our random dynamics assumption explicitly, we represent the complex system evolution as a trajectory in time $\v{q}(t)$, where the configuration vector $\v{q}$ captures the properties of the entire many-particle system. Our messiness assumption amounts to approximating the full complex dynamics between two points $\v{q}(t)$ and $\v{q}(t+\delta t)$ by a random diffusion process. To this end, we take the amplitude of the noise fluctuations $D(\v{q})$ to locally reflect the amplitude of the true configuration dynamics: $|\v{q}(t+\delta t)-\v{q}(t)|^2 \propto D(\v{q}) \delta t$ for short rollouts $\v{q}(t\rightarrow t+\delta t)$ (i.e., samples of system trajectories) of duration $\delta t$ initialized in configuration $\v{q}(t)=\v{q}$ (see supplementary materials for details). Through this approximation, our dynamics are effectively reduced to diffusion in $\v{q}$-space, which then allows us to locally estimate the steady-state probability of system configurations from $D(\v{q})$ as in thermophoresis. Hence, the global steady-state distribution may be predicted from the properties of short-time, local system rollouts.

For rare orderly configurations to be strongly selected in a messy dynamical system, the landscape of local fluctuations must vary in magnitude over a large range of values. While in thermophoresis these fluctuations are directly imposed by an external temperature profile, in driven dynamical systems the effect results from the way a given pattern of driving can affect system configurations differently. The $D(\v{q})$ landscape is emergent from the interplay between the pattern of driving, and the library of possible $\v{q}$-dependent system response properties. In practice, we observe that the amplitudes of system responses to driving do often vary over several orders of magnitude (see Fig.~\ref{fig:systems}). We see this phenomenology in many well-known examples of active matter self-organization~\cite{redner2013phase-sep_colloids,cates15Teff_motilityPhaseSepar,corte2008chaikin_spun_colloids}. For example, the crystals that form in suspensions of self-propelled colloids in~\cite{palacci936living_crystals} may be seen as the collective configurations that respond least diffusively to driving by precisely balancing the propulsive forces among individual particles. This illustrates how the low $D(\v{q})$ configurations are selected in the steady-state by an exceptional matching of their response properties to the way the system is driven.

We apply these ideas in real complex driven systems whose response to driving we cannot predict analytically, such as our robotic swarm of smarticles. In this case, we require an estimator for the local value of $D(\v{q})$ based on observations of short rollouts of system behavior. The estimator of the local diffusion tensor that we choose here is the covariance matrix~\cite{michalet2012singTrajDlim}
\begin{equation}
    \label{eq:covariance}
    \mathcal{C}(\v{q}) = \mbox{cov}[\tilde{\v{v}}_{\v{q}},\tilde{\v{v}}_{\v{q}}]
\end{equation}
where $ \tilde{\v{v}}_{\v{q}} $ is seen as a random variable with samples drawn from $\{(\tilde{\v{q}}(t)-\tilde{\v{q}}(0))/\sqrt{t}\}_{\tilde{\v{q}}(0)=\v{q}} $ at various time-points $ t $ along one or several short system trajectories $\tilde{\v{q}}(t)$ rolled out from $ \tilde{\v{q}}(0)=\v{q} $. We assume these rollouts $\tilde{\v{q}}(t)$ to be long enough to capture fluctuations in the configuration variables under the influence of a drive, but short enough to have $\tilde{\v{q}}(t)$ stay near ${\v{q}}$ (see supplementary materials for details).

While the covariance matrix reflects the amplitude of local fluctuations, in estimating effective diffusivity we are instead interested in a measure of their disorder. This follows from the observation that high-amplitude ordered oscillations do not contribute to the rate of stochastic diffusion~\cite{Chv18least_ratt}. We suggest that the degree of disorder of fluctuations may be captured by the entropy of the distribution of $\tilde{\v{v}}_{\v{q}}$ vectors, which is how we define ``rattling'' $\mathcal{R}(\v{q})$. Physically, vectors $\tilde{\v{v}}_{\v{q}}$ capture the statistics of the force fluctuations experienced in configuration $\v{q}$, and so rattling measures the disorder in the system's driven response properties at that point. By approximating the distribution of $\tilde{\v{v}}_{\v{q}}$ as Gaussian, we can express its entropy (up to a constant offset) simply in terms of $\mathcal{C}(\v{q})$ as:
\begin{equation}
    \label{eq:rattling}
    \mathcal{R}(\v{q})=\frac{1}{2}\text{log}\ \text{det}\ \mathcal{C}(\v{q}).
\end{equation}

With this definition, we generalize the thermophoretic expression for the steady-state density $p_{ss}(\v{q})\propto 1/D(\v{q})$ and express it in a Boltzmann-like form:
\begin{equation}
    \label{eq:p_ss}
    p_{ss}(\v{q})\propto e^{-\gamma \mathcal{R}(\v{q})},
\end{equation}
where $\gamma$ is a system-specific constant of order 1 (see supplementary materials for derivations). We note that when energy varies on the same scale as rattling, the interaction between the two landscapes can generate strong steady-state currents and may break this relation~\cite{Chv18least_ratt}. This way, using rattling we are able predict the long-term global steady-state distribution based on empirical measurements of short-term local system behavior, which suggests that probability density accumulates over time in low-rattling configurations.

\begin{figure*}[ht]
	\includegraphics[width=1.05\textwidth]{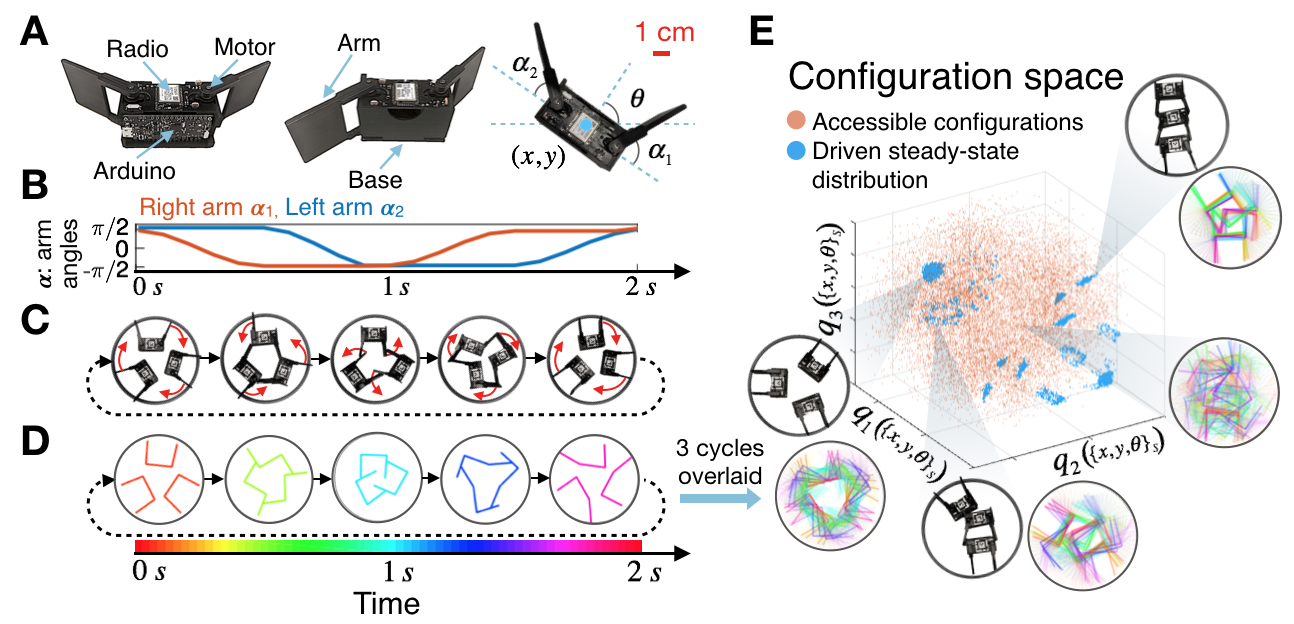}
	\caption{\textbf{Self-organization in a smarticle robotic ensemble}. (\textbf{A})~Front, back and top view of a single smarticle. Of its five degrees of freedom, we consider the time-varying arm angles $(\alpha_1,\alpha_2)$ as ``external'' driving, since these are controlled by a pre-programmed microcontroller, while the robot coordinates $(x,y,\theta)$ are seen as ``internal'' system configuration, since these respond interdependently to the arms. (\textbf{B})~An example periodic arm motion pattern. (\textbf{C})~Top view of three smarticles confined in a fixed ring, all programmed to synchronously execute the driving pattern shown in~(B). The video frames, aligned on the time-axis of (B), show one example of dynamically ordered collective ``dance'' that can spontaneously emerge under this drive (see~(E) and movie~S3 for others). (\textbf{D})~Simulation video, showing agreement with experiment in~(C). We color-code simulated states periodically in time, and overlay them for 3 periods to illustrate the dynamical order over time. (\textbf{E})~shows the system's configuration space, built from nonlinear functions of the three robots' body coordinates $(x,y,\theta)$. The steady-state distribution (blue) illustrates the few ordered configurations that are spontaneously selected by the driving out of all accessible system states (orange). Simulation data shown, see fig.~S5B for experimental data, and fig.~S1 for details on how the configuration space coordinates $ (q_1,q_2,q_3) $ in (E) are constructed from the $3\times(x,y,\theta)$ coordinates.
	}
	\label{fig:smarticle}
\end{figure*}

We study the collective behavior of a simple ensemble of smarticles, aligning ourselves within the tradition of using robotic systems as flexible, physical emulators for self-organizing natural systems~\cite{Li2019,Aguilar2018,rubenstein2014swarm1000,Werfel2014}. Each smarticle (shown in Fig.~\ref{fig:smarticle}A) is composed of three 5.2~cm long links, with two hinges actuated by motors programmed to follow a driving pattern specified by a micro-controller. When a smarticle sits on a flat surface, its arms do not touch the ground, so an individual robot cannot move. However, a group of them can achieve complex motion by pushing and pulling each other (see movie~S1)~\cite{Savoie2019}. The relative coordinates of the middle link of each robot in the ensemble $(x,y,\theta)$ may be thought of as the internal system configurations that dynamically respond to an externally-determined driving force arising from the time-variation of arm angles $(\alpha_1,\alpha_2)$~\cite{materials_and_methods}.

This robotic active matter system offers substantial flexibility in choosing both the programmed patterns of driving, and the properties of internal system dynamics, such as friction coefficients, weights, etc. Additionally, the smarticle system has a flat potential energy landscape, allowing one to focus on the contributions of the drive-induced fluctuations to the collective behavior, making our findings broadly applicable to other strongly-driven systems. When the smarticles are within contact range (as ensured by a confining ring, Fig.~\ref{fig:systems}D), the forces experienced throughout the collective for a given pattern of arm movement are an emergent function of all system coordinates. This configuration-dependent forcing gives rise to varying rattling values, which we refer to as the ``rattling landscape,'' and which we see to be a hallmark property in many far-from-equilibrium examples. The rattling landscape then leads to some system configurations being dynamically selected over others and allowing for self-organization, just as the diffusivity landscape does in thermophoresis. Finally, the combined effects of impulsive inter-robot collisions, nonlinear boundary interactions, and static friction lead to a large degree of quasi-random motion~\cite{Savoie2019}, making this a promising candidate system for exploring our theory.

Reasoning that our fundamental assumption of quasi-random configuration dynamics would be most valid in systems with many degrees of freedom, we also built a simulation that would allow us to study the properties of larger smarticle groups and explore different system parameters~(see fig.~S9). In this regime, we used simulations to gather enough data to sample the high-dimensional probability distributions for our analysis. In a simulation of 15 smarticles, we observed the tendency of the ensemble to reduce average rattling over time after a random initialization. For this 45-dimensional system ($x,y,\theta$ for 15 robots), the configuration-space dynamics are well-approximated by diffusion, and so Eq.~\ref{eq:p_ss} holds, as seen in Fig.~\ref{fig:systems}C. In addition, we noted the emergence of metastable pockets of local order when groups of 3-4 nearby smarticles self-organized into regular motion patterns for several drive cycles (movie~S2). A signature of such dynamical heterogeneity can be seen in the spectrum of the covariance matrix $\mathcal{C}(\v{q})$ from Eq.~\ref{eq:covariance}, as described in supplementary materials and fig.~S10.

The transient appearance of dynamical order in subsets of smarticle collectives raises the question of whether our rattling theory continues to hold for smaller ensembles. For the remainder of the paper we focus on ensembles of three smarticles (as in Fig.~\ref{fig:systems}D), which allows for exhaustive sampling of configurations experimentally, and easier visualization of the configuration space (as in Fig.~\ref{fig:smarticle}E). Both in simulation and experiment, we found that this regime exhibits a variety of low-rattling behaviors that manifest as distinct, orderly collective ``dances'' (movie~S3 and Fig.~\ref{fig:smarticle}, C and D). Despite its small size, this system is well-described by rattling theory, as evidenced by the empirical correlation between rattling and the steady-state likelihood of configurations (see Fig.~\ref{fig:systems}D, bottom).

We consider self-organization as a consequence of a system's landscape of rattling values over configuration space.  This rattling landscape is specific to the particular drive forcing the system out of equilibrium, since different drives will generally produce different dynamical responses in the same system configuration. When the three-smarticle ensemble is driven (under the pattern in Fig.~\ref{fig:smarticle}B), the range of observed rattling values is so large that the lowest-rattling configurations---and consequently highest likelihood---account for most of the steady-state probability mass. Over 99\% of probability accumulates in these spontaneously selected configurations, which represent only 0.1\% of all accessible system states (Fig.~\ref{fig:smarticle}E). Moreover, in these configurations the smarticles exhibit an orderly response to driving (see movie~S4, and Fig.~\ref{fig:smarticle}, C and D). In practice, the ensemble spends most of its time in or nearly in one of several distinct dances, with occasional interruptions by stochastic flights from one such dynamical attractor to another (movie~S5).

From the above observations, we can begin to understand self-organization in driven collectives. In equilibrium, order arises when its entropic cost is outweighed by the available reduction of energy. Analogously, a sufficiently large reduction in rattling can lead to dynamical organization in a driven system.  Moreover, such a reduction can require matching between the system dynamics and the drive pattern.

\begin{figure*}[ht]
	\includegraphics[width=1.05\textwidth]{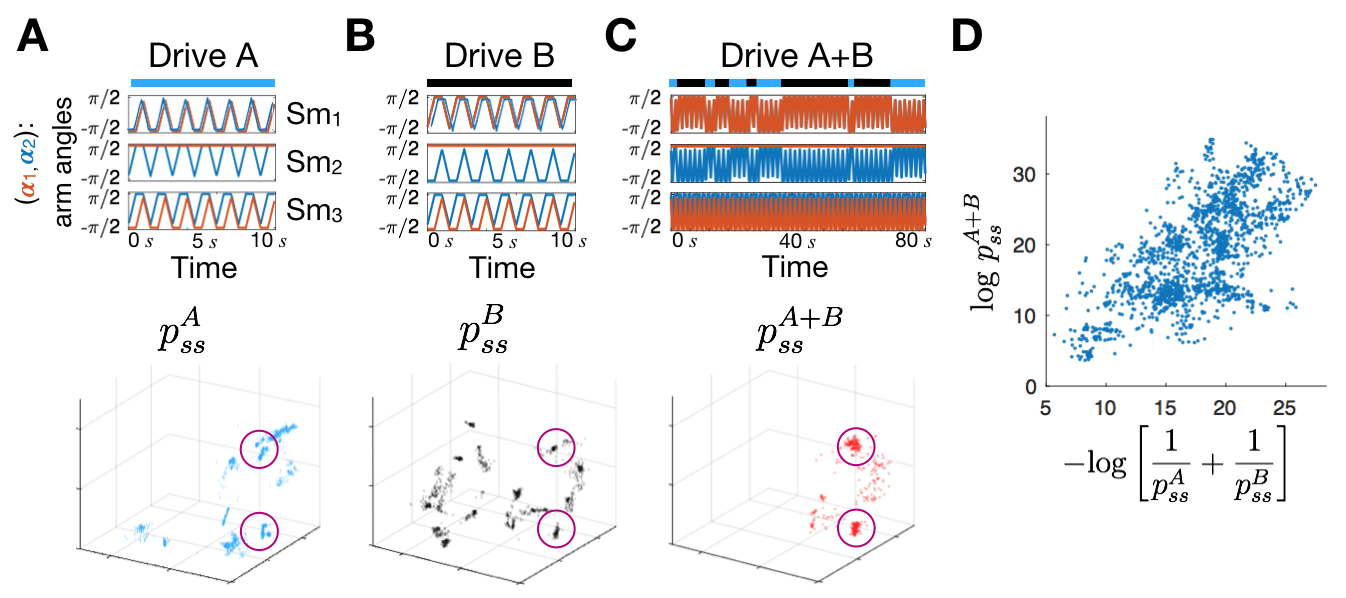}
	\caption{\textbf{Self-organized behaviors are fine-tuned to drive pattern}. (\textbf{A}~and~\textbf{B})~show that changing the arm motion pattern slightly (top) affects which configurations self-organize in the steady-state (bottom, same 3D configuration space as in Fig.~\ref{fig:smarticle}E). (\textbf{C})~By mixing drives A and B as shown (top), we can isolate only those configurations selected in both the steady-states (circled in purple, see movie~S6). This is an analytical prediction of the theory, and (\textbf{D})~further verifies its quantitative formulation. All data shown are experimental, and are reproduced in simulation in fig.~S7, along with derivations in the supplementary materials.
	}
	\label{fig:drive-specific}
\end{figure*}

Through rattling theory we can predict how self-organized states are affected by changes in the features of the drive. We expect the structure of the self-organized dynamical attractors to be specific to the driving pattern, as each drive induces its own rattling landscape. To test this, we programmed the three smarticles with two distinct driving patterns (Fig.~\ref{fig:drive-specific}, A and B, top), which we ran separately. The two resulting steady-state distributions, while each being highly localized to a few configurations, are largely non-overlapping (Fig.~\ref{fig:drive-specific}, A and B, bottom). This indicates that by tuning the drive pattern, it may be possible to design the structure of the resulting steady-state, and hence control the self-organized dynamics (see also~\cite{kedia2019drivespecific,fineberg04surf_waves,Pradillo2019}).

As a proof of principle for such control, we developed a methodology for selecting particular steady-state behaviors by combining drives. By randomly switching back and forth between drives A and B in Fig.~\ref{fig:drive-specific}, we define a compound drive A+B (Fig.~\ref{fig:drive-specific}C and movie~S6). We predicted that this drive would select only those configurations common to both A and B steady-states (Fig.~\ref{fig:drive-specific}, A and B, bottom), since having low rattling under this mixed drive requires having low rattling under both constituent drives. Our experiments confirmed this (Fig.~\ref{fig:drive-specific}C), and we were further able to quantitatively predict the probability that a configuration would appear under the mixed drive based on its likelihood in each constituent steady-state according to
\begin{equation}
    1/p_{ss}^{A+B} \propto 1/p_{ss}^{A} + 1/p_{ss}^{B},
    \label{eq:mix_drive}
\end{equation}
as shown in Fig.~\ref{fig:drive-specific}D and fig.~S7 (see supplementary materials for derivation). This simple relationship suggests that by composing different drives in time, one can single out desired configurations for the system steady-state.

Moreover, we show that we can analytically predict and control the degree of order in the system by tuning drive randomness (in Fig.~\ref{fig:melt}), as well as internal system friction (in supplementary materials, see movie~S7 and fig.~S8). As driven self-organization arises when the system has access to a broad range of rattling values, tuning it requires modulating the rattling of the most ordered behaviors relative to the background high-rattling states.

We can directly manipulate the rattling landscape by modulating the entropy of the drive pattern. This is done by introducing a probabilistic element to the programmed arm motion. At each move, we introduce a probability of making a random arm movement not included in the prescribed drive pattern. Increasing this probability results in flattening the rattling landscape: ordered states experience an increase in rattling due to drive entropy, while states whose rattling is already high do not (see Fig.~\ref{fig:melt}A). Correspondingly, the steady-state distributions become progressively more diffuse (see Fig.~\ref{fig:melt}B), causing localized pockets of order to give way to entropy and ``melt'' away---just as crystals might in equilibrium physics (movie~S8, see also~\cite{goldman2003grains_lattice}).

Even as the range of accessible rattling values in the system shrinks, the predictive relation of Eq.~\ref{eq:p_ss} continues to hold (Fig.~\ref{fig:melt}C), enabling quantitative prediction of how self-organized configurations are destabilized. By calculating the entropy of the drive pattern as we tune its randomness, we derive a lower-bound on rattling for the system. Thus, we can analytically predict how steady-state probabilities change as a function of drive randomness, as shown in Fig.~\ref{fig:melt}D (up to normalization and $\gamma$, see supplementary materials for derivation). This result confirms the simple intuition that more predictably-patterned driving forces offer greater opportunity for the system to find low-rattling configurations, and self-organize (see also fig.~S6).

Our findings suggest that the complex dynamics of a driven collective of nonlinearly interacting particles may give rise to a situation in which a new kind of simplicity emerges. We have shown that when quasi-random transitions among configurations dominate the dynamics, the steady-state likelihood can be predicted from the entropy of local force fluctuations, which we refer to as rattling. In what we term a ``low-rattling selection principle,'' configurations are selected in the steady-state according to their rattling values under the given drive.

More significantly, low-rattling provides the basis for self-organized dynamical order that is specifically selected by the choice of driving pattern. We see analytically and experimentally that the degree of order in the steady-state distribution reflects the predictability of patterns in driving forces. Thus, driving patterns with low entropy pick out fine-tuned configurations and dynamical trajectories to stabilize. This makes it possible for one collective to exhibit different modes of ordered motion depending on the fingerprint of the external driving. These modes differ in their emergent collective properties, which suggests ``top-down'' alternatives to control of active matter and metamaterial design, where ensemble behaviors are dynamically self-selected by the choice of driving, rather than microscopically engineered~\cite{sigmund2009topolMetamaterials, Pradillo2019}.

\paragraph*{Supplementary Materials}
\begin{flushleft}
Materials and Methods \\
Supplementary Text \\
Figures S1-S10. \\
Movies S1-S8. \\
Movie Captions.\\
References~\textit{(33-46)}.
\end{flushleft}

\paragraph*{Acknowledgments}
We thank Paul Umbanhowar, Hridesh Kedia and Jeremy Owen for helpful discussions. \textbf{Funding:} P.C. and J.L.E. were funded by ARO W911NF-18-1-0101, and the James S. McDonnell Foundation Scholar Grant 220020476. Funding for T.A.B., A.S., and T.D.M. provided by ARO MURI Award W911NF-19-1-0233, and NSF CBET-1637764. Support for A.V., W.S., and D.I.G. was provided by NSF PoLS-0957659, PHY-1205878, DMR-1551095, and ARO W911NF-13-1-0347. Funding support for K.W. provided by NSF PHY-1205878. \textbf{Author contributions:} P.C. derived all theoretical results, performed simulations, data analysis, and contributed to writing. T.A.B. performed all experiments in main text, and contributed to writing and data analysis. A.V. and W.S. performed supplementary experiments. A.S. aided in robot hardware and software fabrication. J.L.E., D.I.G., K.W., and T.D.M. secured funding and provided guidance throughout. \textbf{Competing interests:} The authors declare no competing interests. \textbf{Data and materials availability:} All files needed for fabricating smarticles, as well as representative data, can be found in~\cite{git_repo}.

\begin{figure*}[ht]
    \centering
	\includegraphics[width=0.75\textwidth]{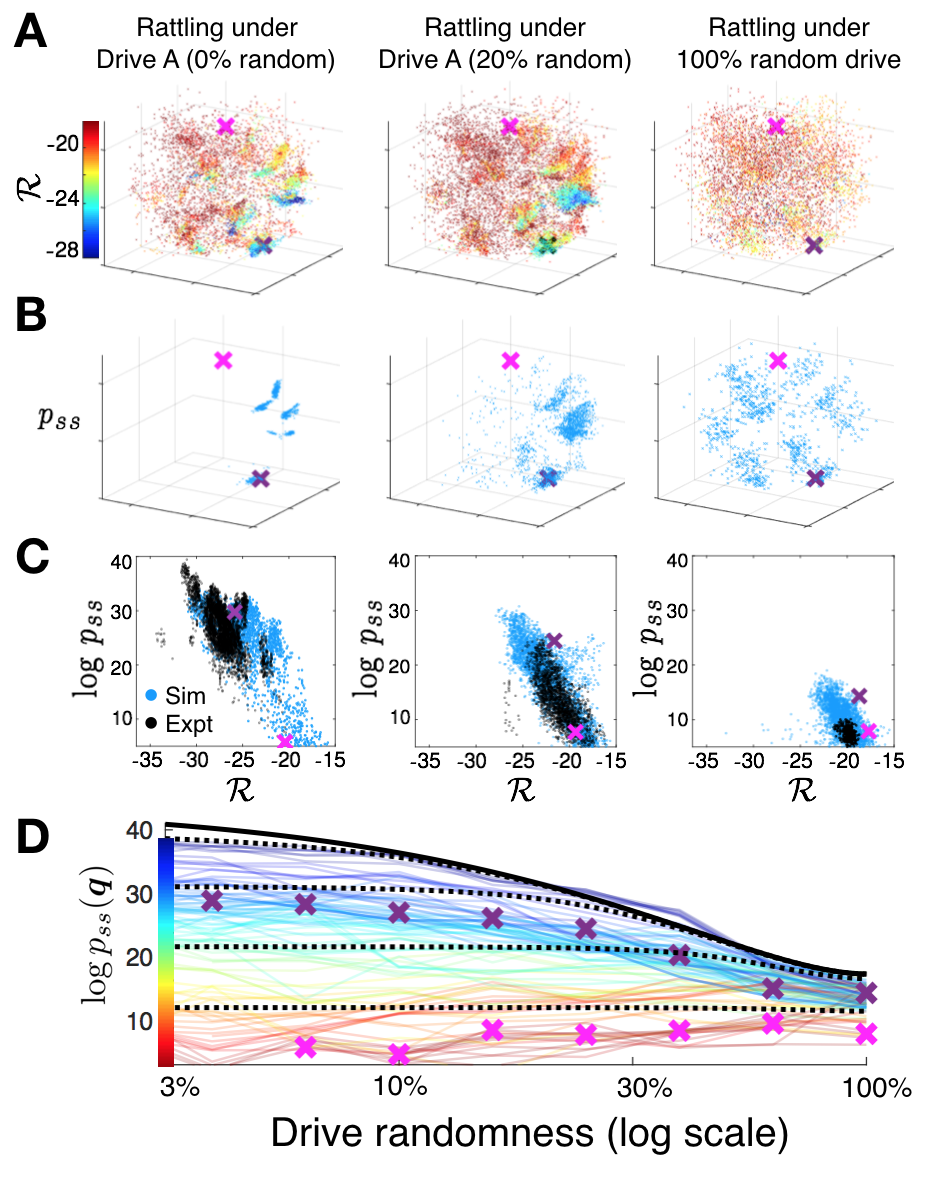}
	\caption{\textbf{Tuning self-organization by modulating drive randomness}. Self-organization relies on the degree of predictability in its driving forces, in a way that we can quantify and compute analytically. As the drive becomes less predictable (left to right, all panels), (\textbf{A})~low-rattling configurations gradually disappear. (\textbf{B})~The corresponding steady-states, reflecting the low-rattling regions of (A), become accordingly more diffuse (panels A and B show simulation data, and use the same 3D configuration space as Fig.~\ref{fig:smarticle}E). (\textbf{C})~verifies that our central predictive relation Eq.~\ref{eq:p_ss} holds for all drives here, as all three correlations fall along the slope of the same line (blue: simulation, black: experiment). The diminishing range of rattling values thus precludes strong aggregation of probability, and with it self-organization. (\textbf{D})~shows our theoretical prediction (solid black line) indicating how the most likely configurations are destabilized by drive randomness. Colored lines track the probability $ p_{ss} $ at 100 representative configurations $ \v{q} $ in simulation, and dashed black lines analytically predict their trends. (movie~S8, see supplementary materials for derivation). Two specific configurations marked by $\times$-s are tracked across analyses.
	}
	\label{fig:melt}
\end{figure*}

\renewcommand*{\theHsection}{S\the\value{section}}
\setcounter{section}{0}

\renewcommand*{\thefigure}{S\the\value{figure}} 
\setcounter{figure}{0}

\renewcommand*{\theequation}{S\the\value{equation}} 
\setcounter{equation}{0}


\includepdf[pages=-]{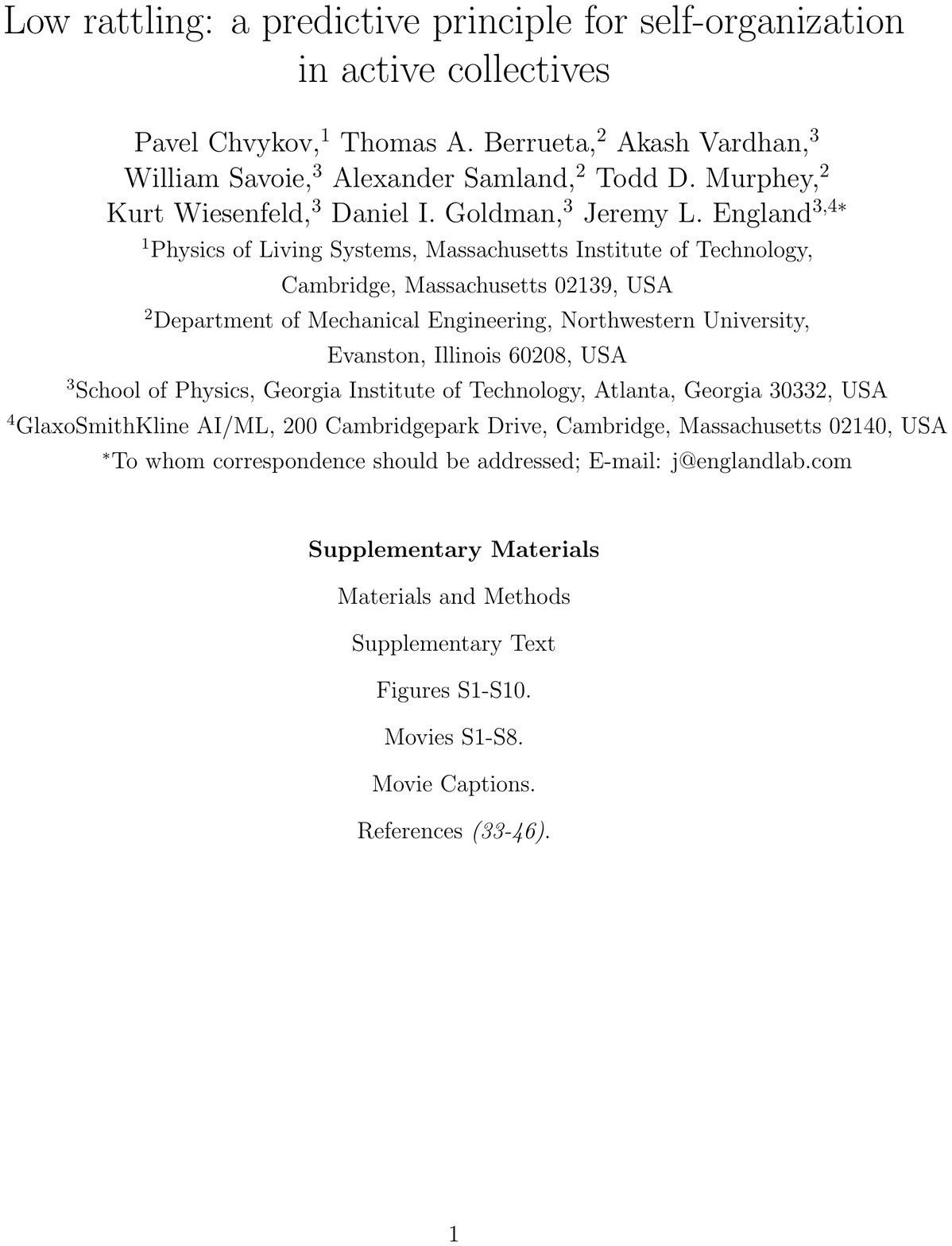}

\begin{center}
    \section*{Materials and Methods}
\end{center}

\section{Experiment Setup} \label{app:experiment}
\subsection{Smarticle Hardware Design}
All files necessary towards fabricating smarticles---\textit{i.e.,} mechanical CAD files, PCB schematics and files, as well as a comprehensive bill of materials---can be found in~\cite{git_repo}.

Smarticles are simple 3-link robots actuated at the hinges of the links. Each of the links is approximately 5.2 \textit{cm} in length and the hinges are actuated by two servomotors. The mechanical design of the smarticle is chosen such that when an individual smarticle sits on a flat surface in the absence of other interactions, the arms cannot propel the smarticle. Beyond this simple design principle, the smarticle bodies also ensure that when smarticles do interact, the circuit boards are protected from collisions. This way, when smarticles do make contact it is a largely smooth plastic-on-plastic interaction.

The primary electrical components enabling the smarticle capabilities highlighted in this work are the ATMEGA328PB micro-controller, and a Digi XBee3 wireless radio for communications. The micro-controller is used to control the drive patterns of smarticles. Once a motion pattern is specified, the micro-controller sends commands to the servomotors to execute them. The wireless radio's purpose is two-fold: first, it enables remote specification of experimental parameters from a central computer; second, it ensures that the smarticle drives remain in-phase relative to one another by sending a synchronizing pulse up to sub-millisecond accuracy. Outside of its synchronizing function, the radio was not utilized for any closed-loop feedback and was largely not critical to the results presented, given another source of synchronization between smarticles. As a result of the large footprint of components like the radio, we split the smarticle electrical components across two boards. Finally, these components are powered by a 150mAh LiPo battery, which enabled experiments up to 2-3 hours in length. 

We note that there are additional components included in the designs (\textit{e.g.}, accelerometers, current sensors, photoresistors, \textit{etc.}) that are not required for replicating any of the results in this study, and so may be removed from the PCB to minimize unit costs. These components were added for potential use in future studies, and they may also be of interest to an experimenter.

\subsection{Smarticle Software Design}
The smarticle software was divided into two primary modules. The first is a high-level communications module that sends and receives commands from the wireless radio. These libraries were written in Python for ease of use. Using these libraries we are able to specify driving patterns and other experimental parameters from a remote computer. Some examples of modifiable parameters relevant to the experiments presented in this work are arm velocity, arm angle noise, random arm move rate, and rate of switching between drive patterns. 

The second module is a low-level module that decodes incoming messages from the wireless radio and implements them directly on the micro-controller. This set of libraries is written in C++ using the Arduino library for ease of access to servomotor and sensor drivers. With these libraries we handle the low-level implementation of the experimental parameters chosen with the Python modules. An important function performed by this module is the seeding of random number generators. We take in ambient noise as an input and use it to seed the generation of pseudorandomness used in several experimental parameters, such as arm angle noise and random arm moves. We note that the design of the software architecture, as well as many of the included features, does not represent a minimal implementation of the software required to replicate the results in this work. Hence, we hope that this more general implementation could be of use to experimenters.   These modules are included in the smarticle fabrication repository~\cite{git_repo}.

\subsection{Smarticle Tracking Setup}
In order to track the smarticles in an experiment in real time we make use of AprilTags~\cite{Olson2011}. These tags are similar to QR tags but are effective at smaller sizes, making them ideal for robotic platforms with a limited footprint. These tags are supplemented by a position tracking library compatible with Python that extracts timestamped smarticle positions at a rate of up to 20~\textit{Hz}.

To capture the live experiment footage, we made use of a Logitech BRIO web-cam capable of 4K resolutions at 60~\textit{fps}. However, increasing the resolution of the image results in a slow down for the AprilTag detection library, so we intentionally capped the resolution of the camera using OpenCV~\cite{opencv_library} at 720p to have a stable tracking rate. We mounted the camera onto a solid structure machined out of aluminum 80/20, and calibrated the camera point of view of the smarticles to be directly from above.

\subsection{Experimental Procedure}
In order to replicate the main results of the paper there are 3 different parameters we vary across experiments: drive pattern, random move rate, drive switching rate. We begin by specifying a drive pattern as a sequence of arm angle pairs. For example, the periodic pattern in Fig.~\ref{fig:smarticle}B is specified by the set of arm angle tuples, $\{(\pi/2,\pi/2),$ $(\pi/2,-\pi/2),$ $(-\pi/2,-\pi/2),$ $(-\pi/2,\pi/2)\}$, as the arm angle sequence, and linearly interpolating arm angles between these points at a given arm velocity. While this example drive pattern is the same for all smarticles, we can also implement unique drive pattern for each smarticle (as in Fig.~\ref{fig:drive-specific}). For all experiments in this manuscript we set the arm move duration to 500~\textit{ms} for any single step. Additionally, for a given choice of drive pattern we can set a random move rate at which we stochastically replace a given arm angle tuple with a random one sampled such that each arm is equiprobably and independently at either $\pi/2$ or $-\pi/2$. Similarly, if one has multiple drives specified in software, the drive switch rate indicates the rate at which the smarticles stochastically switch drives as in a Poisson process. 

In addition to the software setup, in preparation for an experiment we place the smarticles in the ring within the frame of the web-cam and rigidly fix the ring in place. The surface on which all experiments took place is a large foam core board, itself secured in place and verified to be level within 1 degree. Experiments were run in sets of randomly initialized 10 minute long runs. Once the smarticles are in the secured ring and the experimental parameters are set, we begin by randomizing the initial conditions of each run. To do this, we sent the smarticles independently random commands for a duration of 2 minutes prior to starting the 10 minute run of the specified drive pattern. 

Experimental parameters used for specific experiments, as well as time-series data from the smarticles are included in the linked repository.

\section{Simulation Setup} \label{app:simulation}
For easier exploration of hypothesis space of this system, we have also constructed a numerical simulation, whose algorithm is described below. This implementation was chosen so as to optimize speed and scalability to larger swarm sizes, at the cost of some quantitative agreement with experimental details. This was additionally justified as in this work, we are interested in generalizable effects that do not depend on all microscopic system details. As such, instead of using a fully-featured physics engine and implementing the smarticle design parameters exactly, we used MATLAB to implement a simpler abstracted version of the system, choosing simulation parameters by qualitatively benchmarking against experimentally observed behaviors.

Our algorithm was designed as follows. We approximate smarticles by thin three-segment lines (as shown in Fig.~\ref{fig:systems}C). At each time-step (``tick,'' chosen to be about 10~\textit{ms} real time for our specific setup), the algorithm moves the arms slightly according to the chosen arm-motion pattern, and then iteratively cycles through all smarticle pairs in random order, checking for collisions, moving one in each pair slightly according to the net interaction force. If there are multiple points of contact for a given pair, the move is a translation in the direction of the total force, otherwise, it is a rotation about a pivot point chosen so as to balance the forces and torques (based on a general analytical force-balance calculation). Choosing which smarticle in each pair moves is random, weighted by their relative friction coefficients (as motivated by difficulty of predicting static friction). Note that since a move can create new collisions with other smarticles, it is important to take small steps and iterate. The algorithm continues looping through pairs until all collisions are resolved, then proceeding to the next tick with the motion of the arms. While this describes the core of the algorithm, there are a number of additions necessary to improve its stability and reliability:
\begin{itemize}
	\item  If two arms are near-parallel when they approach each other, they can pass through each other between ticks without ever intersecting or registering a collision. To prevent this, along with collision detection, we must explicitly test for parallel arms in each pair of smarticles. We then store the relative position of each such pair of arms for a few ticks into the future to prevent them passing through each other in any of those times.
	\item If a smarticle with small friction coefficient gets trapped between two others, it might move back and forth on each iteration of the collision-resolution loop, with no net effect, creating an infinite loop. To prevent this, we temporarily (until the next tick) increase its friction each time a smarticle moves, so that it is less likely to move again as collisions continue being resolved.
	\item In experiment, when resolving collisions is too hard, the motor simply does not move (\textit{i.e.}, it jams up momentarily). This can happen quite often when smarticles are in a tight confinement, as was the case in many of these experiments. To allow for that possibility in simulation, we add an exit condition in the collision-resolving loop. It triggers when any one smarticle's temporary friction (from last bullet) becomes very large, since this serves as a proxy for how much force a motor must provide. In that case, we ``rewind'' the most-colliding arm back to before the last tick, and try collision-resolving again from scratch. If everything resolves, that arm will then have a chance to catch up to where it needed to be over the following ticks (its speed being capped at some $\omega_{max}$ to prevent discontinuous jumps).
	\item Interactions with the ring boundary are implemented similarly to interactions with smarticles, and collisions with it are resolved in the same loop. 
	\item It is easy to adjust the simulation to give the smarticle inertia: at each tick, we simply move the smarticle according to last step's velocity, scaled by a discount factor, before resolving collisions.
	\item If inter-smarticle friction is 0, then each interaction force is directed normal to one smarticle's surface. To include effects of such friction, we can add a small lateral component to these forces that depends on the interaction angle according to force-balance equations.
\end{itemize}

Even with all these additions, many differences remain between simulation and experiment: smarticles have non-zero thickness in experiments, there are relief features on smarticle body not present in simulation that can get caught, the precise force-response profile of the motors is not captured, etc. Including more of these corrections, while possible, will slow the simulations down, and is not generally desirable as we do not want our results to depend on exact system-specific details. 

\section{Data Analysis}

\subsection{Constructing Configuration Space} \label{app:configSpace}
The raw data generated from both the simulation and the experiment are sets of time-series $ (x_s(t), y_s(t), \theta_s(t)) $, where $s$ indexes the smarticles. First, we want to use these coordinates to construct a set of observables that faithfully parametrize the space of distinct swarm configurations---since, \textit{e.g.}, configurations related by a global rotation should be counted as identical. Constructing such configuration spaces rigorously for many-particle systems is known to produce complicated topological spaces with non-smooth geometries~\cite{han09paramConfigs}. On the other hand, any practical description of an active-matter system will generally be a simple heuristic choice of coarse-grained variables that respect the symmetries of the problem (such as, \textit{e.g.}, permutation symmetry of indistinguishable particles), while distinguishing the configurations of interest. This will generally represent an over-compression, and the distances among configurations in the resulting space will not always faithfully capture their objective distinguishability. Similarly here, rather than trying to construct coordinates that capture precisely the correct geometry of our swarm's space of distinct configurations, we heuristically choose a set of observables that respects the relevant symmetries, and captures enough of the system complexity to allow distinguishing a wide variety of behaviors. 

\begin{figure} 
	\centering\includegraphics[width=1.\textwidth]{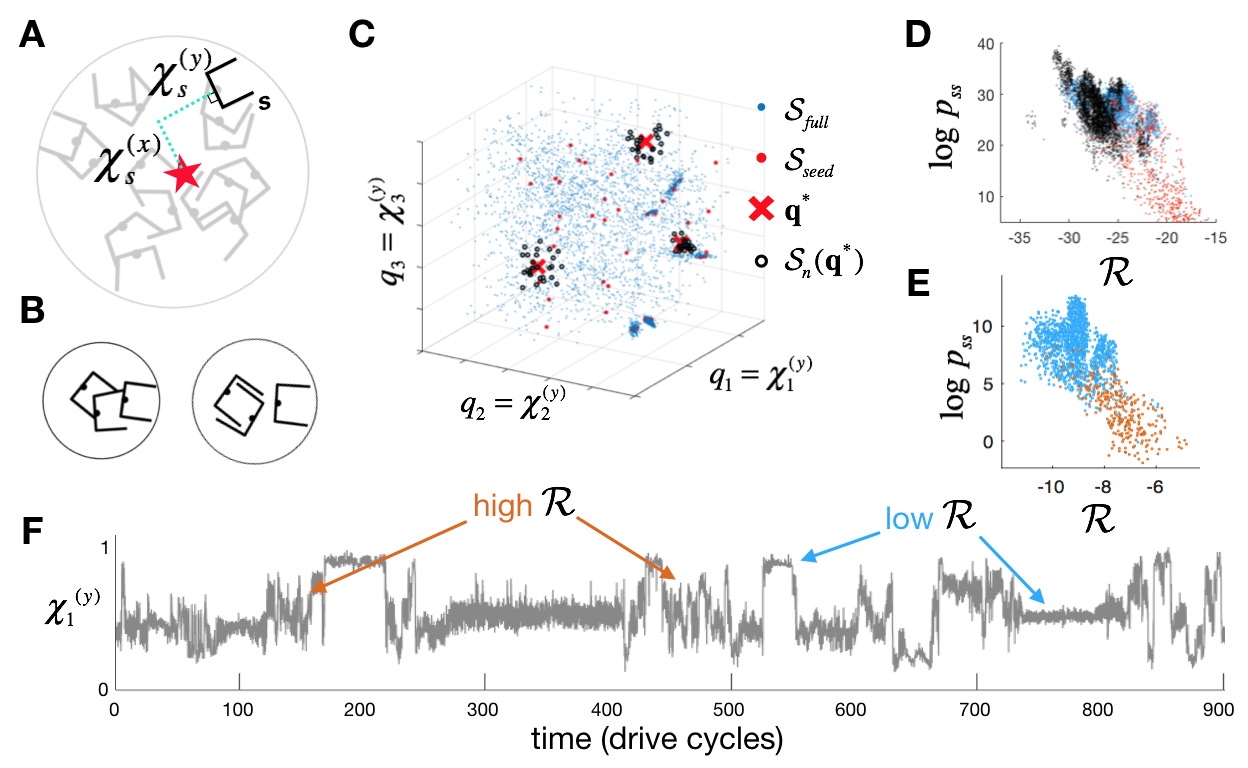}
	\caption{
\textbf{Data analysis steps}.
(\textbf{A})~shows the construction of the observables we use to parametrize the swarm configuration space, chosen to be invariant under global rotations (red star is the swarm center of mass). Note that we stroboscopically select the time-points when all the arms are in U-shape $(\alpha^{(1)}_s,\alpha^{(2)}_s) = (\frac{\pi}{2},\frac{\pi}{2})$, as shown here. (\textbf{B})~two configurations that are indistinguishable by the coordinates constructed in~(\textbf{A}), and are distinguished by adding another observable $\chi^{(\theta)} = \abs{\< \e{i \theta}\>_s}$. (\textbf{C})~shows the configuration space plotted in the main text (Fig.~\ref{fig:smarticle}, \ref{fig:drive-specific}, \ref{fig:melt}) for the three-smarticle swarm. Blue shows the set $ \mathcal{S}_{full}$ (here chosen as $ \mathcal{S}_{p} \cup \mathcal{S}_{rnd}$, see text) sampling all regions of interest in the configuration space. In red we show its subset $\mathcal{S}_{seed}$ of seed points where $p_{ss}$ and $\mathcal{R}$ are to be calculated. 3 of these seed points $\v{q}^*$ are highlighted with red $\times$-s, and for each, their 30 nearest neighbors $ \mathcal{S}_n(\v{q}^*) \subset \mathcal{S}_{full} $ from the blue set are shown in black. Note that these black points define variable size and shape neighborhoods from which $p_{ss}$ and $\mathcal{R}$ are estimated. (\textbf{D})~shows the resulting correlation for the 3-smarticle swarm (see Fig.~\ref{fig:melt}C), with blue indicating the values sampled from the neighborhood of the steady-state distribution $\v{q}^* \in \mathcal{S}_{p}$, and red---from randomly sampled configurations $\v{q}^* \in \mathcal{S}_{rnd}$. Note that experimental data (black) only samples the steady-state configurations, and hence reproduces only the blue part of the correlations. (\textbf{D}~and~\textbf{E})~show the same analysis repeated for two different descriptions of the configuration space: (D) for the 7-dimensional $\left(\{\chi_s^{(x)},\chi_s^{(y)}\}_{s=1,2,3}, \chi^{(\theta)}\right)$ and (E) for the 3-dimensional $\left(\chi_1^{(y)},\chi_2^{(y)}, \chi_3^{(y)}\right)$ space. Their qualitative agreement illustrates that our results will hold for a variety of coarse-grained descriptions of an active matter system. (\textbf{F})~shows a sample trajectory of the $\chi_1^{(y)}$ over time, where high-rattling motion is seen as regions of stochastic exploration, while low-rattling motion shows systematic oscillations. 
}
\label{fig:dataAnal}
\end{figure}

Explicitly, we construct a set of observables invariant under global rotation symmetry: $ \v{\chi}_s \equiv R(\theta_s) \cdot (\v{X} - \v{x}_s) $, where $R(\theta_s) $ is the rotation matrix for $ s $-th smarticle body orientation, $ \v{X}= \frac{1}{N_s} \sum_s \v{x}_s $ is the swarm's center of mass coordinates, $\v{x}=(x,y)$ of each smarticle center in units of its body length, and $N_s$ is number of smarticles in the ensemble (see fig.~\ref{fig:dataAnal}A). This choice projects our number of tracked variables from $ 3N_s $ down to $ 2N_s $, but still distinguishes among most relevant configurations, especially for small $N_s$. We found it helpful to further add one more observable to this set $\chi^{(\theta)} = \abs{\frac{1}{N_s} \sum_s \e{i \theta_s}}$ to help distinguish between the two configurations pictured in fig.~\ref{fig:dataAnal}B. Nonetheless, we also verify that our results do not depend on this specific choice of observables by running the below analysis on various subsets of these observables, as shown in fig.~\ref{fig:dataAnal}, D and E. One caveat to this is that for our rattling calculation to be reliable, our choice of the configuration space must be sufficiently high dimensional so that ordered motion does not appear as space-filling. Practically, we want to have three or more dimensions in our description of the configuration space, which is in contrast to the one or two-dimensional order parameters often used for describing active matter. 

Having constructed trajectories over time of rotation-symmetric observables $\chi$ (one of which is shown in fig.~\ref{fig:dataAnal}F), we need to choose discrete sets of points along our trajectories that we will use to calculate probability densities. With this, we also want to allow for a fair comparison among configurations of swarms running different drive-patterns. Since in addition to creating external forcing, arms also present steric constraints on the allowed configurations, two swarms with different arm angles will generally have a hard time exhibiting the same configurations. To isolate the effect of driving history on the steady-state distribution, we must then only compare swarms having all the same arm angles: here we choose $ (\alpha^{(1)}_s,\alpha^{(2)}_s) = (\frac{\pi}{2},\frac{\pi}{2})$ (U-shape, for all smarticles, fig.~\ref{fig:dataAnal}, A and B). Some of the drives used in this work were periodic, in which case we made sure that this configuration is visited once per period, and could thus select our configurations stroboscopically at the corresponding time-points. For stochastic arm-motion, on the other hand, we programmed the arms to return to this configuration deterministically once every 24 moves---rare enough so as not to introduce significant predictable correlations. While we cannot track the arm angles directly in experiment, we can use the time-stamps of the synchronization pulses sent to the smarticles by radio each time all arms get into the U-shape configuration to know exactly which configurations to compare. 

This way, we end up with the discrete time-series of the $2N_s + 1$ configuration observables at stroboscopic times: $\left(\{\chi_s^{(x)}(t_n),\chi_s^{(y)}(t_n)\}_s, \chi^{(\theta)}(t_n)\right)$. All 3D configuration-space plots in the paper (\textit{i.e.}, Fig.~\ref{fig:smarticle}E, Fig.~\ref{fig:drive-specific}, A-C, bottom, and Fig.~\ref{fig:melt}, A and B) show the $(\chi_1^{(y)}, \chi_2^{(y)}, \chi_3^{(y)})$ subspace for the 3-smarticle swarm---so chosen for visual clarity.

\subsection{Estimating Steady-State Density and Rattling} \label{app:measure_pss,R}

Since in self-organizing systems such as ours, the probability density over configurations can span many orders of magnitude, we cannot accurately sample all the regimes of interest with a uniform sampling scheme. Our analysis must thus use adaptive neighborhood sizes, and be robust to poor sampling of the distributions, as we are working in high-dimensional spaces. To ensure that our key results were not artifacts of such an adaptive algorithm, we benchmarked them across several qualitatively different analysis techniques, including exhaustive uniform sampling of the configuration space. 

We start by outlining our entire algorithm abstractly and generally, after which we explain the practical implementation of generating the various figures in the main text. We call our $ d $-dimensional configuration space coordinates $ q_i $,  indexed by $ i,j\in \{1,...,d\} $---here, we primarily use the set $\{q_i\}= \left(\{\chi_s^{(x)}(t_m),\chi_s^{(y)}(t_m)\}_s, \chi^{(\theta)}(t_m)\right)$  defined in the last section, but subsets and other choices were also tested. Such parametrization allows us to compute distances using the Euclidean metric, as these observables were already chosen to be faithful to the distinguishability of smarticle configurations.

\begin{enumerate}
	\item Begin by defining two sets of points on the configuration space, $ \mathcal{S}_p $, sampling the steady-state probability distribution $p_{ss}$, and $ \mathcal{S}_{full} $, sampling all regions of interest in this configuration space, as best we can. For example, in our simulations we often chose $ \mathcal{S}_{full} = \mathcal{S}_{p} \cup \mathcal{S}_{rnd}$, where $\mathcal{S}_{rnd}$ sampled all swarm configurations uniformly at random, while $\mathcal{S}_{p}$ ensured good sampling of the self-organized regions.
	\item Choose a random subset $ \mathcal{S}_{seed} \subseteq \mathcal{S}_{full}$ of ``seed'' configurations whose neighborhoods we want to evaluate $ p_{ss} $ and $ \mathcal{R}$ in. We can sub-sample the full available set of configurations $\mathcal{S}_{full}$ as it may be computationally redundant to use every point. 
	\item For each seed point $ \v{q}^* \in \mathcal{S}_{seed} $, find the subset $ \mathcal{S}_n(\v{q}^*) \subset \mathcal{S}_{full} $ of its $ n $ nearest-neighbors (with $ n \gg d $: here we use $n \sim 3d$, see fig.~\ref{fig:dataAnal}C). We will estimate $ p_{ss}(\v{q}^*) $ and $ \mathcal{R}(\v{q}^*) $ from the neighborhood covered by this set as follows:
	\item $ p_{ss}(\v{q}^*)$
	\begin{enumerate}
		\item Compute the variance tensor $ V_{ij}(\v{q}^*) = \< (q-q^*)_i (q-q^*)_j\>_{\v{q} \in \mathcal{S}_n(\v{q}^*)}$ and the volume 
		$ \mathcal{V}(\v{q}^*) = \sqrt{\det V(\v{q}^*)} $ of the region occupied by $ \mathcal{S}_n(\v{q}^*) $ (here we need $ n \gg d $ to give sensible estimates).
		\item Find the steady-state probability of this region $ P[\mathcal{S}_n(\v{q}^*)] $ as the fraction of points $\v{q}_p \in \mathcal{S}_p $ that satisfy $ \sum_{i,j} (q_p-q^*)_i (q_p-q^*)_j\, \left[V^{-1} (\v{q}^*)\right]_{ij} < 5 $. This selects the steady-state configurations that are within $ 2.2 \sigma $ of $ \v{q}^* $, where $ \sigma $ is taken with respect to the distribution of $ \mathcal{S}_n(\v{q}^*) $ points.
		\item The steady-state probability density is thus estimated as $ p_{ss}(\v{q}^*) = P[\mathcal{S}_n(\v{q}^*)] / \mathcal{V}(\v{q}^*) $.
	\end{enumerate}
	\item $ \mathcal{R}(\v{q}^*)$
	\begin{enumerate}
		\item To find the rattling $ \mathcal{R} $ of any given configuration $\v{ q}^*$ under the action of some drive, we first sample a short trajectory, or ``rollout,'' $\v{ q}(t)$ starting from $\v{ q}(t=0)=\v{ q}^*$ and evolving under that drive for some short but representative time. Here, we use 2 to 5 seconds, or 4 to 10 discrete arm moves. Generally, this time-horizon may depend on the system in question, and in particular must be longer than the drive-pattern, yet should be kept consistent for all calculations in a given system. 
		\item For every time-point along this rollout, we compute the vector $ \v{v}(t) = (\v{q}(t)-\v{q}^*)/\sqrt{t} $, and calculate the covariance matrix of all these 
		 \begin{align} \label{eq:Cdef}
		 \mathcal{C}_{ij}(\v{q}^*)\equiv \<v_i(t), v_j(t)\>_t 
		 \end{align} (notation: $\<i,j\> \equiv\<i\, j\> - \<i\>\<j\> $ component-wise, or equivalently $\mathcal{C}(\v{q}) = \mbox{cov}[\tilde{\v{v}}_{\v{q}},\tilde{\v{v}}_{\v{q}}]$, as in Eq.~\ref{eq:covariance}). This may be seen as a particular choice of a data-driven estimator for the effective diffusion tensor $ D_{ij}(\v{q}^*) $ locally~\cite{michalet2012singTrajDlim}. 
		  \begin{itemize}
		     \item As we are talking about a stochastic process here, we can average over both, the one rollout duration, and/or over several rollouts initialized in $\v{ q}^*$. Abstractly, we view these $\v{v}(t)$ for various time-points $t$ or for various rollouts as samples of the same random variable $\tilde{\v{v}}_{\v{q}}$, used in the main text. Thus we implicitly assume a sort of local ergodicity, such that average over the length of a short run and average over stochastic realizations give similar results. Still, all the final figures in this paper were generated using just a single rollout for each point, as that approach is more practical for future applications.
		 \end{itemize}
		\item We then define rattling, same as in Eq.~\ref{eq:rattling}, via $\mathcal{R}(\v{q}^*)\equiv \frac{1}{2}\log \det \mathcal{C}(\v{q}^*)$. This is an estimate for the entropy of the distribution of vectors $ \v{v}(t) $ that takes them to be Gaussian distributed. 
		\item Finally, since $ p_{ss}(\v{q}^*)$ was computed for the neighborhood $\mathcal{S}_n(\v{q}^*)$, we similarly want to average rattling over that set of points: $\mathcal{R}_n(\v{q}^*) = \<\mathcal{R}(\v{q})\>_{\v{q} \in \mathcal{S}_n(\v{q}^*)}$. While we drop the subscript $ n $, this is the quantity we plot whenever we compare rattling against $ p_{ss} $. Note that this choice of averaging is more reliable than if we instead averaged the covariance matrices $ \mathcal{C} $, as $\mathcal{R}_n(\v{q}^*) = \frac{1}{2}\log \det \<\mathcal{C}(\v{q})\>_{\v{q} \in \mathcal{S}_n(\v{q}^*)}$. In that case, our adaptive neighborhood size could introduce artificial correlation with $ p_{ss} $ as $ \det\<\mathcal{C}(\v{q})\> $ would tend to grow with neighborhood size. Averaging outside the determinant avoids this problem.
	\end{enumerate}
\end{enumerate}

This setup now allows a clear explanation of how all our figures were generated. All the 3D plots showing the steady-state distribution $\mathcal{S}_p $ (\textit{i.e.}, Fig.~\ref{fig:smarticle}E, Fig.~\ref{fig:drive-specific}, A-C, bottom, and Fig.~\ref{fig:melt}B) are constructed by taking 10 to 30 long runs ($>200$ drive cycles, or equivalent) initialized in some random configurations, and clipping the initial transient section  (typically around 20 first periods of the drive). The plots then show the $(\chi_1^{(y)}, \chi_2^{(y)}, \chi_3^{(y)})$ stroboscopic configuration observables (see Section~\ref{app:configSpace}). In contrast, to uniformly sample the set of all possible swarm configurations, shown in red in Fig.~\ref{fig:smarticle}E, which we will call $\mathcal{S}_{rnd} $ here, we subjected smarticles to translational and rotational random Brownian noise, with their arms held fixed in U-shape $(\alpha_1,\alpha_2)=(\frac{\pi}{2},\frac{\pi}{2})$. This procedure could only be done in simulation, although a similar outcome could be achieved in experiment by moving all the smarticle arms at independent random times between $+\frac{\pi}{2}$ and $-\frac{\pi}{2}$ positions.

For the $\<\mathcal{R}\>$ over time plots in Fig.~\ref{fig:systems}, C and D, we started with $M$ random initial configurations (2000 in simulation, 20 in experiment), and calculated rattling for short snippets along the length of the subsequent evolution run. Averaging was then done over the equal-time configuration ensembles. 

For all the $p_{ss}$ vs. $\mathcal{R}$ correlation plots (bottom of Fig.~\ref{fig:systems}, C and D, and Fig.~\ref{fig:melt}C), we show the sets of $(\mathcal{R}(\v{q}^*), p_{ss}(\v{q}^*))$ tuples calculated as above. To get a good sampling of both, the self-organized behaviors, and the entire configuration space, we chose $\mathcal{S}_{full} = \mathcal{S}_{p} \cup \mathcal{S}_{rnd}$ in simulations.  In fig.~\ref{fig:dataAnal}, D and E, the blue points show the results for seed points chosen from $ \mathcal{S}_{p} $, and red point---for points from $ \mathcal{S}_{rnd} $. The full distribution of seed points $\mathcal{S}_{seed} \sim \mathcal{S}_{full}$ further allowed us to plot the rattling landscape $\{\mathcal{R}(\v{q}^*)\}_{\v{q}^* \in \mathcal{S}_{seed}}$ across the entire configuration space in Fig.~\ref{fig:melt}A.
	
Experimentally, we only had access to the $\mathcal{S}_{full} \sim \mathcal{S}_{p}$ part of the configuration space, as our data consisted of 10-20 long runs of the dynamics, which thus spent most of their time in the steady-state regions. This allowed us to reproduce only the high-$p_{ss}$ part of the correlation, as shown by the black points in fig.~\ref{fig:dataAnal}D and fig.~\ref{fig:sqGait}B, rather than the full range accessible in simulations.

Finally, to generate Fig.~\ref{fig:melt}D, we chose 100 seed configurations sampling the entire space $\mathcal{S}_{seed} \subset \mathcal{S}_{p} \cup \mathcal{S}_{rnd}$, and estimated the steady-state density in their neighborhoods resulting from 8 different drives. These drives were generated by taking a base deterministic arm-motion pattern drive A (see Fig.~\ref{fig:drive-specific}A), and adding a different fraction of random arm moves in each case (still only going to arm angles $\alpha = \pm \frac{\pi}{2}$). This set of simulations thus allowed tracking exactly how drive entropy affects the $p_{ss}$ of these 100 representative configurations.

\newpage

\renewcommand*{\theHsection}{S\the\value{section}}
\setcounter{section}{0}
\begin{center}
\section*{Supplementary Text}
\end{center}

\section{Random dynamical systems}

We begin by examining toy constructs of ``random dynamical systems''---ones where the evolution equations are set up with many randomly chosen parameters. The motivating hypothesis of this approach is that some real many-body dynamical systems might be so complex that their behavior is closer to that of random motion than to some predictable deterministic trajectory. Solutions of this kind have already been found in stochastic approaches to modeling complex systems, such as using random matrices to approximate the Hamiltonians of large atomic nuclei~\cite{Wigner1967}. Subsequently, in contexts ranging from bacterial ecology~\cite{marsland2019randEcosystems} to the study of social networks~\cite{newman2002randomSocialNet}, models that assume random interactions among a system's many components have enabled effective predictions of system-level properties. Rather than assuming the system behaves deterministically with perturbations caused by noise, these approaches take the system to be primarily random while preserving some of its structure.

Instead of starting with some simple linearized dynamics and building out perturbative nonlinear corrections, we start with random dynamics and proceed to gradually introduce different correlations and structures that might matter in a real system. This way, as we make our system ``less random,'' we can monitor which features of the solution are sensitive to the details of our random construct, and which seem to persist universally, despite the strength and structure of the correlations. We can also begin to identify the ``typical'' behaviors of complex dynamical systems that arise generically, and do not rely on any system-specific details. 

While our constructs are often quite simple and may admit some analytic tractability, most of our results presented here are numerical. This is because such numerical explorations are much easier to carry out and tune in various ways, while also being highly reliable, as they easily permit good sampling of the chosen distributions (at least for the simple scenarios tested here).

\subsection{Random Markov processes} \label{app:randMarkov}
We begin by looking at the ``null model'' of a random dynamical system: $ N $ discrete system states connected to each other by independent identically distributed (i.i.d.) random transition rates. In its simplest version, we can solve it analytically. Consider the master equation 
\begin{align} 
\dot{P}_m= \sum_{n \neq m} R_{mn}\, P_n - \sum_{n \neq m} R_{nm}\, P_m 
\label{eq:masterEq}
\end{align}
with $ m,n\in\{1,...,N\} $ labeling the discrete states. In general, the steady-state probability distribution $ P^{ss}_m $ is the null eigenvector of the transition-graph Laplacian matrix $ L_{mn}= -R_{mn} + \delta_{mn}\,\sum_k R_{km} $, where $ \delta_{mn} $ is the Kronecker delta. Thus, $ P^{ss}_{m} $ will depend in a complicated way on all the $ N(N-1) $ transition rates (less $ N $ for the diagonal). 

Now, as we want to understand the typical behaviors of large disordered dynamical systems, we let the transition rates $ R_{mn} $ be i.i.d. random variables, with some mean $ \bar{R} $ and standard deviation $ \sigma $, and let $ N \rightarrow \infty$. We now want to try finding the steady-state as an asymptotic series in powers of $ \frac{1}{N} $: $ P^{ss}_m = P^{(0)}_m + P^{(1)}_m +...\ $. Plugging this into the master equation, Eq.~\ref{eq:masterEq}, we will go order by order, ensuring that $ \dot{P}_m $ asymptotically vanishes as $ N\to\infty $, progressively faster with each additional correction. To set up, we note that by the central limit theorem, we can write the total entrance and exit rates for state $ m $ respectively as:
\begin{align}
&\mathcal{Z}_m=\sum_{n \neq m} R_{mn} = N\left(\bar{R} + \frac{\sigma\, \zeta_m}{\sqrt{N}}\right) \qquad \text{entrance rate of }m\\
&\mathcal{E}_m =  \sum_{n \neq m} R_{nm} = N\left(\bar{R} + \frac{\sigma\, \xi_m}{\sqrt{N}}\right) \qquad \text{exit rate of } m 
\label{eq:exitR}
\end{align}
respectively, where $ \zeta_m $ and $ \xi_m $ are univariate Gaussian random variables specifying the deviation of $ m $-th entrance and exit rate from the mean. With this, plugging in a constant background probability $ P^{(0)}_m =\frac{1}{N} $ into the master equation proves to be a self-consistent choice at leading order:
\begin{align}
\dot{P}_m = \sum_{n \neq m} R_{mn}\, \frac{1}{N} - \sum_{n \neq m} R_{nm}\, \frac{1}{N} =  \bar{R} + \frac{\sigma\, \zeta_m}{\sqrt{N}} - \bar{R} - \frac{\sigma\, \xi_m}{\sqrt{N}}
\end{align}
which vanishes for large $ N $. Next, we want to choose the first correction $ P^{(1)}_m $ for the steady-state such that it further reduces $ \dot{P}_m $:
\begin{align}
\dot{P}_m =\frac{\sigma\, (\zeta_m-\xi_m)}{\sqrt{N}} +\sum_{n \neq m} R_{mn}\, P^{(1)}_n - \sum_{n \neq m} R_{nm}\, P^{(1)}_m \sim o\left(\frac{1}{\sqrt{N}}\right)
\end{align}
using the little-o notation. We can check that letting $ P^{(1)}_m =  \frac{\sigma\, (\zeta_m-\xi_m)}{\bar{R}\, N^{3/2}}$ accomplishes this since with that:
\begin{align}
\dot{P}_m =\sum_{n \neq m} R_{mn}\, P^{(1)}_n - \sigma\, \xi_m\; P^{(1)}_m\,\sqrt{N} \sim o\left(\frac{1}{\sqrt{N}}\right)
\end{align}
where the second term is explicitly of $ \bigO{\frac{1}{N}} $, and the first term averages out to be of that order if $ R_{mn} $ is uncorrelated with $ P^{(1)}_n $. 

This last assumption is the crucial step of the derivation, and plays a similar role to that of the molecular chaos assumption in equilibrium thermodynamics: while two colliding molecules become correlated after the collision, that correlation is entirely lost before they meet again. Similarly here, our assumption breaks time-reversal symmetry by distinguishing between exit and entrance currents for a given state $ m $: the former are all correlated as they are all $ \propto P^{(1)}_m $, while the latter average out as they are proportional to the independent random variables $P^{(1)}_n $. In reality, this assumption does not exactly hold, as $ P^{(1)}_n $ depends on $n$-th exit rate $ \xi_n $, which correlates with  $ R_{mn} $. Nonetheless, as all the rates are assumed independent, the effect of this correlation is suppressed by an additional factor of $ \frac{1}{N} $. Thus we see that if we introduce some structure into our dynamics, and hence correlations in the rates $ R_{mn} $, this assumption will be the first failure mode of the derivation. This gives a more precise sense in which the central assumption of the rattling theory is so akin to molecular chaos.

\begin{figure}
	\centering\includegraphics[width=1.\textwidth]{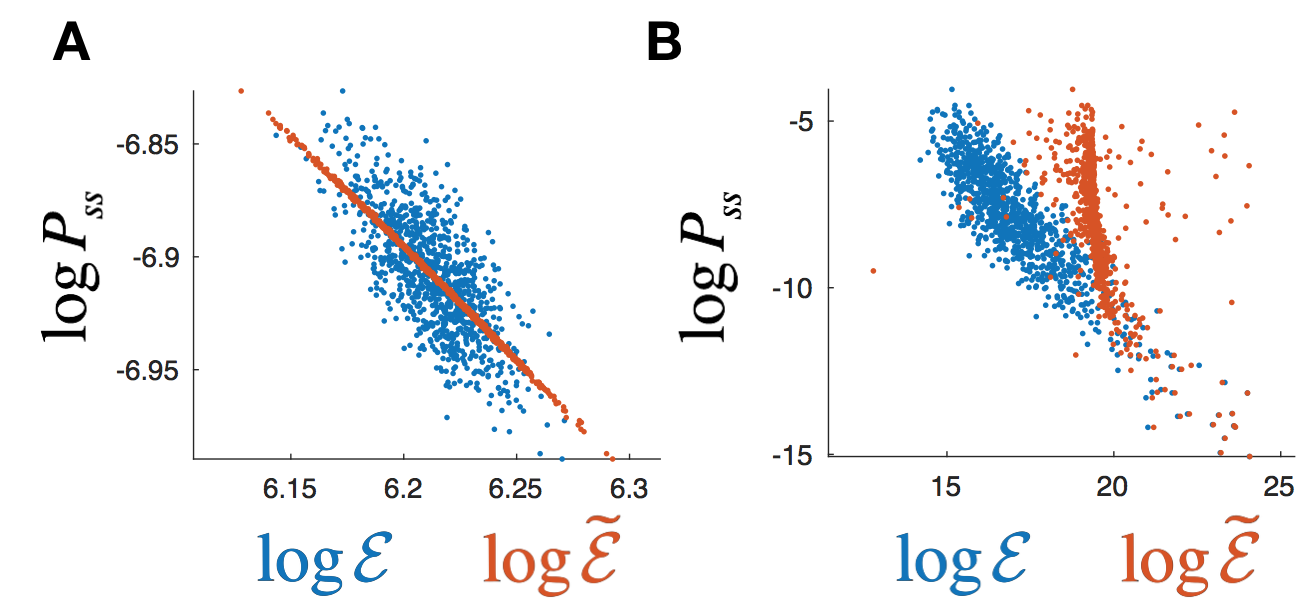}
	\caption{\textbf{Exit rates are predictive of steady-state in discrete Markov process} For each of $ N=1000 $ system states, we plot its steady-state likelihood (according to the master equation Eq.~\ref{eq:masterEq}) versus two different quantities. For blue points, we use that state's total exit rate $ \mathcal{E}_m $, while for orange, we show that exit rate also corrected by entrance rate deviation: $ \tilde{\mathcal{E}}_m \equiv  \mathcal{E}_m - (\mathcal{Z}_m - N \bar{R})$. (\textbf{A})~shows these results for the dynamics generated by sampling the transition rates $ R_{mn} $ uniformly in the range $ (0,1) $. This verifies Eq.~\ref{eq:randMarkRes}, which predicts $ \tilde{\mathcal{E}}_m  $ (orange points) to be perfectly predictive of $ P^{ss}_m $ when $ N \to \infty $. Pure exit rates in blue, while being more noisy, are still clearly correlated with the steady-state. This changes in~(\textbf{B}), where the transition rates $ R_{mn} $ are now sampled from the log-normal distribution: $ \log R_{mn} \sim \mathcal{N}(0,\sigma^2=25) $. While the exit rates (blue) remain robustly predictive of the steady-state, their correction by entrance rates (orange) no longer yields an improvement. Note also that the vast variability in $ R_{mn} $ in (B) allows for a much wider distribution of observed steady-state likelihoods.
	}
	\label{fig:randDynamics}
\end{figure}

Putting the pieces together, we thus get the approximate expression for the steady-state 
\begin{align}
P^{ss}_m= \frac{1}{N} + \frac{\sigma\, (\zeta_m-\xi_m)}{\bar{R}\, N^{3/2}} + \bigO{\frac{1}{N^2}}= 
\frac{1}{N} + \frac{\mathcal{Z}_m - \mathcal{E}_m}{N^2\,\bar{R}} + \bigO{\frac{1}{N^2}} 
\label{eq:randMarkRes}
\end{align} 
We can verify this result numerically, as illustrated by the red points in fig.~\ref{fig:randDynamics}A. One thing we notice about this expression is that the steady-state probability at a state depends in an equal measure on both the exit, and the entrance rates of that state. In practice, however, the total entrance rate may be hard to measure, as it would require initializing the system in all possible configurations and seeing how often it enters $ m $---\textit{i.e.}, it is hardly a ``local'' property of $m$ as much as of the entire system. The exit rate, on the other hand, is a measure of how stable the state $ m $ is and only requires local measurements initialized in that state. While $ \mathcal{E}_m $ is not enough to predict the steady-state exactly, it can already tell us a lot, and will be strongly correlated with it. To highlight this, we can re-write Eq.~\ref{eq:randMarkRes}, at the same order of approximation, as
\begin{align} 
P^{ss}_m = \frac{\bar{R}}{\mathcal{E}_m} +  \frac{\sigma\, \zeta_m}{\bar{R}\, N^{3/2}} + \bigO{\frac{1}{N}}
\label{eq:randMarkExit}
\end{align} 
By plugging in Eq.~\ref{eq:exitR} for $ \mathcal{E}_m $ here and Taylor expanding, we can check that this is the same as Eq.~\ref{eq:randMarkRes} at this order. While the correction to the relation $ P^{ss}_m \propto 1/\mathcal{E}_m $ coming from the entrance rates remains important, even as $N \rightarrow \infty$ (since exit rate inhomogeneities also become small), it can be seen as random noise $\zeta_m$ around this trend (see blue points in fig.~\ref{fig:randDynamics}A).

Another key aspect of this result is that our random transition rates produce only small fluctuations in the steady-state distribution on top of a dominant uniform background. This is in sharp contrast to what we see for the smarticles, where steady-state probability piles up almost entirely in certain stable states. We may wonder if extreme values of the $ P^{ss}_m $ variation here may be comparable to the uniform background -- but from asymptotic expansion of the error function, we can check that for large $ N $ the maximum amplitude of $ \xi_m $ that we can expect only reaches $ \bigO{\sqrt{\log N}} $, short of the $ \bigO{\sqrt{ N}} $ we would need. This way even the largest amplitude $ P^{ss}_m $ variations are small compared to the uniform background.

To address this, make our model more realistic, and specifically, more like the self-organizing systems we are interested in, we can allow the transition rates $ R_{mn} $ to vary over a wide range of magnitudes. This is accomplished by drawing them, still i.i.d., from a log-normal distribution $ \log R_{mn} \sim \mathcal{N}(0,\sigma^2) $, with $ \sigma \gtrsim 4$. Because of its heavy tail, the central limit theorem is no longer a good approximation, and so $ \mathcal{E}_m $ and $ \mathcal{Z}_m $ are no longer normally distributed and can have a large variance that does not vanish as $ N $ grows. In this regime, our above derivation breaks down, but the relation $ P^{ss}_m \propto 1/\mathcal{E}_m $ is numerically shown to persist, and even improve (fig.~\ref{fig:randDynamics}B). Crucially, this allows for $ P^{ss}_m $ to vary over many orders of magnitude for different system states, thus better representing real heterogeneous systems, including our smarticle swarm.  

Counter-intuitively, the variance of the noise we see numerically about the $ \log P_{ss} = -\log \mathcal{E} $ line in fig.~\ref{fig:randDynamics}B does not diminish as $ N $ gets larger, and so can be thought of as an uncertainty inherent to our approximation for any system. Indeed, the correlation coefficient between $ \log P_{ss}$ and $ \log \mathcal{E}$, which is already seen to be quite similar for panels A and B of fig.~\ref{fig:randDynamics}, persists across different values of $ N $ and $ \sigma $. Moreover, this error can no longer be accounted for by the entrance rates, as the same correction as was done in panel A is seen in B to make things worse. This way, we see that while in general we need all $ N(N-1) $ rates to predict the steady-state exactly, the randomness in rates and large system size allow us to robustly approximate it in terms of just $ N $ local measurements of the exit rates.

Finally, we can show that ensemble average of log exit rates tends to go down over time as the system relaxes towards its steady-state from a uniform initial distribution. For this, we solve the master equation, Eq.~\ref{eq:masterEq}, $ P_m(t) = \exp{-t\; L_{mn}} P_n(0) $ for the uniform initial distribution $P_n(0) = 1/N$, and thus compute $ \<\log \mathcal{E}\>(t) = \sum_m P_m(t) \,\log\mathcal{E}_m$. For the log-normal rates $ R_{mn} $, we plot this numerically over time, showing monotonic decay (Fig.~\ref{fig:systems}B). This means that at long times, the system tends to be found in states with relatively low exit rates, and in fact $ \<\log \mathcal{E}\>(t)$ may be a Lyapunov function of such dynamics. 

\newpage
\subsection{Diffusion in random media}
\label{app:randDiff}

The Markov process described above evolved on a fully-connected graph, where any system state could transition to any other in a single step. Many-body dynamical systems, however, usually evolve in a $ d $-dimensional continuous configuration space, where some configurations are closer to each other than others. Thus, instead of using a fully-connected graph in the above Markov model, it would be more accurate to construct a $ d $-dimensional lattice. This essentially amounts to introducing a certain structure on the transition rates matrix $ R_{nm} $, by specifying which of the elements  are allowed to be non-zero. This way, making our model more realistic, requires restricting some of the random parameters, thus introducing correlations.

\begin{figure}[pt]
	\centering \includegraphics[width=1.\textwidth]{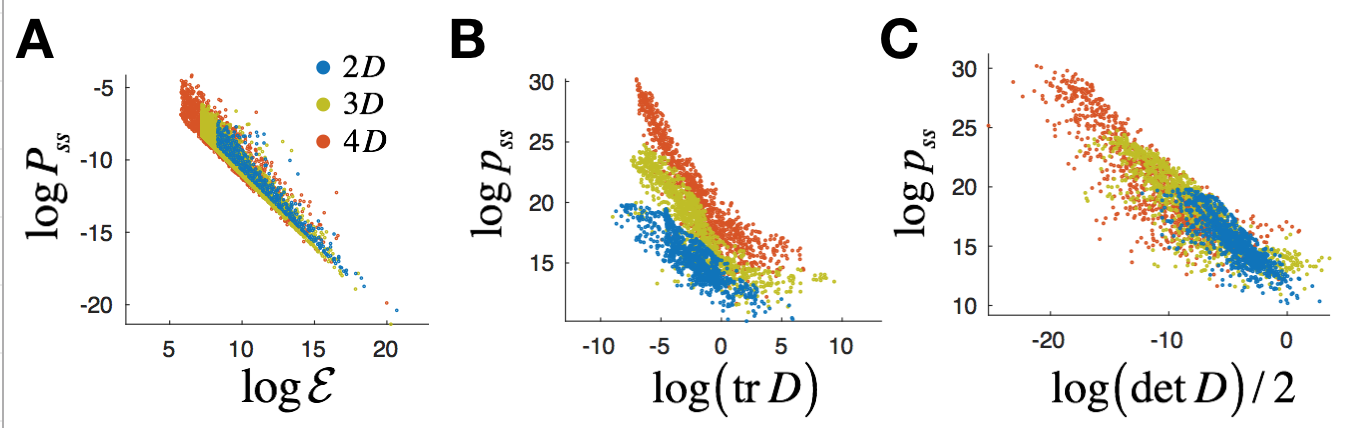}
	\caption{\textbf{Predictive quantities across dimensions}. (\textbf{A})~shows that, like in fig.~\ref{fig:randDynamics}, node exit rates $ \mathcal{E} $ are still predictive of steady-state probabilities $ P_{ss} $ when the Markov process is on a lattice (shown for 2, 3 and 4-dimensional lattices). Transitioning to the continuum in~(\textbf{B}), we see that we still have this correlation when exit rates are replaced by their continuum equivalents given by the trace of the diffusion tensor $ D \equiv \frac{1}{2}\Sigma\,\Sigma^T$ (see text). However, the slope of this correlation now changes with space dimension. (\textbf{C})~the determinant of $ D $ gives a similarly robust correlation as the trace, but also maintains the same slope across dimensions. Note that this is also the same as rattling $ \mathcal{R} = \frac{1}{2}\log \det D $.
	}
	\label{fig:rndDiffCompare}
\end{figure}

Figure~\ref{fig:rndDiffCompare}A numerically shows that despite this additional structure, the node exit rates $ \mathcal{E}_m  = \sum_{n \neq m} R_{nm}$ remain predictive of the steady-state probabilities $ P^{ss}_m$, just as in fig.~\ref{fig:randDynamics}B. The rates along lattice edges here are again chosen i.i.d. according to $ \log R_{nm} \sim \mathcal{N}(0,\sigma^2=25) $. Note that in higher-dimensional lattices, each state has more neighbors, and so in a sense, the limit $ d\to \infty $ recovers our fully-connected graph from previous section. This way, lower-dimensional lattices impose more structure on the transition rates $ R_{nm} $, giving a more stringent test of our framework. 

Moving further towards describing dynamical systems that evolve in continuous configuration spaces, we transition from studying Markov processes on discrete graphs to looking at $ d $-dimensional diffusion processes in the continuum. In a way, we can think of this as introducing the additional constraint of smoothness to our lattice transition rates. On top of this, however, working in continuum requires us to specify the appropriate quantities: instead of node probabilities $ P^{ss}_m $, we use probability densities $ p_{ss}(\v{q}) $; and instead of exit rates $ \mathcal{E}_m $, we will introduce rattling $ \mathcal{R} $ through the discussion below.

Consider a generic $ d $-dimensional diffusion process given by the Langevin equation:
\begin{align} 
\dot{q}_i = \Sigma_{ij}(\v{ q})\cdot \xi_j \quad \mbox{[It\^o]}
\label{eq:MAID1}
\end{align} 
where we sum over the repeated index, and $ D_{ij}(\v{ q}) \equiv \frac{1}{2}\,\Sigma_{ik}(\v{ q})\Sigma_{jk}(\v{ q})$ is some anisotropic diffusion tensor that depends on the particle position $ \v{ q} $, and $ \xi_i(t) $ is univariate white noise process in time given by $ \<\xi_i(t)\>=0 $ and $ \<\xi_i(t) \xi_j(t')\>=\delta_{ij}\, \delta(t-t') $, with $ i,j\in\{1,...,d\} $ indexing the dimensions of the configuration space. Again, as we want to understand the ``typical'' behavior of such dynamical systems, we take $ D_{ij}(\v{ q}) $ to be a random, but smoothly varying landscape in $ \v{ q} $. See \cite{bouchaud89randDiff} for a review of related dynamical systems, and how they can give rise to anomalous diffusion, and some more recent work \cite{solon2019-runTumble_disorder}. The choice of It\^o noise is the more appropriate one to later connect with driven dynamical systems (see discussion in \cite{Chv18least_ratt}). 

Here, It\^o noise implies that we are imagining a particle diffusing in an inhomogeneous temperature landscape, with some mean free path following each interaction with the bath, and a local temperature tensor $ \propto D_{ij}(\v{ q})$. This immediately implies that our problem is far from equilibrium. The Fokker-Planck equation that gives the corresponding evolution of probability density is:
\begin{align}
\dot{p}(\v{q}) = \partial_i\partial_j\left(D_{ij}(\v{q}) \, p(\v{q})\right) \label{eq:diffFP}
\end{align}
While a general analytical solution does not exist, there are two simple limiting regimes that we can solve to build some intuition. In the case of isotropic noise, when $ D_{ij}(\v{q})  = \delta_{ij} \,D(\v{q}) $, we can easily check that the steady-state distribution is $ p_{ss}(\v{q}) \propto 1/D(\v{q}) $. Keeping in mind our correlation plots, this means that $ \log p_{ss} $ is correlated with $ \log\left(\tr D\right) $ in this case. On the other hand, when different dimensions of $ \v{q} $ act as independent systems $ D_{ij}(\v{q}) =  \delta_{ij}\, D_i(q_i)$, then the probability factors along each dimension, such that $ p_{ss}(\v{q}) = \prod_i p_i(q_i) = 1/\prod_i D_i(q_i) = 1/\det D$. In this case, we see that $ \log p_{ss} $ is instead correlated with $ \log\left(\det D\right) $.

In general, however, we will not be able to find the steady-state of Eq.~\ref{eq:diffFP} analytically, and the solution $ p_{ss}(\v{q}) $ may not even be local. This can be understood by recognizing that this stochastic process is generally not detailed-balanced, and so $ p_{ss}(\v{q}) $ can depend non-locally on values of $ D_{ij}(\v{q}) $ anywhere. Here, we will find the steady-state numerically by directly simulating Eq.~\ref{eq:MAID1} and computing the probability densities according to the algorithm described in Materials and Methods. For this, we must first generate a smooth, random diffusion landscape. As we are looking for results that hold generally, we need not be careful with ensuring any ``nice'' properties of the random distribution for $ D_{ij}(\v{q}) $, such as isotropy or homogeneity in $ \v{q} $. We do, however, want to ensure that it is continuous and reasonably smooth in $ \v{q} $, and that it provides a vast diversity of noise amplitudes, to allow large range over which to see the correlation with $ p_{ss} $. 

To this end, for each of the $ d^2 $ entries of $ \Sigma_{ij} $,  we independently generate a $ d $-dim grid of random numbers (representing the configuration space $ \v{q} $). These are once again sampled according to a log-normal distribution, $ \log \Sigma_{ij} \sim \mathcal{N}(0,\sigma^2) $. Note that we must take care to keep the noise variance small enough to allow for reasonable step-sizes in our simulation, and so practically $ \sigma $ values around 3 and 4 were used. We then set up an interpolating function that returns the tensor $ \Sigma_{ij} $ for any input coordinates $ \v{q} $ provided by the time-integration loop of the simulation. This interpolator then ensures the smoothness and continuity of our diffusion landscape. Figure~\ref{fig:systems}A shows $ \log \left(\det D\right) $ for a typical such landscape in 2D.

\begin{figure}[pt]
	\centering \includegraphics[width=1.\textwidth]{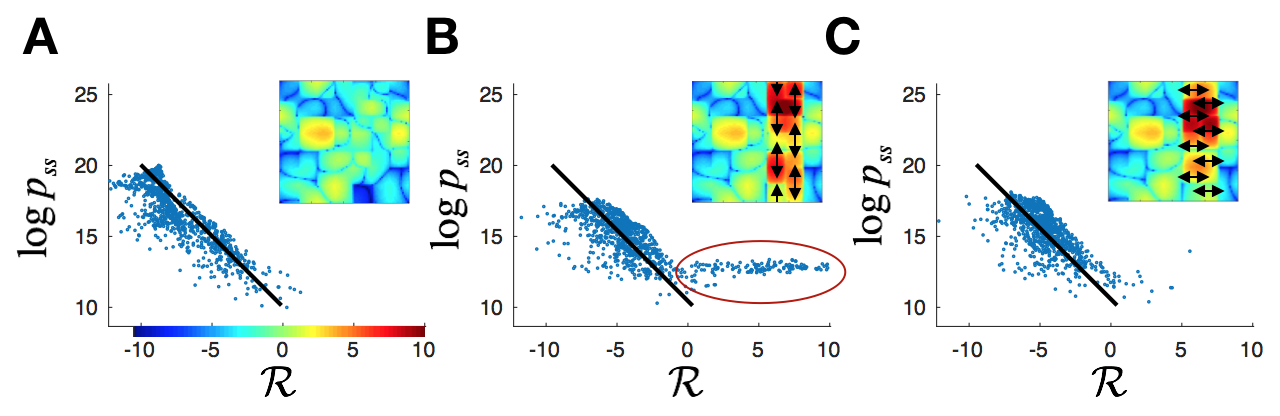}
	\caption{\textbf{Breaking the correlation requires specific global structure}.
	Starting with a typical 2D random diffusion landscape in~(\textbf{A}), we look at the effect of adding two different globally coordinated perturbations in (B~and~C). The insets show corresponding diffusion landscapes, colored by $ \mathcal{R} = \frac{1}{2}\log{\det D}$. In~(\textbf{B}), we enhance the vertical components of the diffusion along the vertical band shown (which actually forms a closed loop, due to periodic boundary conditions). This breaks our predictive correlation (black line) in that region, as illustrated by the circled line of points. In~(\textbf{C}), we instead enhance the horizontal diffusion component along the same band, and see that the correlation is largely restored everywhere. This illustrates that breaking this correlation requires a specifically adversarial global structure.}
	\label{fig:rndDiffBreak}
\end{figure}

With this, we can simulate the stochastic process in Eq.~\ref{eq:MAID1} in a smooth random diffusion landscape, and measure the resulting probability density $ p_{ss}(\v{q}) $. For every configuration we also know the diffusion tensor $ D_{ij}(\v{q}) $, but to see if it is predictive of the probability density, we must first construct a scalar from it. The two reasonable options that we saw in the discussion above are to take the trace or the determinant. Since analytically there is no strong argument for one or the other in general, we compare them numerically in fig.~\ref{fig:rndDiffCompare}, B and C. While we see that both show a clear correlation with $ p_{ss} $, the slope remains more consistent across dimensions if we choose the determinant. This reproduces our desired correlation between $ \log p_{ss} $ and $ \mathcal{R} $ in Eq.~\ref{eq:p_ss}, since the covariance matrix from Eq.~\ref{eq:Cdef} in this case evaluates to $ \mathcal{C}_{ij}(\v{q}^*)=\<\dot{{q}}_i(t), \dot{{q}}_j(t) \,dt\>_{\v{q}(t)=\v{q}^*}=\Sigma_{ik}(\v{q}^*)\Sigma_{jl}(\v{q}^*)\,\delta_{kl}\,\delta(t-t')\,dt =2 D_{ij}(\v{q}^*)$, and so rattling is $ \mathcal{R} = \frac{1}{2}\log \det D$ by Eq.~\ref{eq:rattling} (up to a constant offset). For the present case of diffusion in random media, we do not develop this point further, leaving a more in-depth and general discussion in the context of driven dynamical systems to Section~\ref{app:rattScalar}. 

As a last point, we also test our assumption that the correlation between $ p_{ss} $ and $ \mathcal{R} $ relies on the disorder and ``typicality'' of our chosen dynamics. This way, we should be able to break it by introducing some fine-tuning or specific structure into the diffusion landscape. In particular, if we have strong diffusive current running along a closed loop, then while $ \mathcal{R} $ will be large there, it will not contribute to any suppression of $ p_{ss} $, as this does not cause the trajectories to leave the loop any more frequently. We implement this scenario in 2D (for better visualization, as it works in the same way in higher dimensions), as shown in fig. \ref{fig:rndDiffBreak}B. This indeed causes a corresponding failure of the correlation: we see the appearance of states with high rattling but not a sufficiently low $ p_{ss} $. In fig. \ref{fig:rndDiffBreak}C, we see that if the direction of enhanced diffusion is not aligned along the loop, such that the added currents are not confined, then the correlation is restored. This illustrates that breaking our correlation in Eq.~\ref{eq:p_ss} really requires strong adversarial fine-tuning.

\newpage
\section{Rattling Theory}
\label{app:rattThy}

At its foundation, rattling theory arises from the suggestion that sufficiently ``messy'' dynamical systems may be more suitably described by diffusion in their configuration space, than by the many complex subsystem interactions. In this section, we develop this main hypothesis in several different ways, each starting with its own set of assumptions. In all cases, we are looking for a local quantity that can approximately predict the steady-state likelihood $ p_{ss} $ of any configuration in a large complex nonequilibrium system. The hallmark of such systems is that there are few general constraints that they need to obey. As they may be driven by some energy input, which gets dissipated into some thermal bath, energy is not conserved. Due to high dimensionality and ubiquitous nonlinearities, the system dynamics at one time are hardly predictive of its future evolution. Even the configuration space in which the system should be described is often not entirely known or accessible, as there may be microscopic or other hidden variables.

The one remaining rule still obeyed by these systems is locality in their configuration space: the system cannot discretely jump among states, it must evolve smoothly according to the structure of its configuration space. Our general approach will be to take this remaining constraint, and to assume that nothing else about the dynamics can be known or predicted.

In the rest of this section, we present four distinct arguments to frame this approximation more precisely, which can be viewed either as independent motivations, or as a unified derivation. First, in Section~\ref{app:rndForceField}, we give a quick motivation for why diffusive exploration of configuration space may be a broadly reasonable way to view such messy dynamical systems. In Section~\ref{app:max-ent} we show more carefully that if all we know about our system is the amplitude of local fluctuations throughout the configuration space, then our best guess for the dynamics (our ``null hypothesis'' of sorts) should be a diffusion process. Conversely, Section~\ref{app:inferT} starts off by assuming a diffusive approximation for the dynamics, and uses Bayesian inference to describe the appropriate estimator for the diffusion tensor based on a trajectory $ \v{q}(t) $. In Section~\ref{app:reparSym}, we further argue that if a local expression for steady-state probability density $ p_{ss}(\v{q}) $ exists, its form is largely constrained by the requirement that it must hold regardless of what coordinates we choose to describe our configuration space. We close in Section~\ref{app:rattScalar} with a discussion on how to then best define a scalar rattling $ \mathcal{R} $ value consistent with physical considerations of applying these ideas practically to driven dynamical systems. Note also that in~\cite{Chv18least_ratt}, we developed some of these ideas more rigorously for driven systems with two strongly separated time-scales, which we no longer assume here. Additionally, we note that the rattling quantity $ \mathcal{R} $ was defined slightly differently in that work. 

\subsection{Random first-order dynamics} \label{app:rndForceField}
A quick way to motivate why complex high-dimensional dynamics might often lead to approximately diffusive behavior comes from literature on motion in random force fields~\cite{bouchaud89randDiff}. If we write our dynamical evolution as a first-order system $ \dot{\v{q}} = \v{F}(\v{q},t) + \v{\xi}(t) $ in some high-dimensional configuration space (of dimension $ d $), then the assumption of ``complex and messy'' dynamics allows us to view the vector field $ \v{F}(\v{q},t) $ as basically random. Being smooth, it will generally have correlations in space and time, such that $ \overline{{F}_i(\v{q},t)\, {F}_j(\v{q}',t')}= G_{ij}(\v{q}-\v{q}', t-t')$, where overbar denotes average over random realizations of the field $ \v{F} $, and we take $ \overline{{F}_i(\v{q},t)}=0 $. While it may not be surprising that the resulting dynamics $ \v{q}(t) $ will be approximately diffusive when these correlations are short-range in space and time ($ G_{ij} $ decaying exponentially with $ \|\v{q}-\v{q}'\| $ and $ t-t' $), this turns out to hold much more generally. For example, if we let $ G_{ij} $ be entirely independent of time (infinite temporal correlation), and also have long-range power-law correlations in space $ G_{ij}(\v{q}-\v{q}')\propto \|\v{q}-\v{q}'\|^{-a}$, one can show that a trajectory starting from the origin will move in a diffusive fashion $ \|\v{q}(t)\|\propto t^{1/2} $, as long as $ \min(a,d)>2 $, see~\cite{bouchaud89randDiff}. Allowing time-dependence, this constraint may be relaxed further, thus suggesting that the diffusive approximation may indeed be reasonable for a wide array of dynamical systems.

\subsection{As max-entropy modelling} \label{app:max-ent}

Rather than trying to solve our complex system dynamics directly, which is often impossible analytically, we can try to first approximate the full behavior by something simpler. In some sense, the most ``natural'' such approximation can be found using maximum entropy modelling (or more precisely, ``maximum caliber,'' as in~\cite{dill18max_caliber}). We ask, if we knew nothing about our dynamics besides the amplitude of local fluctuations (two-point correlators) throughout the configuration space, what would be our most unbiased guess at what the dynamics really are? We can frame this question precisely by finding the maximum-entropy probability distribution $ P[\v{q}(t)] $ over the space of possible trajectories $ \{\v{q}(t)\} $, under the constraint that two-point correlators match some empirically known values. As with all maximum-entropy modeling, the result depends crucially on what we choose to constrain---even with two-point functions, we have the freedom to choose between equal-time correlator $ \<\dot{q}_i(t)\, \dot{q}_j(t)\>_{\v{q}(t)=\v{q}^*} $, integrated correlator $ \int ds \<\dot{q}_i(t)\, \dot{q}_j(t+s)\>_{\v{q}(t)=\v{q}^*} $, different choices of averaging, \textit{etc}. To begin, we will first try to constrain the equal-time correlator at every point in configuration space. Thus the entropy functional we want to maximize with respect to $ P[\v{q}] $ is (note that we use $ P[\v{q}] $ and $ P[\v{q}(t)] $ interchangeably here):
\begin{align}
&\mathcal{S} = P[\v{q}] \;\log\(P[\v{q}]\) + \lambda_0 \(1- \int \mathcal{D} \v{q} \;P[\v{q}]\) + 
\int d\v{q}^* \; \lambda_{ij}(\v{q}^*) \(\< \dot{q}_i\,\dot{q}_j\> _{\v{q}^*} - \mathcal{C}_{ij}(\v{q}^*)\)\\
&\mbox{with}\qquad \< \dot{q}_i\, \dot{q}_j\>_{\v{q}^*} \equiv \int \mathcal{D}\v{q} \; P[\v{q}] \,\int dt \; \dot{q}_i(t) \dot{q}_j(t) \,\cdot\,\delta(\v{q}(t) - \v{q}^*) 
\label{eq:noiseConstr}
\end{align}
where $\mathcal{D}\v{q}$ stands for integration over all possible trajectories $\v{q}(t)$, and we take the multiplication by the Dirac $ \delta $ function on second line in the It\^o sense. Here, we imply summation over all repeated indices, and $ \mathcal{C}_{ij}(\v{q}^*) $ is the empirical measurement of $ \< \dot{q}_i\, \dot{q}_j\>_{\v{q}^*} $ for the actual system. $ \lambda_0 $ and $ \lambda_{ij}(\v{q}^*) $ are Lagrange multipliers for the normalization and fluctuation constraints respectively. Note that, interestingly, in constraining the two-point function, we get a mixture of time-integration, functional integral over trajectory space, all on top of an integral over configuration space. Setting the variation $ \fd{\mathcal{S}}{P[\v{q}]} $ to 0, we get:
\begin{align}
&1+ \log(P[\v{q}])+\lambda_0 + \int d\v{q}^* \; \lambda_{ij}(\v{q}^*) \int dt \; \dot{q}_i(t) \dot{q}_j(t) \,\delta(\v{q}(t) - \v{q}^*) =0 \nonumber\\
&\Rightarrow \qquad P[\v{q}(t)]=\exp{-1-\lambda_{0}-\int dt \; \lambda_{ij}(\v{q}(t))\,\dot{q}_i(t) \dot{q}_j(t) } 
\label{eq:maxEnt}
\end{align} 
We must then set the Lagrange multipliers such that our constraints are satisfied. This means that $ \lambda_0 $ is chosen to give proper normalization, and from Eq.~\ref{eq:maxEnt} we get $ \< \dot{q}_i\, \dot{q}_j\>_{\v{q}^*} = \frac{1}{2}\lambda^{-1}_{ij}(\v{q}^*) $, which we must then set to $ \mathcal{C}_{ij}(\v{q}^*)$, giving us the result
\begin{align}
P[\v{q}(t)]=\frac{1}{Z}\; \exp{-\frac{1}{2}\int dt \; \mathcal{C}^{-1}_{ij}(\v{q}(t)) \,\dot{q}_i(t) \dot{q}_j(t)}
\end{align}
But this is precisely the probability distribution for diffusive dynamics! Just as in Eq.~\ref{eq:MAID1}, this is $ \dot{q}_i = \Sigma_{ij}(\v{ q})\cdot \xi_j $ with $ \mathcal{C} = \Sigma\, \Sigma^T $. 

In a sense, this result suggests that the least-informative Bayesian prior over the system's dynamics is configuration-space diffusion, once constraints over local fluctuations $ \mathcal{C}(\v{q}) $ are taken into account. This can be seen as the ``null hypothesis'' we should have about system dynamics if all we know are the local fluctuations. But we have already studied such diffusion processes in Section~\ref{app:randDiff}---which means that it is reasonable to suggest that results from that section will apply to studying complex dynamical systems.

This derivation turns out to be remarkably fragile to the choice of constraint. If instead of the equal-time correlator we chose to constrain almost anything else, such as the connected or the integrated correlator, we would either have no closed-form expression for $ P[\v{q}(t)] $ at all, or have no way to easily express $ \lambda_{ij}(\v{q}^*) $ in terms of $ \mathcal{C}_{ij}(\v{q}^*) $---and in particular, the dependence would be non-local. This is because the distribution in Eq.~\ref{eq:maxEnt} is non-Gaussian in $ \v{q} $, and so the two-point correlator is not simple in general.  The only other interesting local constraint that would still allow the derivation to go through is that of $ \tr\mathcal{C} $, which would give isotropic diffusion. While on the one hand, these are merely practical limitations of this method, they may indicate that no other simple dynamical process can be seen as a max-entropy approximation of a complex system dynamics. This can already meaningfully restrict our space of allowed constraint choices, and hence, of the resulting diffusive approximations.

A particularly important instance of such a restriction is the fact that our result here yields an It\^o diffusion process~\cite{yang13ExtLang_Tgrad}. Similar to the above discussion, our derivation only gives a closed form local dynamics if we choose to constrain the fluctuations in the It\^o sense in Eq.~\ref{eq:noiseConstr}. For driven dynamical systems, this result is further motivated in~\cite{Chv18least_ratt} for a restricted context. Physically, we can motivate this by seeing that as the system moves through the configuration space, its fluctuations when it is at $ \v{q}(t+dt) $ are still determined by the response properties of configuration $ \v{q}(t) $, due to a finite relaxation time, thus leading to the non-anticipating It\^o noise convention.

\subsection{As inference problem} \label{app:inferT}

In the last section, we saw that approximating system dynamics by configuration-space diffusion is a natural choice. Here we will ask the inverse question: assuming that the diffusive approximation $ \dot{q}_i = \Sigma_{ij}(\v{ q})\cdot \xi_j $ (in the It\^o sense) is reasonable, we ask how should we empirically best estimate the diffusion tensor $ D_{ij}(\v{ q}) \equiv \frac{1}{2} \Sigma_{ik}\, \Sigma_{jk}$? Let's assume that for some observed dynamical process $ \v{q}(t) $ (generated by a black box), the underlying motion is indeed governed by our diffusion equation. We can then use Bayesian inference to find the maximum-likelihood estimation of $ D_{ij}(\v{ q}) $ from observations of trajectories~\cite{zdeborova2016inferencePhys,boyer2012singTrajD}. The inference problem is thus: $ P\[D(\v{q})\,|\,\v{q}(t)\]= P\[\v{q}(t)\,|\,D(\v{q})\]\, P\[D(\v{q})\]/Z $ (where $ Z = P[\v{q}(t)] $, normalization). For our prior distribution $ P\[D(\v{q})\] $, we assume only that the diffusion tensor varies smoothly over $ \v{q} $---otherwise it would have infinite degrees of freedom and we could never hope to estimate it from any finite observation of a trajectory. Explicitly, this can be implemented by biasing the spatial derivatives of $ D $ to smaller values: $ P\[D(\v{q})\] \propto \exp{-\epsilon \int d^d \v{q}\; \partial_k D_{ij} \partial_k D_{ij}}$, with some coefficient $ \epsilon $. Note that $ \epsilon \to 0 $ conveniently restores the uniform prior. From our diffusion equation above, we can write the distribution for Gaussian white noise $ \xi(t) $ at every time point, which gives $ P\[\v{q}(t)\,|\,D(\v{q})\] = \prod_t \[ \sqrt{\frac{1}{\det(4\pi D)}}\,\exp{-\frac{1}{4} D^{-1}_{ij}\dot{{q}}_i \dot{{q}}_j} \]$. This way we can write our inference problem as
\begin{align}
 P\[D(\v{q})\,|\,\v{q}(t)\] = \frac{1}{Z[\v{q}(t)]} \; \mbox{exp} \Bigg[-\int dt \(\frac{1}{2}\, \tr \log D(\v{q}(t)) + \frac{1}{4}\, D^{-1}_{ij}(\v{q}(t))\;\dot{{q}}_i(t)\,  \dot{{q}}_j(t)\) \nonumber\\
 	-\;\epsilon \int d^d \v{q}'\; \partial_k D_{ij}\partial_k D_{ij}\Bigg]
\end{align}
As in the section above, we see an interesting mix of integrals over configuration space and over trajectory duration in the exponent. We then want to choose some estimator for $ D(\v{q}) $. The easiest to evaluate here is the maximum-likelihood, which we can get by setting the variation $ \fd{\quad}{D_{ij}(\v{q})} $ of the exponent to 0:
\begin{align}
\fd{\quad}{D_{i'j'}(\v{q}^*)} \int d\v{q}' &\[\epsilon\; \partial_k D_{ij}\partial_k D_{ij} 
+ \frac{1}{2}\, \tr \log D(\v{q}') 
+ \frac{1}{4} D^{-1}_{ij}(\v{q}') \int dt \;\dot{{q}}_i(t)\,  \dot{{q}}_j(t) \; \delta(\v{q}(t)-\v{q}')\] = 0\nonumber\\
&-\;2\epsilon\,\partial_k \partial_k D_{ij} (\v{q}^*)
+ \frac{1}{2} D^{-1}_{ij}(\v{q}^*) - \frac{1}{4} D^{-1}_{ik}(\v{q}^*) \<\dot{{q}}_k(t)\,  \dot{{q}}_l(t)\>_{t\; |\;\v{q}(t)=\v{q}^*} D^{-1}_{lj}(\v{q}^*) =0\nonumber\\
\Rightarrow \qquad& 4\epsilon\, \left. D_{in}D_{jm} \partial_k^2 D_{mn}\right|_{\v{q}^*} = D_{ij}(\v{q}^*) - \frac{1}{2}\<\dot{{q}}_i(t)\,  \dot{{q}}_j(t)\>_{t\; |\;\v{q}(t)=\v{q}^*}
\label{eq:inferDiff}
\end{align}
where the average on the last line is over all the times when trajectory $ \v{q}(t) $ passes through $ \v{q}^* $. So if we had a uniform prior $ \epsilon\rightarrow 0 $, then we would have our result, stating that $ D_{ij}(\v{q}^*) = \frac{1}{2}\<\dot{{q}}_i(t)\,  \dot{{q}}_j(t)\>_{t\; |\;\v{q}(t)=\v{q}^*}$. For $ \epsilon>0 $, any deviations from this rule become sources for the Poisson equation Eq.~\ref{eq:inferDiff}, thereby smoothing out the $ D(\v{q}^*) $ landscape. Thus for small $ \epsilon $, we simply get that the average $ \<\cdot\> $ above is taken not only at the point $ \v{q}^* $, but over its neighborhood: 
\begin{align} 
D_{ij}(\v{q}^*) = \frac{1}{2}\<\dot{{q}}_i(t)\,  \dot{{q}}_j(t)\>_{t\; |\;(\v{q}(t)-\v{q}^*)^2 \,<\, \epsilon}\,,
\label{eq:Cij}
\end{align} 
which basically reproduces our covariance estimator $ \mathcal{C}_{ij} $ in Eq.~\ref{eq:Cdef} (see Section \ref{app:eval_R} for details about the practical choice for the construction of that estimator).

Now, we can use this result to take our observed trajectories $ \v{q}(t) $, which came from some non-diffusive dynamical system, and infer the effective diffusion landscape that might have created such trajectories. This will always produce some result, regardless of how poor of a model diffusive behavior is for our dynamics---but if we do want to use the diffusive approximation, then this gives the optimal expression for $ D_{ij}(\v{q}) $. 

Note that if we went through these steps starting instead with the assumption of isotropic diffusion $ \dot{q}_i = \sqrt{2 D(\v{ q})}\cdot \xi_i $, then our estimator for $ D(\v{q}) $ would have been the trace of the result in Eq.~\ref{eq:Cij}. Also, while maximum likelihood estimators are not always reliable, it does not seem analytically tractable here to calculate others, such as minimal mean-squared error. Our result in this section motivates the general approach to calculating the effective local diffusion tensor at any configuration for a driven dynamical system, as used in Eq.~\ref{eq:rattling}.

\subsection{By reparametrization symmetry} \label{app:reparSym}

To get an expression for $ p_{ss}(\v{q}) $ we then need to solve the resulting diffusion problem. As we saw empirically in Section~\ref{app:randDiff}, for such diffusion $ \log p_{ss} $ correlates both, with $ \log \tr \mathcal{C} $ and $ \log \det \mathcal{C} $, where $ \mathcal{C}(\v{q}) $ is the noise covariance matrix, an estimator for the diffusion tensor, as seen in Eq.~\ref{eq:Cij}. 
On the other hand, the result we presented in the main text can be written, by combining Eq.~\ref{eq:rattling} and Eq.~\ref{eq:p_ss}, as
\begin{align}
&p_{ss}(\v{q})\propto \(\det \mathcal{C}(\v{q})\)^{-\gamma/2} 
\label{eq:p_ssC}
\end{align}
In this and the following sections, we will focus on motivating this expression.

For the argument of this section, we need only to assume that there exists some expression for $ p_{ss}(\v{q}) $ in terms of the system's local dynamics at $ \v{q} $. Since both, $ p_{ss}(\v{q}) $ and the dynamics $ \v{q}(t) $ are described with respect to the particular coordinates arbitrarily chosen on the configuration-space, the form any such expression relating the two must be invariant under any coordinate transformations. This will be true as long as there is no physically-selected parametrization, such that any choice of coordinates is arbitrary. In other words, our expression is constrained by local diffeomorphism symmetry of the configuration space. This turns out to be a strong restriction, which almost entirely constrains us to Eq.~\ref{eq:p_ssC}---without even needing to directly appeal to the diffusive approximation of the previous sections.

More concretely, the parametrization chosen for the smarticle system in Materials and Methods, as well as the coarse-grained descriptions often used in other active matter systems, are typically arbitrary, while also being highly nonlinear and inhomogeneous with respect to the microscopic system coordinates. For any such choices of description to give the same form of expression for $ p_{ss}(\v{q}) $, we must impose invariance not only under global, but also under arbitrary local coordinate changes, as $ \v{q} \to \v{q}'=\v{f}(\v{q}) $. We know that under such transformation, $ p_{ss} $ will pick up a Jacobian determinant factor, since $ d^d\v{q}\; p_{ss}(\v{q}) = d^d\v{q}'\; p_{ss}'(\v{q}') $, giving $p_{ss}'(\v{q}') = p_{ss}(\v{q}) / \det J(\v{q}) $, with the Jacobian matrix $ J_{ij}(\v{q})=\pd{f_i(\v{q})}{q_j} $. 

Thus, if we assume that a local expression for $ p_{ss}(\v{q}) $ exists, it must transform in this same way. The only local information about a complex dynamical system we can have, in general, at some configuration $ \v{q}^* $ is some statistics about how it tends to behave in this neighborhood, which we can write as $ P[\v{q}(\tau) \;|\; \v{q}(0)=\v{q}^*] $. For this information to be truly local to $ \v{q}^* $, we should only look at short times $ \tau \to 0 $, which means that this local behavior can be fully captured by the set of derivatives as  $ P[\{\dot{\v{q}},\ddot{\v{q}},...\}_t \;|\; \v{q}(t)=\v{q}^*] $. At this point, the simplest expressions we can construct from this information, consistent with our local diffeomorphism symmetry (and hence transforming the same way as $ p_{ss} $ under coordinate changes), are of the form
\begin{align} 
p_{ss}(\v{q}^*) \propto \( \det\<\partial_t^n q_i \,,\,\partial_t^m q_j\>_{\v{q}(t)=\v{q}^*}\)^{-1/2}.
\label{eq:reparExpr}
\end{align}
We see that under the change of coordinates, the Jacobian factors out from the determinant, and the $ -\frac{1}{2} $ power then results in the same $ 1/\det J(\v{q}) $ factor as on the left hand side. The order of time-derivatives $ n $ and $ m $ remains unconstrained by this, as does the question of whether the correlator $ \<\cdot,\cdot\> $ is connected or not. Moreover, any sums of such terms as in Eq.~\ref{eq:reparExpr}, with different values of $ n $ and $ m $, are also allowed. Not only this, but such summation can also be performed inside the parenthesis, or even inside the determinant, without breaking reparametrization invariance. All these possibilities end up giving us a series of a combinatorially large number of possible factors for $ p_{ss}(\v{q}) $. 

Thus, to further constrain this result and arrive at Eq.~\ref{eq:p_ssC}, we must use other arguments. First, we note that since $  \det\(\<\partial_t^n q_i \>\<\partial_t^m q_j\>\)=0 $, we know that averaging must be outside of the product. Second, for highly chaotic noisy motion $ \v{q}(t) $, higher time-derivatives will generally be large. Indeed, for an ideal diffusion process $ \dot{\v{q}}=\sqrt{2D(\v{q})}\cdot\v{\xi} $ all but the first time-derivative will diverge. While physical complex dynamics are often smooth, it may still be reasonable to assume that
\begin{align} 
\dot{\v{q}} \ll \tau^n\,\partial_t^{n+1} \v{q} \qquad \forall n>0
\label{eq:reparAss}
\end{align}
The time-scale $ \tau $ should be chosen relative to the resolution with which we measure $ p_{ss}(\v{q}) $. Since probability density can only be defined over some finite neighborhood $ \epsilon $, it is reasonable to pick $ \tau $ as the time it takes $ \v{q}(t) $ to traverse this neighborhood. Of course, on sufficiently microscopic scales of resolution in $ \v{q} $ (small $ \epsilon $, and so $ \tau \to 0 $), our chaotic dynamics may start looking smooth, and this approximation breaks down. However in that regime, our fundamental assumption that the system is highly complex and chaotic no longer applies, and so it is reasonable that our predictive correlation between $ p_{ss} $ and $ \mathcal{R} $ should fail. On the other hand, such smooth microscopic regime may be tractable with the more standard direct solution techniques. Thus, in many dynamical systems we may see the scenario that while the microscopic local dynamics can be described in terms of usual perturbation theory, on larger scales rattling theory becomes the more appropriate description. This view suggests that our theory could emerge as the infra-red fixed-point of renormalization group flow under configuration space coarse-graining, which may in the future allow to quantify its universality properties.

For now, we see that the assumption in Eq.~\ref{eq:reparAss} implies that all terms in Eq.~\ref{eq:reparExpr} that contain any number of higher derivatives will be negligible compared to the one dominant term $ \( \det\<\dot{q}_i ,\,\dot{q}_j\>_{\v{q}(t)=\v{q}^*}\)^{-1/2} $. This recovers our main result in the paper, as stated in Eq.~\ref{eq:p_ssC}.

More than that, however, this expression further requires $ \gamma=1 $, which we know empirically does not exactly hold in some cases (see Fig.~\ref{fig:systems}D). Such violations can arise if the effective noise landscape $ \mathcal{C}(\v{q}^*)\equiv \<\dot{q}_i ,\,\dot{q}_j\>_{\v{q}(t)=\v{q}^*} $ has some regularity or structure that biases towards a preferred parametrization of the space. This is consistent with the observation that the only deviation from $ \gamma=1 $ we saw here was in the mesoscopic three-robot swarm, which is small enough that such regularity is not unlikely.

Note that the local reparametrization symmetry used here is roughly equivalent to the assumption that nothing but locality can be known about the system's complex dynamics, discussed in the beginning of Section~\ref{app:rattThy}. The presence of any additional structure, such as finite relaxation time or persistence length, breaks the reparametrization symmetry at the corresponding scales---as we saw above with $ \tau $. The extent to which this symmetry fails gives the extent to which a truly local expression for $ p_{ss}(\v{q}) $ does not exist, making our results approximate. At the same time, for the equilibrium Boltzmann distribution $ p(\v{x}) \propto \exp{-\beta\,U(\v{x})} $ reparametrization symmetry is not obeyed for a different reason. There, while all nonlocal effects are irrelevant, a particular physical coordinate system gets selected by the requirement that thermal noise amplitude be uniform throughout space $ \v{x} $. This way, local reparametrization invariance is quite unique to our method in this work. 

\subsection{Defining rattling in driven dynamical systems} \label{app:rattScalar}
In this section we discuss some physical considerations about active matter systems, thus constructing an appropriate definition of rattling for practical applications. We will further argue why the determinant of the covariance matrix $ \mathcal{C} $ may be the more appropriate choice for our theory, even though as mentioned above, trace works better with the derivations in Sections~\ref{app:max-ent} and~\ref{app:inferT}. Finally, we will motivate our specific definition for $ \mathcal{R} $ in Eq.~\ref{eq:rattling} from practical considerations.

\subsubsection{Composite systems}
Similar to the requirement of reparametrization symmetry in the last section, here we discuss another consistency condition that must be satisfied for our theory to apply to many-body systems. In that context, we may often see the scenario where some groups of agents become sufficiently separated so as to become entirely independent. In this case, the steady-state probability of such configurations should factor as a product of the probabilities for the constituent sub-system
\begin{align}
p_{ss}(\v{q}) = p_{ss}(\v{q}^{(1)}) \;p_{ss}(\v{q}^{(2)})
\end{align}
Here, we take $ \v{q} \in \mathcal{Q}$ to be an element of the full system configuration space, while $ \v{q}^{(1)} \in \mathcal{Q}_1$ and $\v{q}^{(2)} \in \mathcal{Q}_2$---of their respective subsystem spaces, such that $ \mathcal{Q} = \mathcal{Q}_1 \oplus \mathcal{Q}_2 $. This way, the full $ \mathcal{C}_{ij}(\v{q}) $ will be block-diagonal (up to spurious correlations, which we assume to be small in a large system), and thus the determinant factors as needed
\begin{align}
\det \mathcal{C}(\v{q}) =  \det\mathcal{C}_1(\v{q}^{(1)}) \; \det \mathcal{C}_2(\v{q}^{(2)})
\end{align}
As this must hold for any decomposition of the full space $ \mathcal{Q} $ into any number of subsystems that may at one time or another decouple from each other, we get another strong motivation for choosing the determinant of $ \mathcal{C} $ over its trace. Note that unlike reparametrization symmetry, this does not put restrictions on the power $ \gamma $ in Eq.~\ref{eq:p_ssC}. 

\subsubsection{Filtering ordered motion}
Another important consideration for driven systems, which does not come up in actual diffusion processes as in Section~\ref{app:randDiff}, is that any component of the dynamics that is ordered should not contribute to raising the value of $ \mathcal{R} $. This is because in estimating the exit rate from a system state, we should ignore any motion that predictably returns soon after, as this does not contribute to suppressing the steady-state likelihood of that state. Thus, \textit{e.g.}, for periodic drives, we want to ignore any similarly periodic dynamics of the system. In general, however, this is a very subtle point, as such ``ordered behavior'' may be tough to identify in a given system, or even to define. 

In particular, we note that $ \tr \mathcal{C}(\v{q}) $ actually evaluates to the mean-squared velocity $ \<{\dot{q}}_i^2\>_{\v{q}} $, and thus is affected by stochastic and ordered motion in exactly the same way. In contrast, the determinant can be seen as the volume of configuration space into which the system dynamics starting from $ \v{q} $ may spread under the drive. If this spread predominantly extends along just a few dimensions of the configuration space, its volume will be close to 0. Thus, insofar as ordered motion does not explore all dimensions of the configuration space, its contribution to $ \det \mathcal{C}(\v{q}) $ will be small. This way, the determinant can heuristically tease apart the stochastic (or chaotic) component of the dynamics based on its space-filling property. Note that for this reason, while we can use coarse-grained descriptions of our complex system, the configuration space must be sufficiently high dimensional so as to make space-filling ordered trajectories unlikely.

We can better ground this discussion in physical intuition by formulating it in terms of the entropy of local forces. Indeed, for the diffusion process $ \dot{q}_i = \Sigma_{ij}(\v{q})\cdot \xi_j $, the entropy of random forces acting on configuration $ \v{q} $ is given by (up to a constant offset)
\begin{align} 
\mathcal{S}(\v{q}) = \log \det \Sigma = \frac{1}{2} \log \det \<\dot{q}_i ,\, \dot{q}_j\>_{\v{q}}
\label{eq:Snorm}
\end{align}
This comes from the expression for entropy of a normal distribution, since here $ \dot{\v{q}} \sim \mathcal{N}(0,\Sigma\,\Sigma^T)$. More generally, it seems that this entropy is precisely what we want to measure to quantify the continuum version of state exit rates, or as the ``stochastic component'' of the dynamics. For this reason, we define rattling $ \mathcal{R}(\v{q}) $ as this force entropy. Additionally, this choice allows us to easily relate rattling to information-theoretic properties of the drive, which turns out crucial for understanding self-organization (see Fig.~\ref{fig:melt} and fig.~\ref{fig:adapt}).

\subsubsection{Practical evaluation of $ \mathcal{R} $} \label{app:eval_R}
Finally, we must discuss the practicalities of measuring $ \mathcal{R}(\v{q}) $ from limited observations of trajectories $ \v{q}(t) $. Since we will need to estimate $ \mathcal{R} $ for all configurations we are interested in, the calculation for any one configuration must be very light in terms of the required data. As such, measuring the actual distribution of local forces for each $ \v{q} $ to then evaluate its entropy is impractical---and likely unnecessary as our entire framework is approximate anyway. We can instead quickly get a rough estimate of these entropies by assuming the force distribution to be Gaussian, which allows us to simply use the expression in Eq.~\ref{eq:Snorm}. One important difference between this and the rest of the above discussion, is that here we use the connected correlator in the definition of $ \mathcal{C}_{ij}(\v{q}) = \<\dot{q}_i ,\, \dot{q}_j\>_{\v{q}} \equiv \<\dot{q}_i \, \dot{q}_j\>_{\v{q}} - \<\dot{q}_i \>_{\v{q}}\< \dot{q}_j\>_{\v{q}}$. This distinction did not come up before because for a diffusion process $ \<\dot{q}_i \>_{\v{q}}=0 $. In contrast here, this change helps to further reduce the effect of ordered motion on $ \mathcal{R} $.

There is one other approximation that we make in this work from practical considerations. The averages that are taken in evaluating $ \mathcal{C}_{ij}(\v{q}^*) $, such as in Eq.~\ref{eq:Cij}, require us to know the distribution $ P(\dot{\v{q}} \;|\; \v{q}=\v{q}^*) $ under the influence of the drive. First, measuring this empirically requires many re-initializations of the system in configuration $ \v{q}^* $, which is often very impractical---single-shot measurements of $ \mathcal{R} $ would be much preferred. Second, as the drives we use in our experiments often have temporal correlations that we are interested in (\textit{e.g.}, periodicity), capturing their effect requires average over appropriate time-windows. Fortunately, we can use these two considerations to resolve each other by assuming something like ``local ergodicity.'' As such, we assume that a single short rollout $ \v{q}(t) $ starting from $ \v{q}(0)=\v{q}^* $ and evolving under the influence of the drive for a representative time-segment, can give us a sampling of the distribution $ P(\dot{\v{q}} \;|\; \v{q}=\v{q}^*) $. As this distribution talks about velocities at the point $ \v{q}^* $, we approximate them by finite differences $ \v{q}(t)-\v{q}^* $ for the collection of times $ t $ along this trajectory. Moreover, as we are estimating our local force entropy by that of a diffusion process (Eq.~\ref{eq:Snorm}), we need to divide each finite difference by $ \sqrt{t} $, rather than the usual $ t $, so that our ergodicity approximation can be stated as
\begin{align} \label{eq:locErg}
\Big\{\dot{\v{q}} \;|\; \v{q}=\v{q}^*\Big\}_{\text{noise realizations}} \sim \left\{\frac{\v{q}(t)-\v{q}(0)}{\sqrt{t}} \;\Big|\; \v{q}(0)=\v{q}^*\right\}_{\text{single short run}}
\end{align}
which is, almost certainly, a very rough approximation. Nonetheless, in practice, it allows efficient empirical estimate of local rattling values $ \mathcal{R} $ even in real-time, and which, as demonstrated, are robustly predictive of the steady-state probabilities. This way, evaluating Eq.~\ref{eq:Snorm} using this estimate, we finally recover the definition of rattling $ \mathcal{R} $ stated in Eq.~\ref{eq:rattling}.

One comment about the approximation in Eq.~\ref{eq:locErg} is worth making. As rattling is closely related to the local diffusion coefficient $ D(\v{q}) $, its evaluation here should be considered in the context of other studies estimating the diffusion constant from single particle tracking data~\cite{boyer2012singTrajD,michalet2012singTrajDlim}. If we take $ \v{q}^*=0 $ to be the origin, then the estimator used here may be written as $ D_{ij}(\v{q}^*=0) \propto \frac{1}{\tau}\int_{0}^\tau dt \;t^{-1}\,q_i(t) q_j(t)$. This choice is less common than, \textit{e.g.}, the maximum-likelihood estimator $D_{ij} \propto \frac{1}{\tau^2}\int_{0}^\tau dt \,q_i(t) q_j(t) $, or the linear-least-squares estimator $D_{ij}\propto \frac{1}{\tau^3}\int_{0}^\tau dt \,t\, q_i(t) q_j(t) $~\cite{boyer2012singTrajD}, but we see that the only difference between these is the relative weighting of the data at early vs. late times~\cite{michalet2012singTrajDlim}. Since the main challenge for us is to estimate the diffusion tensor associated to the specific location $ \v{q}^* $, it therefore makes sense to up-weigh the early data when the trajectory $ \v{q}(t) $ is closest to $ \v{q}^* $. Note one consequence of this choice is that for diffusion where $ D(\v{q})=D $ is uniform, our estimator will not recover the exact diffusion tensor $ D $ from a single trajectory, even if the trajectory is long $ \tau \to \infty $: averaging over multiple rollouts would be necessary. 

\newpage
\section{Self-organization with 3 smarticles}

Since rattling arises due to the interplay of the internal and applied forces on the system, we can view every applied drive pattern as creating its own ``rattling landscape'' on the configuration space. For us, rattling plays the same role as energy does in the Boltzmann distribution, and so the intuition of energy landscapes applies quite well here. For the drive illustrated in Fig.~\ref{fig:smarticle}B, when all 3 smarticles are executing the same synchronous motion, the resulting rattling landscape is shown in fig.~\ref{fig:sqGait}A. Panel B then shows the corresponding steady-state distribution, which is seen to be localized in the low-rattling regions of A. This correspondence is then quantitatively confirmed by plotting steady-state density versus local rattling values. 

\subsection{Spontaneous fine-tuning to drive properties}

\begin{figure}
	\centering\includegraphics[width=0.7\textwidth]{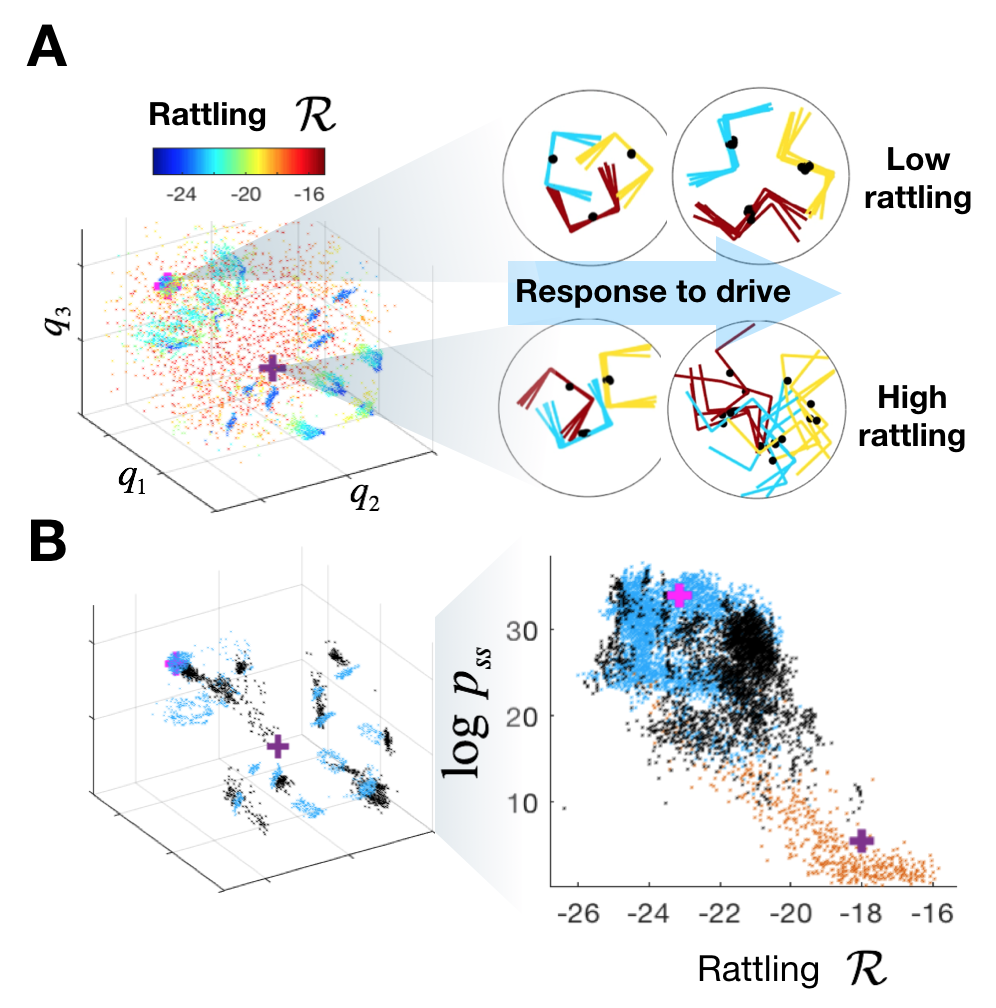}
	\caption{\textbf{Rattling landscape predicts steady-state}.
	(\textbf{A})~illustrates the rattling landscape for the arm-motion pattern shown in Fig.~\ref{fig:smarticle}B (also in fig.~\ref{fig:adapt}, drive 3). Two configurations, one low-rattling and one high-rattling, are labeled with pink and purple crosses, respectively. To illustrate the contrast, we show the short-time response of 5 nearby configurations to the drive pattern (see movie S4). In this setup, rattling can be measured as the entropy of the distribution of resulting configurations. (\textbf{B})~shows the steady-state distribution, which is concentrated primarily in the low-rattling regions of the configuration space. This observation is then quantified by plotting steady-state density vs. local rattling values around 5000 randomly chosen configurations: in red we sample these uniformly at random, and in blue and black we sample from the steady-state. Black shows experimental data. All other data in this figure is from simulations.
	}
	\label{fig:sqGait}
\end{figure}

\begin{figure}[pt]
	\centering\includegraphics[width=1.\textwidth]{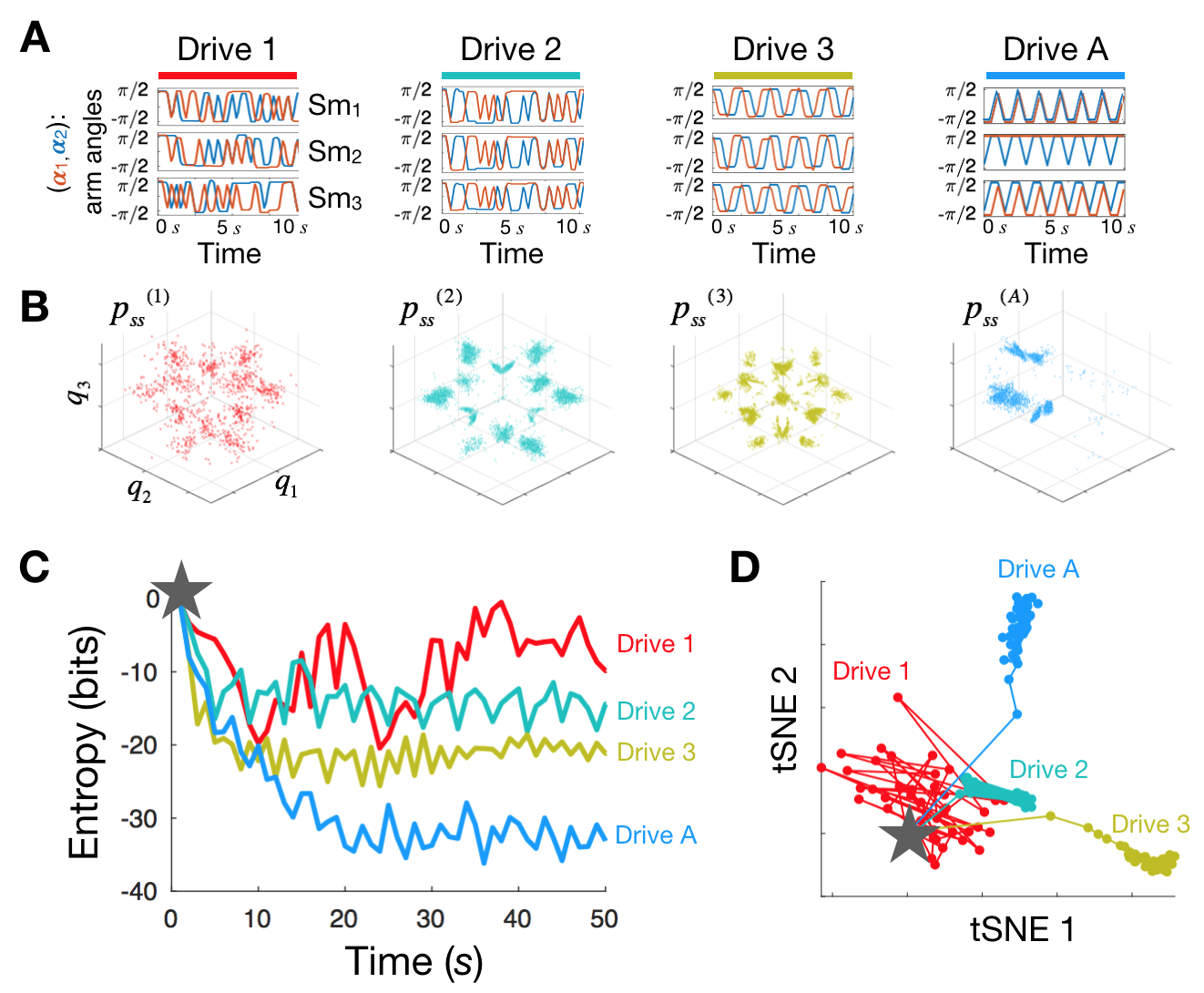}
	\caption{\textbf{Steady-state distributions encode drive properties}.
		(\textbf{A})~shows 4 distinct arm-motion patterns, in order of increasing predictable information content, from fully random to an intricate periodic form. (\textbf{B})~shows their corresponding steady-state distributions (note different orientation of the 3D axes compared to Fig.~\ref{fig:smarticle} and fig.~\ref{fig:sqGait}). Reflecting drive properties, these get concentrated into smaller fraction of configuration space for more intricate drives. (\textbf{C})~quantifies this effect by showing distribution entropies. Applying each of the 4 drives to the  uniform initial distribution (grey star), we can see how drive information gets gradually imprinted on it, causing the low-entropy fine-tuning in the steady-state, with more intricate drives causing more fine-tuning. Finally, to quantify the specificity of each distribution to its drive, we calculate ``distances'' between pairs of distributions as relative entropies $D_{KL}$. We then use these in~(\textbf{D}) to visualize the path that distributions take as they evolve from uniform initial conditions under the influence of their respective drive. Starting from the same point, we can see precisely how their shapes diverge over time.
		}
		\label{fig:adapt}
\end{figure}

In order to understand how drive properties inform the resulting rattling landscape, and hence self-organized states. We pick 4 distinct arm-motion patterns shown in fig.~\ref{fig:adapt}A. For drive 1, each smarticle moves both arms independently at random between the limiting angles $\pm \pi/2$ (this is the drive in Fig.~\ref{fig:melt}, right column). For drive 2, the motion is still random in time, but synchronized across the 3 smarticles. Drive 3 is also identical across smarticles, but now deterministic in time (this drive is illustrated in Fig.~\ref{fig:smarticle}B and used for fig.~\ref{fig:sqGait}). Finally, drive A assigns distinct motion patterns to the three smarticles, while still being fully deterministic (this one is used in Fig.~\ref{fig:drive-specific}A). 

This way, each of these drives has more ``predictive information'' than the last, defined as the amount of information about drive history that can help to predict its future~\cite{marzen16predictive_infMarkovProc}. Since drive 1 is random, no knowledge of the past will help to know the future---other than that arms always go to angles $\pm \pi/2$. For drive 2, we have the information that all smarticles do identical motion---the drive is permutation invariant. For drive 3, we further have the periodic time-sequence information. And finally to predict drive A, we must store the individual drive-pattern of each smarticle.

Empirically we see that the harder it is to predict the drive, the fewer low-rattling configurations we will have in our landscape. This intuition may be understood by viewing low-rattling determinism as having ``low surprise'' in the presence of the drive, which requires storing predictive information in the configurational degrees of freedom. The more such information we need, the fewer system configurations will have it. This means that for drive 1, which is impossible to predict, no configuration has lower rattling than others, and hence we have a near-uniform steady-state (fig.~\ref{fig:adapt}B). Next, the information about permutation-invariance of drives 2 and 3 gets clearly reflected in the flower-like symmetry of the rattling landscape (fig.~\ref{fig:sqGait}A), and of the steady-states shown (from a different 3D orientation in fig.~\ref{fig:adapt}B). For drive A this symmetry is broken, requiring individualized information, and we are left with only a couple low-rattling regions, which then absorb most of the probability density, causing the strongest fine-tuning (fig.~\ref{fig:adapt}B). 

Note that as we increase drive complexity, besides the steady-state becoming more fine-tuned, it also takes longer to reach (see fig.~\ref{fig:adapt}C). This indicates that for yet more complicated drives, this relaxation time may diverge, and there may once again be no low-rattling configurations, destroying all self-organization. We may view this regime as one where no system configuration can store all the information needed to match the drive well. All these considerations thus provide useful upper and lower bounds on drive complexities needed to engineer self-organized dynamics. The specifics of this for a given system may, nonetheless, be difficult to predict \textit{a priori}.

It is further instructive to consider the temporal emergence of self-organization under the influence of a drive. For this, we start with a distribution uniformly sampling all possible swarm configurations, and watch it evolve under our 4 different drive patterns. As it relaxes to the respective steady-state distribution, we track its entropy over time relative to the uniform distribution at initial time (fig.~\ref{fig:adapt}C). As the distribution entropy drops, we can think of it as encoding information about its drive, which gets ``imprinted'' on the distribution over time. This is akin to how in equilibrium, the distribution ``learns'' about its energy landscape by relaxing into the Boltzmann distribution---except here, since the rattling landscape is encoding the drive properties, the distribution is, by extension, learning about its drive. Figure~\ref{fig:adapt}C shows that the amount of information ultimately attained by each distribution scales with the amount of predictive information in each drive, reinforcing the idea of learning. Moreover, this trend hints at the exciting possibility to \textit{a priori} predict the degree of self-organization in the system based only on the properties of the drive. 

We can better quantify how distributions tune to their respective drives over time by tracking their motion through an abstract ``shape-space''---fig.~\ref{fig:adapt}D.  Taking the 4 temporal sequences of distributions, this plot represents each distribution by a single point $ p $, with pair-wise distances between points given by $ d(p_1,p_2) = D_{KL}(p_1 || p_2) + D_{KL}(p_2 || p_1)$, where $ D_{KL} $ is the Kullback-Leibler divergence. As we cannot embed so many points in 2D while reproducing all pair-wise distances exactly, we use an approximation algorithm tSNE: t-distributed stochastic neighbor embedding. Starting from the uniform initial distribution (labeled by a grey star), we see that under drive 1 the distribution remains near the uniform one, though with large random fluctuations. Under drive A, on the other hand, it shows clear directed evolution into its own unique shape. For drives 2 and 3, both of which are symmetric under smarticle permutations, the distributions evolve in a similar direction, but drive 3 having more predictive information moves further from the uniform starting point. This once again illustrates that the steady-state distributions, and hence the self-organized dynamics, tend to encode symmetries and other predictive information of their drives.

\subsection{Engineering steady-states}

\begin{figure}[pt]
	\centering\includegraphics[width=1.\textwidth]{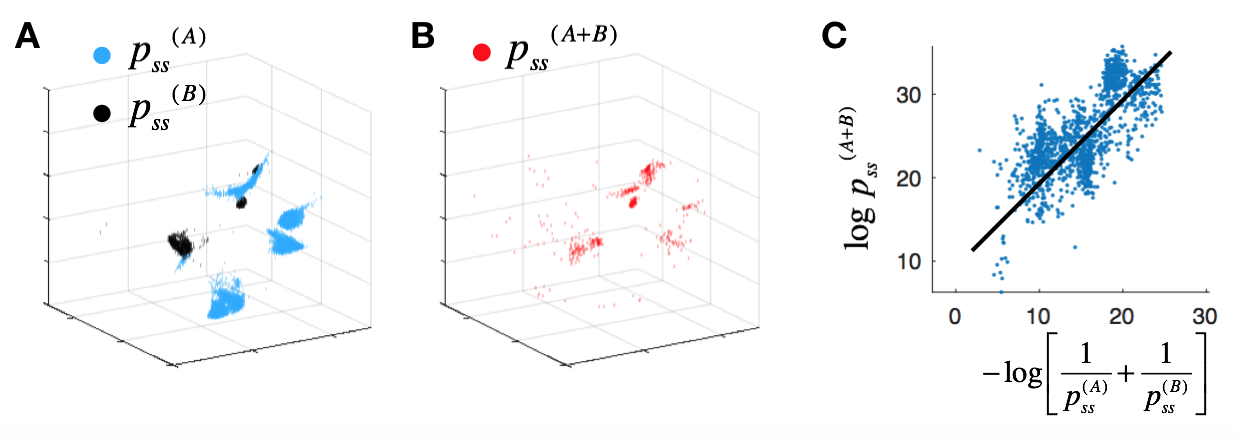}
	\caption{\textbf{Steady-state can be controlled by engineering the drive}.
	We reproduce the experimental results of Fig.~\ref{fig:drive-specific} in simulation, for the same drive-patterns A and B shown there. (\textbf{A})~illustrates the two corresponding steady-states, overlaid to highlight the self-organized states that are common to the two drives. (\textbf{B})~shows the steady-state of the mixed-drive protocol A+B (Fig.~\ref{fig:drive-specific}C), which concentrates primarily in the overlap regions of A and B. (\textbf{C})~quantifies this intuition by plotting the actual mixed-drive probability density $p_{ss}^{(A+B)}$ vs. that predicted by the overlap of $p_{ss}^{(A)}$ and $p_{ss}^{(B)}$ (black line gives the prediction up to normalization).
	}
	\label{fig:mixDrive_sim}
\end{figure}

Since self-organized states encode drive properties, we should be able to control the self-organization by astute choices of drive pattern. In particular, if we mix two distinct drives by stochastically switching back and forth between them, we can predict that the resulting steady-state will be the ``intersection'' of the steady-states under constituent drives. This is because predicting this mixed drive will require information about both constituent drives. By switching stochastically, according to a Poisson process with a mean rate of one switch every 5 drive cycles, we make sure not to introduce any new information into the mixed drive. Thus, only those configurations that have low rattling under both individual drives (thus encoding information about both) will be selected in the mixed-drive steady-state (Fig.~\ref{fig:drive-specific} and fig.~\ref{fig:mixDrive_sim}, also movie~S6).

To be more precise, we can quantify this prediction. We begin by defining the covariance matrix
\begin{align}
&\mathcal{C}_{ij}^{(drive)}(\v{q}_0) \equiv \<\frac{\delta q_i}{\sqrt{\delta t}},\frac{\delta q_j}{\sqrt{\delta t}}\>_{(drive) \,|\, \v{q}(t=0)=\v{q}_0} 
\label{eq:defC}
\end{align}
Using the definition Eq.~\ref{eq:rattling}, which here gives $ \mathcal{R}^{(drive)} = \frac{1}{2} \log\det \mathcal{C}^{(drive)} $, and Eq.~\ref{eq:p_ss}: $ p_{ss} = \exp{-\gamma\, \mathcal{R}} / Z $, with $ Z $ being a normalization constant, we can express steady-state probability density in terms of $ \mathcal{C} $ as
\begin{align}
&p_{ss}^{(drive)}(\v{q})=\frac{1}{Z} \(\det \mathcal{C}^{(drive)}(\v{q})\)^{-\gamma/2} \label{eq:pSS_C}
\end{align}
Thus, to predict probabilities under the mixed drive $ (A+B) $, we need to find $ \mathcal{C}^{(A+B)} $. Since the Poisson switching process keeps effects of the two constituent drives uncorrelated, the averaging over drive realizations in Eq.~\ref{eq:defC} decouples, giving
\begin{align}
{C}_{ij}^{(A+B)}(\v{q}) =  \frac{1}{2} \mathcal{C}_{ij}^{(A)}(\v{q}) + \frac{1}{2} \mathcal{C}_{ij}^{(B)}(\v{q})
\end{align}
for any configuration $ \v{q} $. Note that here we can work with these quantities abstractly, without worrying about what measuring these $ \mathcal{C} $ matrices would entail. Now, since the covariance matrices are positive-definite and symmetric, we have the inequality: $ \det\(\frac{1}{2} \mathcal{C}^{(A)} + \frac{1}{2} \mathcal{C}^{(B)}\) \ge \det \(\frac{1}{2} \mathcal{C}^{(A)}\) + \( \frac{1}{2} \mathcal{C}^{(B)}\) $. With that, and using Eq.~\ref{eq:pSS_C} above, we get
\begin{align}
\(p_{ss}^{(A+B)}\, Z^{(A+B)}\)^{-\frac{2}{\gamma}} \ge \(p_{ss}^{(A)}\, Z^{(A)}\)^{-\frac{2}{\gamma}}  + \(p_{ss}^{(B)}\, Z^{(B)}\)^{-\frac{2}{\gamma}} 
\end{align}
Now, because $ 2/\gamma \sim \bigO{1} $, and the values of $ p_{ss} $ vary over many orders of magnitude, we can approximate this by ignoring the powers and relative normalizations $ Z^{(A)} $ and $ Z^{(B)} $ (assuming they are of roughly the same order), leaving the simple expression:
\begin{align}
p_{ss}^{(A+B)} \le \frac{1}{Z}\(\frac{1}{p_{ss}^{(A)}} +\frac{1}{p_{ss}^{(B)}}\)^{-1}
\end{align}
which is an upper bound on the likelihoods under mixed drive, with some normalization $ Z $. This establishes the intuition that the only configurations that may be stable under mixed drive are ones that are stable under both of the pure drives. Moreover, as we see from empirical tests shown in Fig.~\ref{fig:drive-specific}D for experiments and fig.~\ref{fig:mixDrive_sim}C for simulations, in practice this bound tends to be saturated, giving us a way to predict the mix-drive steady-state $ p_{ss}^{(A+B)} $ up to normalization (see Eq.~\ref{eq:mix_drive}).

Note that this result represents a key milestone in the development of both, the theory of rattling, and its applications. For the theory, this is the first quantitative analytical prediction we show that---while making use of $\mathcal{R}(\v{q})$ at the conceptual level in the course of the derivation---arrives at a result that directly relates measurable quantities of practical interest. On top of this, it illustrates one key property of rattling. In equilibrium, to predict the likelihood of a system configuration, we need not measure its energy directly, as we can often predict it by adding up all the energies of system parts. Our result here indicates that rattling may sometimes be similarly predicted by decomposing either the drive or the system into constituent parts whose behavior is known. This suggests that a key property that makes the Boltzmann distribution so powerful for equilibrium systems may have a counterpart in the rattling theory.

Our result suggests that, at least in principle, it may be possible to single out configurations corresponding to desired behaviors in the steady-state of the collective by carefully engineering drive patterns. Normally, such ordered and coordinated behaviors in robots are the result of high-dimensional non-convex optimization that becomes intractable as the number of agents in the collective increases. Moreover, even when such optimized control schemes are found, they are specific to the dynamics of the collective under consideration, and would not be applicable generally. Here, we have introduced a general principle scalable to large collectives enabling the emergence of finely-tuned and engineered behaviors that exists beyond the toolset of traditional robot control. 

\subsection{Controlling the degree of organization}

We saw that self-organization arises when the response to a given drive varies widely across the configuration space of the system. This leads to a wide variety of accessible rattling values, causing most of the probability to accumulate in the least-rattling configurations. Destabilizing these states, and thus destroying self-organization altogether, amounts to flattening the rattling landscape. In this section, we present two qualitatively different approaches to this, derive our analytical prediction for destabilization curves, and verify them against simulation results. Since rattling is the entropy of local forces, both methods work by increasing amount of entropy in the system. The first of these approaches gradually increases external drive entropy, while the second tunes the system's friction coefficient in order to increase internal entropy generation. 

\subsubsection{Tuning drive entropy}
To tune drive entropy, we start with a base deterministic drive (here we use drive A from Fig.~\ref{fig:drive-specific} and fig.~\ref{fig:adapt}), and introduce some small error rate $ \epsilon $ (see movie S8, where $p = 2 \,\epsilon$). Specifically, since at every time step each arm is programmed to go to $ \pm \pi/2 $ according to the pre-set pattern, we now give it $ \epsilon $ probability of making the wrong move (for each arm on every smarticle independently). This means that at every step, there is a discrete probability distribution over the possible next arm moves for the swarm, with $ p_0=(1-\epsilon)^6 $ probability of all the 6 arms doing the ``right'' move, and $ p_k = {6\choose k} \,\epsilon^k (1-\epsilon)^{6-k}$ probability of making $ k $ mistakes. We can easily calculate the entropy of this distribution $ S_d(\epsilon) $ (which is a long expression derived in Mathematica), thus predicting the amount of entropy that such drive injects into the system. 

Different system configurations will be affected differently by this change in drive entropy. We know that rattling is the entropy of local forces acting on a configuration. Moreover, we can define least-rattling configurations for a given drive $ \v{q}_{l.r.} $ as those that generate no additional entropy internally under that drive. It is thus reasonable to posit that their rattling values $ \mathcal{R}(\v{q}_{l.r.}) $ will scale with $ S_d(\epsilon) $ as we tune $ \epsilon $. Since the relation between rattling and probability density $ \log p_{ss} = -\gamma\, \mathcal{R} - \log Z $ persists for all values of $ \epsilon $ (as is verified explicitly in experiment and simulation in Fig.~\ref{fig:melt}C), we get: 
\begin{align} 
\log p_{ss}^{(A+\epsilon)}\(\v{q}_{l.r.}\) = -\gamma\, S_d(\epsilon) -\log Z 
\label{eq:drEntrBound}
\end{align}
for the least-rattling states. As the rattling of all the other configurations will be higher, this will be an upper bound on probability in any state (see solid black curve in Fig.~\ref{fig:melt}D).

One key prediction of our framework is that the stochasticity coming from dynamical chaos inherent to our system, and stochasticity coming from explicit drive randomness, while of very different origins and dynamical properties, are indistinguishable in their effect on the steady-state. This is reminiscent of universality in statistical mechanics, where most microscopic details of interactions and dynamics become irrelevant for determining some global property. This prediction is confirmed in Fig.~\ref{fig:melt}C, where all three shown plots fall on the same line, illustrating that rattling $ \mathcal{R} $ remains predictive of $p_{ss}$ whether the noise comes directly from external drive or from chaotic interactions. This way, increasing $ \epsilon $ suppresses the likelihood of the ordered states, and collapses the range of $ p_{ss} $ variation.

Thus, to find the probability decay of any configuration, we can combine its original rattling value with the drive entropy. As in the last section, we use the covariance matrix $ \mathcal{C}_{ij}^{(drive)}$, defined in eq.~\ref{eq:defC}. If we let $ \v{q}^{(A+\epsilon)}(t) $ to be the system response under the perturbed drive, $ \v{q}^{(A)}(t) $---that under the deterministic base drive, then we can define $ \v{q}^{(\epsilon)}(t) \equiv \v{q}^{(A+\epsilon)}(t)-\v{q}^{(A)}(t) $ as the ``response to drive noise.'' This way, 
\begin{align}
\mathcal{C}_{ij}^{(A+\epsilon)} = \<\frac{\delta q^{(A)}_i + \delta q^{(\epsilon)}_i}{\sqrt{\delta t}},\frac{\delta q^{(A)}_j + \delta q^{(\epsilon)}_j}{\sqrt{\delta t}}\>
=\mathcal{C}_{ij}^{(A)} + \mathcal{C}_{ij}^{(\epsilon)} +  \<\frac{\delta q^{(A)}_i}{\sqrt{\delta t}},\frac{\delta q^{(\epsilon)}_j}{\sqrt{\delta t}}\> + (i \leftrightarrow j) 
\label{eq:crossCorrC}
\end{align}
Since the noise we introduced to the drive is entirely uncorrelated with the structure of the drive pattern itself, the cross-correlation terms vanish.  The original rattling landscape $ \mathcal{R}^{(A)}(\v{q}) $ is thus modified by the addition of drive noise as: 
\begin{align}
 \mathcal{R}^{(A+\epsilon)}(\v{q}) =  \frac{1}{2} \log \det \(\mathcal{C}_{ij}^{(A)} + \mathcal{C}_{ij}^{(\epsilon)}\) \ge \frac{1}{2} \log \(\det \mathcal{C}_{ij}^{(A)} + \det\mathcal{C}_{ij}^{(\epsilon)}\) = \frac{1}{2} \log \(\e{2\mathcal{R}^{(A)}(\v{q})} + \e{2 a\,S_d(\epsilon)}\)
\end{align}
where the inequality is true for any two positive definite symmetric matrices. In the last step, we again use our above hypothesis, which now says that the entropy of $ \v{q}^{(\epsilon)} $ fluctuations, given by $ \frac{1}{2} \log \det \mathcal{C}_{ij}^{(\epsilon)} $, is well approximated by drive entropy $ S_d(\epsilon) $, up to a proportionality constant $ a $. This gives a prediction for how the steady-state probability of any given configuration $ \v{q} $ will be modified by $ \epsilon $: 
\begin{align}
\log p_{ss}^{(A+\epsilon)}(\v{q}) \le -\frac{\gamma}{2} \log \(\e{2\mathcal{R}^{(A)}(\v{q})} + \e{2 a\, S_d(\epsilon)}\) - \log Z
\end{align}
which is a tighter bound than Eq.~\ref{eq:drEntrBound}. We know that this bound is saturated for the low-rattling configurations, and trivially for $ \epsilon=0 $. Empirically, it also seems to hold as a good predictive principle throughout the configuration space. 

In Fig.~\ref{fig:melt}D, we plot the decay of $ \log p_{ss}^{(A+\epsilon)}(\v{q}) $ values with increasing $ \epsilon $ around 100 randomly sampled configurations $ \v{q} $ from simulation data. These lines are colored by values of $ \log p_{ss}^{(A)}(\v{q}) $ to allow distinguishing them visually. Note that ``drive randomness'' labeled in that figure is actually $ 2\epsilon $, since a 50\% error rate corresponds to fully random drive. The solid line shows the upper bound coming from low-rattling states $ \e{2\mathcal{R}^{(A)}(\v{q})} \sim 0 $, while the dashed lines show how this gets modified for other states. These predictions use two global fitting parameters: the vertical offset, corresponding to the normalization $ Z $, and vertical scaling, for the proportionality constant $ a $. Note also that these predictions were done with the assumption that the effect of $\epsilon$ on the normalization $ Z $  is much weaker than on the $p_{ss}$ of lower-rattling configurations that are affected most strongly. For the higher-rattling states, this may be wrong, and so the predictions in the regime when $ \epsilon $ is large and self-organization is already destroyed are not reliable. 

\subsubsection{Tuning friction}

Besides injecting entropy externally, we can achieve a similar effect by increasing the rate of internal entropy production in a configuration-independent way. Note that the rattling landscape itself arises through internal entropy generation, but in a way that depends on the state and the drive. By lowering friction, on the other hand, we can increase the time between energy injection by the drive, and its dissipation through damping. In that time-window, this energy has a chance to thermalize via the many system nonlinearities, thereby creating an effective temperature bath that uniformly affects all configurations. It will then play a similar role to drive noise in the previous section, raising the rattling of the self-organized state, and flattening the rattling landscape.

\begin{figure}
	\centering \includegraphics[width=1.0\textwidth]{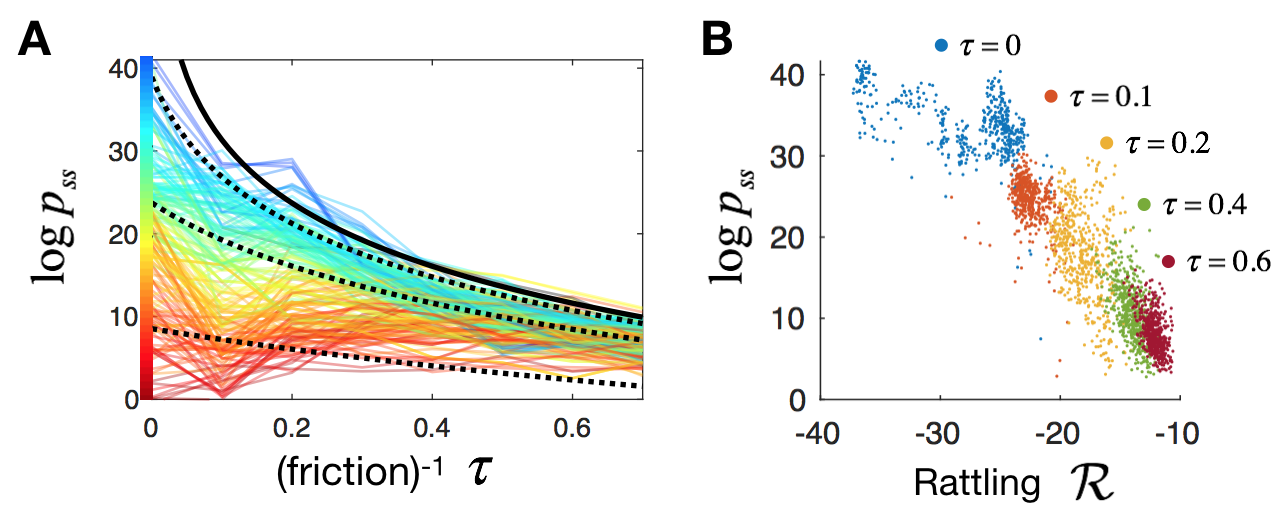}
	\caption{\textbf{Destroying self-organization by reducing friction}. 
		In~(\textbf{A}), we plot the steady-state probabilities at 200 different configurations under drive A (shown in Fig.~\ref{fig:drive-specific}, fig.~\ref{fig:adapt}) as we gradually reduce smarticle friction in simulation ($\tau $ is the velocity decay time-scale). Lines are colored according to state likelihood in the over-damped regime ($ \tau \sim 0 $). The solid black curve is the analytical prediction for the decay of low-rattling states, which also serves as an upper bound for probabilities of other configurations, which are predicted by dotted curves (with fitting parameter $ \gamma=3.1 $). (\textbf{B})~illustrates the robustness of the relation between probability and rattling $ \log p_{ss} \sim -\mathcal{R} $ as friction in the system is changed. This, along with panel A, shows that measuring the overdamped dynamics $ \tau=0 $ is sufficient to predict system behaviors for all lower friction values. (the e-folding velocity decay time $\tau $ is shown in normalized units, where $ \tau=1 $ corresponds to 0.2 of a drive cycle.)
	}
	\label{fig:inertial}
\end{figure}

Experimentally, we can dramatically reduce friction between the smarticles and the stage by filling the ring with a loose single layer of small plastic beads, and putting the smarticles on top of these, thus allowing them to skate around with very low friction (see movie S7). Note that the beads used are too light to back-react on the smarticles, and thus do not produce any effects other than to reduce friction. In simulation, we can run a series of experiments interpolating between these two regimes by gradually reducing damping (fig.~\ref{fig:inertial}). While the exact damping force in experiment is complicated, and besides the rolling friction also depends on collisions among smarticles and the ring, in simulation we simply approximate it by Stokes drag. This may be unrealistic in detail, but guarantees stability as velocities are capped---which may not be the case if we used a constant friction force. Empirically, the resulting behavior agrees qualitatively with experiments.

This way our velocities get a finite decay curve: $ \ddot{x} + \dot{x}/\tau =\delta(t) \quad \Rightarrow \quad v(t) = v(0)\,\exp{-t/\tau}$. We then predict that the associated increase in the mean free path acts to raise the effective noise level everywhere by the same $ \tau $-dependent value $ \mathcal{C}^{(\tau)}_{ij} $ (since $ v(0) $ would be roughly set by motor speeds, independent of configuration). We take $ \mathcal{C}^{(A)}_{ij}(\v{q}) $ to be the baseline noise landscape under drive A with high friction, such that $ \tau \sim 0 $---which is close to the regime in all our other experiments. 

We then artificially separate the full system dynamics into the drive response and the thermalized inertial fluctuations $ \v{q}^{(A+\tau)} =\v{q}^{(A)}+\v{q}^{(\tau)} $, just as in the last section, and run through a similar analysis as that in Eq.~\ref{eq:crossCorrC}. Here, the cross correlation terms $ \<\delta q_i^{(A)},\, \delta q_j^{(\tau)}\> $ again vanish by assuming that due to fast thermalization on the many strong nonlinearities in the system, the inertial noise ends up being uncorrelated with the drive pattern. This way, with Eq.~\ref{eq:pSS_C}, we expect $ p_{ss}^{A+\tau}(\v{q}) \propto \det\(\mathcal{C}^{(A)}(\v{q}) + \mathcal{C}^{(\tau)}\)^{-\gamma/2} $. Thus, as we tune up $ \tau $, configurations that used to be low-rattling ($ \mathcal{C}^{(A)}\sim 0 $) get substantially suppressed, while chaotic configurations are only weakly affected (and may even become slightly more probable as normalization changes). 

We can do more by analytically estimating the dependence of $ \mathcal{C}^{(\tau)}_{ij} $ on $ \tau $. We know that in the steady-state, the rate of heat dissipation $ dQ = dt \sum_i \dot{x}_i^2/\tau $ from inertial smarticle motion must balance the rate of input work $ dW $, which will be related to motor speeds, and so be independent of both $ \tau $  and configuration $ \v{q} $. Note that the heat dissipation here must be written in terms of real-space physical coordinates $ x_i $, rather than any abstract configuration-space observables. For us, the chosen configuration-space parametrization reflects physical motion, so we can approximately use $ q_i $ directly, thus getting $ dQ = \sum_i \<dt\; \dot{q}_i^{(\tau)},\, \dot{q}_i^{(\tau)}\>/\tau = \tr \mathcal{C}^{(\tau)}/\tau $. Balancing this against a constant work input $ dW $, we thus get $ \tr \mathcal{C}^{(\tau)} = \tau\, dW $. Now, if we further assume that as we tune $ \tau $, the inertial noise anisotropy $ \mathcal{C}^{(\tau)}_{ij} $ does not change shape, then we can write $ \det \mathcal{C}^{(\tau)} \propto \tau^d $, where $ d $ is the configuration-space dimension, and the proportionality constant depends on $ dW $ and the precise noise characteristics. This way, for low-rattling states $ \v{q}_{l.r.} $ we get the prediction:
\begin{align} \label{eq:frictBound}
\log p_{ss}^{(A+\tau)}\(\v{q}_{l.r.}\) = -\frac{\gamma\,d}{2} \log \tau +const  
\end{align}
which is the solid black curve shown in fig.~\ref{fig:inertial}A. Note that once again, while rattling was used as a concept in the derivation, this result directly relates experimental quantities of interest in a novel and measurable way. Similarly, using the superadditivity property of the determinant, we can predict the probabilities of other states based on their rattling values in the overdamped regime $ \mathcal{R}^{(A)}(\v{q}) $: 
\begin{align}
\log p_{ss}^{(A+\tau)}(\v{q}) \le -\frac{\gamma}{2} \log\(\e{\mathcal{R}^{(A)}(\v{q})} + \tau^d\) + const
\end{align}
which are shown as the dotted lines in fig.~\ref{fig:inertial}A. All these use two fitting parameters: vertical offset (normalization constant) and vertical scaling ($ \gamma =3.1$ here). Again, $ \tau $ is assumed to affect the normalization much less than it does the probability of the ordered configurations, which is permissible as long as there is a wide variety of $ p_{ss} $ values.

Figure~\ref{fig:inertial}B once again verifies that the steady-state probabilities for various friction values all fall on the same correlation line $ p_{ss} \propto \e{-\gamma\, \mathcal{R}} $. This again illustrates generalizability of the notion of rattling: the non-determinism in response to drive, and noise from thermalized inertial motion have the same effect of state probabilities.

\newpage
\section{Simulations of larger swarms}

\begin{figure}
	\centering \includegraphics[width=0.8\textwidth]{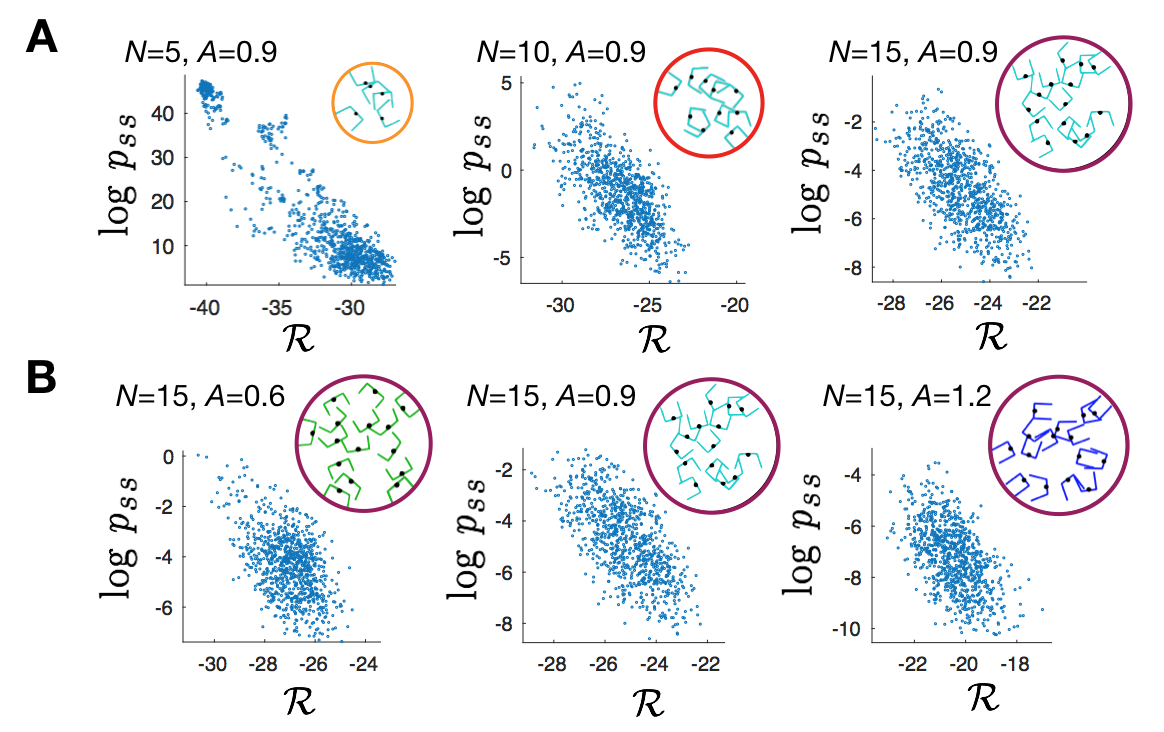}
	\caption{\textbf{Rattling prediction is robust across system parameters}. 
	(\textbf{A})~illustrates that for larger numbers of smarticles $ N $, the correlations between $p_{ss}$ and $\mathcal{R}$ given by Eq.~\ref{eq:p_ss} persists. (\textbf{B})~similarly shows robustness to varying arm-lengths $A$, shown in units where middle link (body) is length 1. Simulation data shown, experiments qualitatively confirmed these trends.
	\label{fig:paramScan}}
\end{figure}

\begin{figure}
	\centering \includegraphics[width=1.\textwidth]{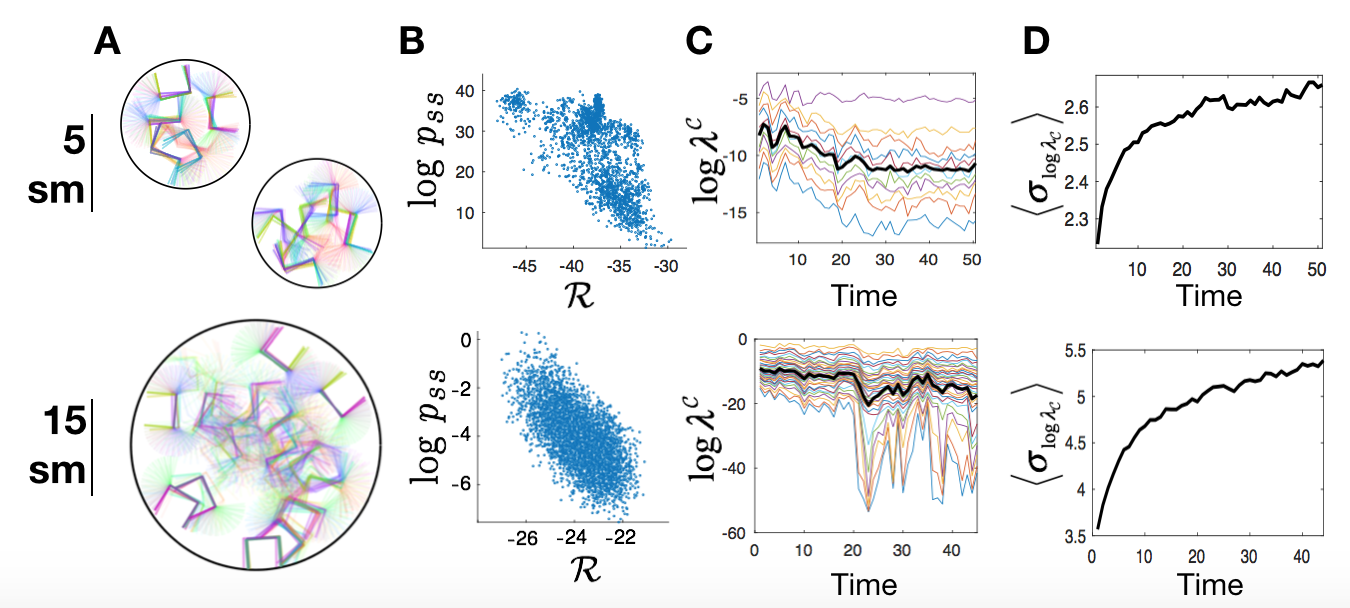}
	\caption{\textbf{Comparing 5 and 15 smarticle swarms}. 
		(\textbf{A})~visualizes the swarm dynamics for 3 drive cycles, same way as in Fig.~\ref{fig:smarticle}E. While globally ordered motion dominates the 5-smarticle behavior, for 15 smarticle swarm we only find organization in local metastable groups of 2-4 smarticles. This way, the correlation in~(\textbf{B}) is still seen in both cases, but the relative likelihoods of swarm configurations vary over a larger range for 5 smarticles. We further probe these behaviors in~(\textbf{C}) by plotting log-eigenvalues of the covariance matrix $ \mathcal{C} $ over time at a particular realization of a self-organizing transition in each swarm size. Qualitatively, we see that a gap in the spectrum appears near the top for 5 smarticles, signifying global organization, and near the bottom for 15. Black line shows the average value, which is proportional to $ \mathcal{R} $ (by our definition in Eq.~\ref{eq:rattEig}). Similarly, (\textbf{D})~shows the standard deviation of $ \log \lambda^{\mathcal{C}} $ averaged over 1000 runs from random initializations. This shows that the spectrum eventually spreads out in most runs, indicating some degree of organization.
		\label{fig:largeN}}
\end{figure}

To check that our results are not restricted to the three smarticle system or other parameters used, we simulated and ran experiments with larger ensembles and various arm-lengths. In fig.~\ref{fig:paramScan} we see that in all cases, rattling remains predictive of the steady state as given by Eq.~\ref{eq:p_ss}. On the other hand, we also see that the range of $ p_{ss} $ variation observed becomes qualitatively smaller in larger ensembles (fig.~\ref{fig:paramScan}A). To understand this, we focus on the 5 and 15 smarticle swarms (fig.~\ref{fig:largeN}). In the former case, as for three smarticles, we observe globally ordered behaviors that self-organize and persist for substantial periods of time until stochastically transitioning. As the configuration space of the swarm is larger, so is the entropic cost of such fine-tuning, which practically means that there are more possible fluctuations that can take the system out of such ordered motion. This makes them generally less robust than in the 3-smarticle swarm. We characterised this behavior in simulations and qualitatively confirmed it in experiments. 

In contrast, for 15 smarticles global self-organization is no longer visibly obvious, but we do see the appearance of local pockets of metastable order in groups of 2-4 smarticles (fig.~\ref{fig:largeN}A). These do not tend to persist for much longer than 5 drive cycles, as they are strongly affected by their chaotic neighbors (see movie~S2). Nonetheless, studying these transient ordered behaviors may allow understanding and predicting the global dynamics of the swarm, as pockets of local order appear frequently and can be seen as controlling the overall behavior. Furthermore, such partial organization is what we typically observe in most many-body active self-organization processes, and so we want to ensure that our framework can address these. 

In both, the 5 and 15 smarticle swarms, we verify that local measures of rattling are still predictive of the steady-state density around those configurations (fig.~\ref{fig:largeN}B). For these larger swarms, it is much harder to reliably sample the steady-state distributions over the entire $ 3N $-dimensional configuration space. As such, we can run the analysis described in Materials and Methods on lower-dimensional coarse-grainings of the configuration space---which is also how many-body active matter is typically characterized. As long as the selected parametrization captures a diversity of possible system behaviors (\textit{e.g.}, can distinguish between ordered and chaotic states well), the correlation between rattling and steady-state density persists (see fig.~\ref{fig:dataAnal}, D and E).

As we move to larger many-body systems, we may want to ask more detailed questions about the dynamics that cannot be addressed by a single scalar rattling value. In particular, when rattling drops, this may be due to strong self-organization in a part of the system, or a slight reduction of noise throughout. We want to distinguish these behaviors, and explicitly to identify our 15-smarticle dynamics with the former regime. We first remember that the motion of each smarticle is realized along its own dimensions in the configuration space. This way, if the dynamics of a few smarticles become regular, the system trajectory through configuration space will be confined to a sub-space corresponding to the remaining chaotic smarticles' exploration. 

Indeed, any degree of organization in a many-body system restricts the configuration-space trajectory to a submanifold, as some configurations become inaccessible. Then, the covariance matrix $ \mathcal{C}_{ij}^{(drive)} \equiv \<\frac{\delta q_i}{\sqrt{\delta t}},\frac{\delta q_j}{\sqrt{\delta t}}\>_{(drive)}$ can be seen as a local estimate of the shape of that submanifold. If some configuration space dimensions are explored much more than others, this matrix will have vast diversity of eigenvalues. In a qualitative sense, the number of ``large'' eigenvalues will give the number of dimensions of the restriction submanifold. This is to say, globally ordered dynamics, being restricted to a one or two dimensional subspace, will have all but one or two eigenvalues be near zero. If only a small part of the system self-organizes, on the other hand, then we expect to see the few corresponding eigenvalues drop to zero, while the rest remain large indicating chaotic exploration by the rest of the system. In either case, when some ordered motion emerges in the system, we expect to see an increased diversity of eigenvalues of $ \mathcal{C} $, and perhaps even a gap appearing in the spectrum. This is in contrast to the scenario where all motion becomes slightly less noisy, shifting the entire spectrum uniformly down. Note also that these eigenvalues capture similar information as Lyapunov exponents, though here we are working with a stochastic coarse-grained system, and so do not necessarily expect exponential divergences of trajectories.

In fig.~\ref{fig:largeN}C, we plot the evolution of these eigenvalues over time for a single realization of a 5 and a 15 smarticle run. Note that by our definition of rattling in Eq.~\ref{eq:rattling}, we have that 
\begin{align} \label{eq:rattEig}
\mathcal{R} = \frac{1}{2}\sum_i \log \lambda^{\mathcal{C}}_i
\end{align}
Thus, the mean value of the curves in fig.~\ref{fig:largeN}C, shown in black, gives the behavior of $ \mathcal{R} $. 

In these plots, we picked a time-window showing a self-organization transition to illustrate the concepts described above. We see that at the onset of global order in 5 smarticles, most eigenvalues drop off, leaving only one larger value, as the swarm motion gets (largely) confined to a one dimensional submanifold of the configuration space. In contrast, for 15 smarticle swarm, most of that space continues being explored, with only a few of the eigenvalues dropping off significantly from the rest, indicating the part of the system that organized. Note also that unlike for 5, the spectral gap for 15 smarticles does not persist stably over time, indicating the instability of these ordered pockets. This agrees with and quantifies the behaviors we observed visually. 

To check that these findings hold generally in an ensemble of system runs, we can look at the mean standard deviation of these eigenvalues, specifically: $ \<\mbox{std}\(\log \lambda^{\mathcal{C}}_i\)\>_{ensemble} $. Thus, while their mean (\textit{i.e.}, rattling), indicates ``how noisy'' the overall system's response to drive is, the standard deviation signifies the degree of anisotropy in the configuration space exploration. This way, the plots in fig.~\ref{fig:largeN}D, along with the knowledge that rattling is decaying over time (Fig.~\ref{fig:systems}C), show that swarm motion typically gets confined over time to a progressively lower dimensional submanifold of possible configurations. This trend indicates that the degree of organization in the swarm grows on average in an ensemble of runs. 

\pagebreak

\begin{center}
\section*{Movie Captions}
\end{center}
\begin{itemize}
    \item \textbf{Movie S1: Smarticle dynamics.} Smarticles are simple 3-link robots comprised of a pair of servomotors programmed by a microcontroller that receives remote commands from a wireless radio. Smarticles cannot locomote on their own because their arms cannot touch the flat ground they rest on. However, a swarm of smarticles can exhibit complex dynamics resulting from purely repulsive interactions with one another. The confining ring ensures that they maintain contact, avoiding the trivial decoupled attractor. 
    
    \item \textbf{Movie S2: 15 smarticle swarm.} To study variations of system parameters, as well as larger smarticle ensembles, we developed a simulation platform in MATLAB. When we simulate the dynamics of larger swarms of smarticles---15 smarticles in this illustration---we discover small pockets of meta-stable emergent order. These regular dynamical states are mostly comprised of groups of 3-4 smarticles, which eventually get broken up by their chaotic neighborhoods. 
    
    \item \textbf{Movie S3: Self-organized behaviors.} When we closely study groups of 3 smarticles at a time, we find that the swarm can exhibit strikingly ordered dynamics, both in experiment and simulation. Moreover, we find that even for a single pattern of driving (\textit{i.e.}, smarticle arm movement pattern) the swarm can self-organize into several distinct orderly ``dances'' that may be explained by the low-rattling selection principle.
    
    \item \textbf{Movie S4: Rattling illustration.} To qualitatively illustrate the difference between high and low rattling configurations, we simulate a system of 3 smarticles first initialized at 5 nearby low rattling configurations and then at 5 nearby high rattling configurations. We observe that the low rattling configurations remain orderly and in close synchrony despite their different initializations, while the trajectories starting from high rattling configurations rapidly diverge from one another. While this may be reminiscent of local Lyapunov exponents, their technical relation to rattling is non-trivial.
    
    \item \textbf{Movie S5: Stochastic flights between dynamical attractors.} For the smarticle system, we have shown that for a given drive pattern there are several corresponding low rattling states that the system can inhabit. By assuming that the collective smarticle dynamics are ``messy,'' we posit that they are well-approximated by configuration-space diffusion. Here, we demonstrate one sense in which this assumption holds. Rather than remaining trapped in a single low rattling state, the smarticle swarm spontaneously diffuses between such states.
    
    \item \textbf{Movie S6: Mixing drive patterns.} To illustrate the predictive power of the rattling theory, we quantitatively predict the nonequilibrium steady-state of the smarticle swarm under a ``mixed-drive'' protocol, which is shown here. Figure~3 in the main text, along with Eq.~4, illustrate that this steady-state contains only those configurations that have low rattling under both drives A and B independently. This suggests one way in which rattling may be used to engineer collective behaviors. 
    
    \item \textbf{Movie S7: Low friction experiment.} We can dramatically lower the friction between the smarticles and the surface they slide on by adding a layer of light plastic beads. The orderly ``dances'' observed without the beads may still be identified in the transients, but are very unstable, making the system behavior largely random. By interpolating between the low and high-friction regimes using simulation, we show in fig.~S8 that rattling quantitatively predicts how these dances are destabilized. This is explained in terms of an increased rate of internal entropy production when friction is low, which flattens the rattling landscape, destroying the low-rattling configurations. 
    
    \item \textbf{Movie S8: Melting low rattling states.} Starting with a deterministic arm motion pattern, we introduce a probability of making a random arm movement, which we then gradually tune from completely deterministic to completely stochastic (in simulation). We note that the self-organized smarticle dances are initially present but gradually become destabilized as we tune up drive randomness. This quantitatively agrees with predictions based on rattling (see Fig.~4). As the last movie showed the role of internal entropy generation, this illustrates that external entropy injection has the same effect on the smarticle swarm behavior. Such agreement validates the rattling framework by showing that steady-states are independent of the details of the dynamics, as long as rattling landscape is the same.
    
\end{itemize}


\begin{thebibliography}{10}

\bibitem{Rothemund2006}
P.~W.~K. Rothemund, {\it Nature\/} {\bf 440}, 297 (2006).

\bibitem{kardar2007particles}
M.~Kardar, {\it Statistical physics of particles\/} (Cambridge University
  Press, 2007).

\bibitem{corte2008chaikin_spun_colloids}
L.~Corte, P.~Chaikin, J.~P. Gollub, D.~Pine, {\it Nature Physics\/} {\bf 4},
  420 (2008).

\bibitem{grosberg2015phase-sep_2T}
A.~Grosberg, J.-F. Joanny, {\it Physical Review E\/} {\bf 92}, 032118 (2015).

\bibitem{Sumino2012}
Y.~Sumino, {\it et~al.\/}, {\it Nature\/} {\bf 483}, 448 (2012).

\bibitem{ramaswamy2010actMatt_review}
S.~Ramaswamy, {\it Annu. Rev. Condens. Matter Phys.\/} {\bf 1}, 323 (2010).

\bibitem{bertini2015macroFluctTheory}
L.~Bertini, A.~De~Sole, D.~Gabrielli, G.~Jona-Lasinio, C.~Landim, {\it Reviews
  of Modern Physics\/} {\bf 87}, 593 (2015).

\bibitem{paoluzzi16critPhenActMatt}
M.~Paoluzzi, C.~Maggi, U.~Marini Bettolo~Marconi, N.~Gnan, {\it Phys. Rev. E\/}
  {\bf 94}, 052602 (2016).

\bibitem{speck16neqThermo_activeMatter}
T.~Speck, {\it {EPL} (Europhysics Letters)\/} {\bf 114}, 30006 (2016).

\bibitem{Chv18least_ratt}
P.~Chvykov, J.~England, {\it Phys. Rev. E\/} {\bf 97}, 032115 (2018).

\bibitem{cates15Teff_motilityPhaseSepar}
M.~E. Cates, J.~Tailleur, {\it Annual Review of Condensed Matter Physics\/}
  {\bf 6}, 219 (2015).

\bibitem{savoie_thesis}
W.~Savoie, Effect of shape and particle coordination on collective dynamics of
  granular matter, Ph.D. thesis, Georgia Institute of Technology (2019).

\bibitem{Aguilar2018}
J.~Aguilar, {\it et~al.\/}, {\it Science\/} {\bf 361}, 672 (2018).

\bibitem{rubenstein2014swarm1000}
M.~Rubenstein, A.~Cornejo, R.~Nagpal, {\it Science\/} {\bf 345}, 795 (2014).

\bibitem{Werfel2014}
J.~Werfel, K.~Petersen, R.~Nagpal, {\it Science\/} {\bf 343}, 754 (2014).

\bibitem{Li2019}
S.~Li, {\it et~al.\/}, {\it Nature\/} {\bf 567}, 361 (2019).

\bibitem{Vicsek2018}
G.~V{\'a}s{\'a}rhelyi, {\it et~al.\/}, {\it Science Robotics\/} {\bf 3} (2018).

\bibitem{Mayya2019nonuniform}
S.~Mayya, G.~Notomista, D.~Shell, S.~Hutchinson, M.~Egerstedt, {\it IEEE
  International Conference on Intelligent Robots and Systems (IROS)\/}  (2019).

\bibitem{Duhr2006}
S.~Duhr, D.~Braun, {\it Proceedings of the National Academy of Sciences\/} {\bf
  103}, 19678 (2006).

\bibitem{van1992stochastic}
N.~G. Van~Kampen, {\it Stochastic processes in physics and chemistry\/}, vol.~1
  (Elsevier, 1992).

\bibitem{Landauer1993}
R.~Landauer, {\it Physica A: Statistical Mechanics and its Applications\/} {\bf
  194}, 551 (1993).

\bibitem{landauer75blowtorch}
R.~Landauer, {\it Phys. Rev. A\/} {\bf 12}, 636 (1975).

\bibitem{redner2013phase-sep_colloids}
G.~S. Redner, M.~F. Hagan, A.~Baskaran, {\it Phys. Rev. Lett.\/} {\bf 110},
  055701 (2013).

\bibitem{palacci936living_crystals}
J.~Palacci, S.~Sacanna, A.~P. Steinberg, D.~J. Pine, P.~M. Chaikin, {\it
  Science\/} {\bf 339}, 936 (2013).

\bibitem{michalet2012singTrajDlim}
X.~Michalet, A.~J. Berglund, {\it Physical Review E\/} {\bf 85}, 061916 (2012).

\bibitem{Savoie2019}
W.~Savoie, {\it et~al.\/}, {\it Science Robotics\/} {\bf 4} (2019).

\bibitem{materials_and_methods}
Materials and methods are available as supplementary materials at the Science
  website.

\bibitem{kedia2019drivespecific}
H.~Kedia, D.~Pan, J.-J. Slotine, J.~L. England, {\it arXiv\/}  (2019).

\bibitem{fineberg04surf_waves}
T.~Epstein, J.~Fineberg, {\it Phys. Rev. Lett.\/} {\bf 92}, 244502 (2004).

\bibitem{Pradillo2019}
H.~Karani, G.~E. Pradillo, P.~M. Vlahovska, {\it Phys. Rev. Lett.\/} {\bf 123},
  208002 (2019).

\bibitem{goldman2003grains_lattice}
D.~I. Goldman, M.~Shattuck, S.~J. Moon, J.~Swift, H.~L. Swinney, {\it Phys.
  Rev. Lett.\/} {\bf 90}, 104302 (2003).

\bibitem{sigmund2009topolMetamaterials}
O.~Sigmund, {\it IUTAM Symposium on Modelling Nanomaterials and Nanosystems\/}
  (Springer, 2009), pp. 151--159.

\bibitem{git_repo}
T.~A. Berrueta, A.~Samland, P.~Chvykov. Repository with all data, hardware,
  firmware, and software files needed to recreate the results of this
  manuscript: https://doi.org/10.5281/zenodo.4056700.

\bibitem{Olson2011}
E.~Olson, {\it Proceedings of the {IEEE} International Conference on Robotics
  and Automation ({ICRA})\/} (IEEE, 2011), pp. 3400--3407.

\bibitem{opencv_library}
G.~Bradski, {\it Dr. Dobb's Journal of Software Tools\/}  (2000).

\bibitem{han09paramConfigs}
L.~Han, L.~Rudolph, {\it Robotics: Science and Systems\/} (2009).

\bibitem{Wigner1967}
E.~P. Wigner, {\it SIAM Review\/} {\bf 9}, 1 (1967).

\bibitem{marsland2019randEcosystems}
W.~Cui, R.~Marsland~III, P.~Mehta, {\it arXiv preprint arXiv:1904.02610\/}
  (2019).

\bibitem{newman2002randomSocialNet}
M.~E. Newman, D.~J. Watts, S.~H. Strogatz, {\it Proceedings of the National
  Academy of Sciences\/} {\bf 99}, 2566 (2002).

\bibitem{bouchaud89randDiff}
J.~P. Bouchaud, A.~Georges, {\it Physics Reports\/} {\bf 195}, 127 (1990).

\bibitem{solon2019-runTumble_disorder}
Y.~Ben~Dor, E.~Woillez, Y.~Kafri, M.~Kardar, A.~P. Solon, {\it Phys. Rev. E\/}
  {\bf 100} (2019).

\bibitem{dill18max_caliber}
P.~D. Dixit, {\it et~al.\/}, {\it The Journal of Chemical Physics\/} {\bf 148},
  010901 (2018).

\bibitem{yang13ExtLang_Tgrad}
M.~Yang, M.~Ripoll, {\it Phys. Rev. E\/} {\bf 87}, 062110 (2013).

\bibitem{zdeborova2016inferencePhys}
L.~Zdeborov{\'a}, F.~Krzakala, {\it Advances in Physics\/} {\bf 65}, 453
  (2016).

\bibitem{boyer2012singTrajD}
D.~Boyer, D.~S. Dean, C.~Mej{\'\i}a-Monasterio, G.~Oshanin, {\it Physical
  Review E\/} {\bf 85}, 031136 (2012).

\bibitem{marzen16predictive_infMarkovProc}
S.~E. Marzen, J.~P. Crutchfield, {\it Journal of Statistical Physics\/} {\bf
  163}, 1312 (2016).

\end{thebibliography}

\newpage
\begin{thebibliography}{10}

\makeatletter
\addtocounter{\@listctr}{32}
\makeatother

\bibitem{git_repo}
T.~A. Berrueta, A.~Samland, P.~Chvykov. Repository with all data, hardware,
  firmware, and software files needed to recreate the results of this
  manuscript: https://doi.org/10.5281/zenodo.4056700.

\bibitem{Olson2011}
E.~Olson, {\it Proceedings of the {IEEE} International Conference on Robotics
  and Automation ({ICRA})\/} (IEEE, 2011), pp. 3400--3407.

\bibitem{opencv_library}
G.~Bradski, {\it Dr. Dobb's Journal of Software Tools\/}  (2000).

\bibitem{han09paramConfigs}
L.~Han, L.~Rudolph, {\it Robotics: Science and Systems\/} (2009).

\bibitem{Wigner1967}
E.~P. Wigner, {\it SIAM Review\/} {\bf 9}, 1 (1967).

\bibitem{marsland2019randEcosystems}
W.~Cui, R.~Marsland~III, P.~Mehta, {\it arXiv preprint arXiv:1904.02610\/}
  (2019).

\bibitem{newman2002randomSocialNet}
M.~E. Newman, D.~J. Watts, S.~H. Strogatz, {\it Proceedings of the National
  Academy of Sciences\/} {\bf 99}, 2566 (2002).

\bibitem{bouchaud89randDiff}
J.~P. Bouchaud, A.~Georges, {\it Physics Reports\/} {\bf 195}, 127 (1990).

\bibitem{solon2019-runTumble_disorder}
Y.~Ben~Dor, E.~Woillez, Y.~Kafri, M.~Kardar, A.~P. Solon, {\it Phys. Rev. E\/}
  {\bf 100} (2019).

\bibitem{dill18max_caliber}
P.~D. Dixit, {\it et~al.\/}, {\it The Journal of Chemical Physics\/} {\bf 148},
  010901 (2018).

\bibitem{yang13ExtLang_Tgrad}
M.~Yang, M.~Ripoll, {\it Phys. Rev. E\/} {\bf 87}, 062110 (2013).

\bibitem{zdeborova2016inferencePhys}
L.~Zdeborov{\'a}, F.~Krzakala, {\it Advances in Physics\/} {\bf 65}, 453
  (2016).

\bibitem{boyer2012singTrajD}
D.~Boyer, D.~S. Dean, C.~Mej{\'\i}a-Monasterio, G.~Oshanin, {\it Physical
  Review E\/} {\bf 85}, 031136 (2012).

\bibitem{marzen16predictive_infMarkovProc}
S.~E. Marzen, J.~P. Crutchfield, {\it Journal of Statistical Physics\/} {\bf
  163}, 1312 (2016).

\end{thebibliography}
\end{document}